%% file: Asymptotic.tex
\documentclass[11pt]{article}
\usepackage{jheppub}
\usepackage[utf8]{inputenc}
\usepackage[english]{babel}
\usepackage[T1]{fontenc}
\usepackage{amsmath,amssymb,graphicx, latexsym, theorem, enumerate,bbm}
\usepackage{booktabs, multirow, nicefrac, float, xspace, comment}
\usepackage{xcolor}
\usepackage{slashed}

\usepackage{tikz}
\usetikzlibrary{positioning, backgrounds, scopes, 3d, calc, shapes.geometric, arrows.meta}

\usepackage[compat=1.1.0]{tikz-feynman}
\tikzfeynmanset{compat=1.1.0, warn luatex=false}

\hypersetup{
    linktocpage,
     colorlinks,
     citecolor=darkgreen,
     linkcolor= darkgreen,
     urlcolor=darkgreen
}

\definecolor{darkred}{rgb}{0.8,0.0,0.0}
\definecolor{darkblue}{rgb}{0.0,0.0,0.9}
\definecolor{darkgreen}{rgb}{0.0,0.6,0.0}

\newcommand{\blue}{\color{darkblue}}
\newcommand{\green}{\color{darkgreen}}
\newcommand{\black}{\color{black}}

\newcommand{\Li}{\ensuremath{\operatorname{Li}}}
\newcommand{\cB}{\mathcal{B}}
\newcommand{\cO}{\mathcal{O}}
\newcommand{\cOe}{\cO(\eps)}

\newcommand{\looseoverset}[3][1ex]{%
  \mathrel{\vcenter{\hbox{%
    \begingroup
      \setlength{\arraycolsep}{0pt}
      $\begin{array}{c}
         \scriptstyle #2\\[#1]  
         #3                      
       \end{array}$%
    \endgroup
  }}}%
}

\newcommand{\nobracket}{}

\newcommand{\oneA}{\ensuremath{\boldsymbol{1_A}}}
\newcommand{\oneG}{\ensuremath{\boldsymbol{1_G}}}
\newcommand{\oneGzero}{\ensuremath{\boldsymbol{1_{G_0}}}}
\newcommand{\epsco}{{\ensuremath{\boldsymbol{\epsilon}}}}
\newcommand{\id}{{\ensuremath{\operatorname{id}}}}
\newcommand{\Dtilde}{\widetilde{\Delta}}
\newcommand{\bdelta}{\ensuremath{\boldsymbol{\delta}}}
\newcommand{\bgamma}{\ensuremath{\boldsymbol{\gamma}}}
\newcommand{\bgammap}{\ensuremath{\boldsymbol{\gamma_P}}}
\newcommand{\bgammaP}{\bgammap}
\usepackage{cancel,ulem}
\newcommand{\msbar}{\ensuremath{\overline{\operatorname{MS}}}}
\newcommand{\Rbar}{\overline{R}}
\newcommand{\eps}{\varepsilon}

\newcommand{\mathbbbl}{%
  \mathchoice{\mathbbbl@{1.00}}{\mathbbbl@{0.85}}{\mathbbbl@{0.75}}{\mathbbbl@{0.60}}}

\newcommand{\circledbbL}{%
  \mathord{%
    \tikz[baseline=(c.base)]{
      \node[draw,circle,minimum size=0.3pt,inner sep=0.5pt,line width=.09pt,scale=0.7] (c) {$\mathbb{L}$};
    }}}
    
\newcommand{\LL}{{\mathbb{L}}}
 \newcommand{\LLL}{{\circledbbL}}

\newcommand{\tripleL}{\mathrel{\ooalign{\hss$\mathbb{L}$\hss\cr
                                     \hss\kern1.0pt$\mathbb{L}$\hss}}}
\newcommand{\third}{ {\frac{1}{3} }}

\definecolor{LuenPurple}{HTML}{800080} 

\input{tikzdiagrams.tex}

\title{\boldmath Asymptotic Behavior of Diagram Classes}

\author{Luen Clingerman and}
\author{Matthew D. Schwartz}
\affiliation{Department of Physics, Harvard University , 02138 Cambridge, MA, USA}
\emailAdd{lmalshi@g.harvard.edu}
\emailAdd{schwartz@g.harvard.edu}

\abstract{
The asymptotic nature of perturbative expansions in quantum field theory can arise 
from the factorial growth in the number of Feynman diagrams with loop order, as with instantons,
 or from a series of individual diagrams whose values grow factorially, as with renormalon chains in QED. 
 Other classes of diagrams are known
 also to grow factorially, such as the Hopf series of graphs in $\phi^3$ theory. 
 This Hopf series was studied using Schwinger-Dyson equations and the Connes-Kreimer Hopf algebra of decorated rooted trees. We review
 the Hopf algebra approach and show that the same results can be obtained using
 analytic QFT techniques as with Hopf-algebraic ones. We present an efficient method to extract the asymptotic behavior and thereby generalize the analysis of the Hopf series to other
 classes of diagrams in other theories.  We confront the question of whether
 these classes correspond to new types of asymptotic growth beyond instantons and renormalons, and find that they appear
 to be incomplete calculations of what would be renormalons in these theories if all diagrams
 were included. Although the Hopf algebra approach is not essential to deriving asymptotic 
 behavior from Schwinger-Dyson equations, it does provide some other insights into quantum field theory.
 We therefore attempt also to provide a map between some relevant aspects of the Hopf algebra
and quantum field theory. }

\begin{document}

\maketitle
\flushbottom

\newpage

\newpage
\section{Introduction}
Understanding the origin of asymptotic growth of series expansions in quantum field theory is critically important to characterize their predictive power.
While much is known about these series, from general arguments~\cite{Dyson:1952,tHooft:1979,Bender:1999box,Marino:2015yie}, from perturbative series~\cite{Bender:1969si,Bender:1973rz}, and from toy integrable systems~\cite{Gross:1974jv,Polyakov:1975rr}, much remains to be
understood~\cite{Beneke:1998ui}. From early work in the 1970s, two sources of asymptotic growth have been observed: instantons, associated with saddle points of the Euclidean path integral, and renormalons, associated with anomalous scale invariance.
Alternatively, instantons arise from the factorial growth of the number of diagrams at a given loop order
($\int d x e^{-x^2} \frac{1}{n!}(x^4)^n \sim n!$), while renormalons arise from the integration over loop momenta $\left( \int d x \ln^n x \sim n! \right)$. 
It is not clear if other sources of asymptotic growth can exist in quantum field theories. 
For example, in~\cite{Broadhurst:1999ys,Broadhurst:2000dq,Borinsky:2021hnd}  a set of ``chain''
self-energy diagrams in $\lambda \phi^3$ theory was found to lead to growth of
the form $\sum_n (-\frac{1}{6})^n n! a^n$ at large $n$ with $a=\lambda^2/(4\pi)^3$ while an alternative series of
``Hopf'' graphs produced stronger behavior $\sum_n(-\frac{1}{3})^n n! a^n$. The
chain graphs look superficially like the set of vacuum polarization graphs
leading to $\beta_0$ and the leading renormalon in QED, while the Hopf graphs are qualitatively different and have stronger asymptotic
growth. Determining how to classify these diagrams: renormalons,
instantons, or something else, was the first motivation for
the current work.

The calculations of the Hopf series in~\cite{Broadhurst:1998ij,Broadhurst:2000dq,Borinsky:2021hnd} rely heavily on techniques from the Connes-Kreimer Hopf algebra of decorated rooted trees~\cite{Kreimer:1997dp,Connes:1998qv,
Connes:1999yr,Connes:1999zw}. This algebra is a beautiful
mathematical framework whose applications in physics include understanding
renormalizability, Schwinger-Dyson equations, and the renormalization group.  
On the mathematical side, there has been steady progress in translating physics concepts such as renormalizability and the renormalization group, into algebraic language~\cite{Connes:1999yr, Connes:1999zw,OlsonHarris:2025dse,Balduf:2025fjp}. 
Unfortunately, after 30 years the Connes-Kreimer Hopf algebra has still not percolated very deeply into the physics community.~\cite{Weinzierl:2022eaz}.
Thus, a second goal of this paper is to dissect what the Hopf algebra can accomplish efficiently enough to impress a physicist, and  conversely, what aspects of it might seem to a physicist to be rather inelegant.
To be concrete, we ask whether or not the Hopf algebra approach simplifies the computation of the leading asymptotic behavior of the Hopf series and the determination of the associated
renormalization group equation.

Historically, the asymptotic growth of the Hopf series has {\it only} been computed with Hopf algebraic techniques. So to begin, we demonstrate that in fact the Hopf algebra is not needed. In Section~\ref{sec:phichi}, we reproduce the results for the sum of the Hopf, chain and rainbow graphs in $\phi^3$ theory without reference to the Hopf algebra at all. 
More concretely, the approach of~\cite{Broadhurst:1999ys,Broadhurst:2000dq} was to combine a renormalization group equations (RGEs) derived from the Lie algebra of the group of Hopf-algebraic characters 
with a combinatorial Schwinger-Dyson equation
to produce an ordinary differential equation for the anomalous dimension which can be solved iteratively. 
We show first of all that the RGE can be derived more directly from the Schwinger-Dyson equation by analyzing its scale dependence without the digression into the Hopf algebra. Second we show that the asymptotic growth can be determined without the RGE at all, using instead leading-logarithmic resummation as in the study of renormalons in QFT.
In fact, all of the  singularities in the Borel plane can be determined this way, not just the one closest to the origin.
This observation allows us to extract the asymptotic behavior in a number of other cases which we present in Section~\ref{sec:other}. These cases include the double-arm $\phi^3$ case in $d=6$, which comprises all $\phi^3$ self-energy graphs without vertex corrections, and various classes of graphs in $\phi^4$ theory in $d=4$. 

After having extracted the asymptotic behavior of a number of examples using QFT rather than Hopf-algebraic techniques, we we proceed to examine to what extent the Hopf algebra simplifies or does not simplify these calculations. For this end, we include in Section~\ref{sec:Hopf} a mini review aimed at an audience of physicists with some sympathy for mathematics but no prior mastery of the Kreimer-Connes Hopf algebra.  Rather than provide rigorous proofs, we attempt instead to emphasize some elements of the Hopf algebra which we find most elegant or
think may be most relevant to physics. We concede at the outset that this review is incomplete,
and hope only that our subjective treatment may slightly lower the barrier to entry for some readers.
With the basics of the Hopf algebra and our notation established, we proceed to compare and contrast some elements of the Hopf algebra with traditional QFT. For example, explore how the
renormalization group arises in both approaches. We examine to what extent the Hopf algebra approach to renormalization, which closely parallels the pioneering work of Bogoliubov, Parasiuk, Hepp and Zimmermann (BPHZ)~\cite{bogoliubov:1957, hepp:1966,zimmermann:1969}, is equivalent to multiplicative
renormalization used in QFT. Although their equivalence in a general local quantum field theory is well known, what was not clear was whether the equivalence still holds if only a subset of all diagrams in a theory are included. This question is explored in Section~\ref{sec:isHM}.
Conclusions are given in Section~\ref{sec:conclusions}.

\section{Chain, Rainbow and Hopf graphs \label{sec:phichi}}
In this section, we study the chain, rainbow and Hopf graphs in scalar field theory in $d=6$. 
We derive various integro-differnetial, partial-differential
and ordinary-differential equations
that they satisfy and
analyze their asymptotic behavior.
Many of the results of this section are contained in Ref.~\cite{Borinsky:2021hnd}
which  builds on the work of Broadhurst and Kreimer~\cite{Broadhurst:1999ys}. 
These references used a Hopf-algebra to construct an ordinary differential equation (see Eq.~\eqref{biggamma} below)
which could be solved iteratively. 
We provide instead a self-contained derivation of the relevant differential equation and Borel singularities using QFT and renormalon methods. 
By confirming the location of the Borel singularities, we 
 validate our methodology which we then use to analyze other diagram sums in Section~\ref{sec:other}.
In addition, we argue that these Borel singularities are more accurately characterized as renormalons than as instantons which is the terminology use in Ref~\cite{Borinsky:2021hnd}.

\subsection{Notation}
Asymptotic series at large order $n$ in
a coupling $a$ have the form
\begin{equation}
  \Sigma (a) = \sum_n c_n a^n, \quad c_n = C n^{- r} A^n n!,
\end{equation}
for some $C,A$ and $r$. 
At large $n$
\begin{equation}
    \ln c_n \sim n \ln n+ n \ln A - n - r\ln n + \ln C+ \cdots,
\end{equation}
so that the value of $A$ determines the dominant behavior at large $n$ and the value
of $r$ affects only subleading behavior. The Borel transform of $\Sigma (a)$ is
\begin{equation}
  B (t) = \sum_{n = 1}^{\infty} \frac{c_n}{n!}  t^n = C \text{Li}_r (A t),
\end{equation}
which has a singularity at $t = A^{-1}$ with the strength of the singularity
depending on $r$. More details on properties of the Borel transform are in Appendix~\ref{app:Borel}. 
For the dominant behavior we only need to determine the locations of singularities in the Borel plane but not their precise form (e.g.
logarithmic or algebraic). For example, the location of renormalon Borel
singularities in QED are determined by the leading order $\beta$ function
coefficient $\beta_0$, while the next-to-leading order $\beta$ function
coefficient $\beta_1$ affects the strength of the singularities but not their
locations~\cite{Grunberg:1995vx}.

We consider in this section a theory with two massless real
scalar fields $\phi$ and $\chi$ and Euclidean action
\begin{equation}
  S = \int d^6 x\left[ \frac{1}{2} (\partial \phi)^2 + \frac{1}{2}
  (\partial \chi)^2 - \frac{\lambda}{2} \phi^2 \chi\right].
\end{equation}
 In $d = 6$ dimensions the coupling $\lambda$ is dimensionless and this theory is renormalizable.
Having two fields instead of a single field (as was used in Ref.~\cite{Borinsky:2021hnd}) allows us to
isolate the diagrams of interest more easily. Although we work in Euclidean signature for simplicity, all the results we derive for the asymptotic behavior hold in Lorentzian signature as well. 

We are interested in the perturbative computation of the renormalized $\phi$ propagator in this theory:
\begin{equation}
  G_{\phi} (p) = \int d^6 x e^{i p x} \langle 0| \phi (x) \phi (0)| 0 \rangle = \frac{1}{p^2} + \cdots.
\end{equation}
We write $G_\phi^0$ for the equivalent bare Green's function.
We define $\Sigma^0$ as the sum of the bare 1PI contributions. So $G^0_\phi = \frac{1}{p^2}\frac{1}{1-\Sigma^0}$. We also define
\begin{equation}
    L=\ln\frac{p^2}{\mu^2},
\end{equation}
so $\partial_L = -\frac{1}{2} \mu \partial_\mu$. 
The anomalous dimension is defined as
\begin{equation}
    \gamma = - \frac{d}{dL} \ln [p^2 G_\phi(L)] = \frac{d}{dL} \ln [1-\Sigma(L)].
    \label{gammadef}
\end{equation}
In the momentum subtraction scheme where $\Sigma(L)=0$ at $L=0$ we also have $\gamma = -(\partial_L \Sigma)_{L=0}$.

At 1-loop, in dimensional regularization the bare graph evaluates to
\begin{align}
\Sigma_1^0 &=\;
\bubblediagramlabel
= \lambda^2 \mu^{6 - d} \int \frac{d^d k}{(2 \pi)^d} 
  \frac{1}{(p + k)^2} \frac{1}{k^2} \frac{1}{p^2} 
  \\
  &= \frac{\lambda^2}{(4
  \pi)^3} \left( \frac{p^2}{4 \pi \mu^2} \right)^{\frac{d - 6}{2}}
  \frac{\Gamma \left( \frac{d}{2} - 1 \right)^2 \Gamma \left( 2 - \frac{d}{2}
  \right)}{\Gamma (d - 2)}.
\end{align}
This graph is quadratically divergent in $d = 6$ as evidenced by the pole at
$d = 4$.  The quadratic divergence can be absorbed into a mass correction for $\phi$ so that its
renormalized mass remains zero. The remaining logarithmic divergence shows up
by expanding in $d = 6 - 2 \varepsilon$ dimensions,
\begin{equation}
  \Sigma_1^0 = \frac{a}{ 6} \left[ - \frac{1}{\eps} + \ln \frac{p^2}{\tilde{\mu}^2} - \frac{8}{3} +
  \mathcal{O} (\eps) \right],
\end{equation}
with $a \equiv \frac{\lambda^2}{(4 \pi)^3}$ and ${\tilde \mu}^2 \equiv \mu^2 4\pi e^{-\gamma_E}$. 
In the $\msbar$ subtraction scheme, the Green's function $G_\phi(p) = \frac{1}{p^2}\frac{1}{1-\Sigma}$ to 1-loop is then
\begin{equation}
  G_{\phi} (p) = \frac{1}{p^2} \left( 1 + \frac{a}{6}  \left(L - \frac{8}{3}
  \right) + \cdots \right) \quad \left( \msbar \right).
\end{equation}
Instead of $\msbar$ it is often convenient to work in the momentum
subtraction scheme  (MOM) defined so that $G_{\phi} (p) = \frac{1}{p^2}$ to all
orders at $p^2 = \mu^2$ $(L = 0)$. Then
\begin{equation}
  G_{\phi} (p) = \frac{1}{p^2} \left(1 + \frac{a}{6} L + \cdots\right) \quad  \text{(MOM)}.
\end{equation}
The leading-order anomalous dimension $\gamma = -\frac{a}{6}$ is scheme independent, but higher order contributions to $\gamma$ are not.

\subsection{Chain Graphs\label{sec:chains2} }
We would now like to consider three infinite series of 1PI Feynman diagrams in this theory, following~\cite{Broadhurst:1998ij,Broadhurst:2000dq,Borinsky:2021hnd}. The first are the chain diagrams. The bare chain diagrams are
\begin{equation}
  \Sigma_C^0 (p^2) = 
  \bubble  +  \doublebubble  +  \chain  + \chainwide + \cdots.
\end{equation}
These diagrams are given by a geometric series of insertions into the
$\phi$ propagator. This series sums into 
\begin{equation}
  \Sigma_C^0 (p^2) = \frac{a}{\pi^3} \int d^6 k \frac{1}{p^2 k^2 (p + k)^2}
  \frac{1}{1 - \Sigma^0_1 (k^2)},
\end{equation}
which is a type of Schwinger-Dyson equation (SDE).

To solve the SDE we note that this is a single scale problem, so $\Sigma_C^0(p^2)$ depends only on $L = \ln \frac{p^2}{\mu^2}$. Then
every integral is the sum of terms like  
\begin{equation}
  \frac{a}{\pi^3} \int d^6 k 
  \frac{1}{p^2 k^2 (p + k)^2} 
  \ln^n\!\frac{k^2}{\mu^2} = \lim_{\rho \rightarrow 0} \partial_{\rho}^n F (\rho, L),
\end{equation}
where $F (\rho, L)$ is the analytically-regulated integral
\begin{equation}
  F (\rho, L) \equiv \frac{a}{\pi^3} \int d^6 k \frac{1}{p^2 k^2 (p + k)^2} \left(
  \frac{k^2}{\mu^2} \right)^{\rho}.
  \label{Fd6}
\end{equation}
In analytic regulation, one often evaluates integrals by contour deformation which requires identifying the convergent regions in the complex $\rho$ plane. This integral converges in the UV (large
$k^2$ region) when $\rho < - 1$ and converges in the IR (small $k^2$ region)
when $\rho > - 2$. So it is convergent in a strip where $- 2 < \text{Re} \rho
< - 1$. 
Doing the integral in this region using Schwinger parameters and
analytically continuing to the physical region where $\rho \approx 0$ gives
\begin{equation}
  F (\rho, L) = \frac{ a \mu^{- 2 \rho} }{ p^2 \Gamma (1 - \rho)} \int_0^1 d x
  x^{- \rho} \int_0^{\infty} d \lambda \lambda^{- \rho - 2} e^{- \lambda x (1
  - x) p^2} = a \frac{e^{\rho L}}{\rho P (\rho)},
\end{equation}
where
\begin{equation}
  P (\rho) = (\rho + 3) (\rho + 2) (\rho + 1).
\end{equation}
The poles in $F(\rho,L)$ at $\rho = 0$ and $- 1$ are of UV origin while the poles at $\rho = - 2$ and $-3$ are of IR origin. In the region with $\rho \approx 0$ the divergences are
UV and removed with counterterms. In the momentum subtraction scheme these
counterterms subtract the amplitude at $L = 0$. Using also that $\Sigma_1 =
\frac{a}{6} L$ 
in MOM we then find
\begin{equation}
  \Sigma_C (L) = \lim_{\rho \rightarrow 0} \frac{a}{\left( 1 - \frac{a}{6}
  \partial_\rho \right)} \frac{e^{\rho L} - 1}{\rho P (\rho)}.
\end{equation}
Acting on both sides with $P (\partial_L) \partial_L$ then gives
\begin{equation}
  P (\partial_L) \partial_L \Sigma_C = \lim_{\rho \rightarrow 0}
  \frac{a}{\left( 1 - \frac{a}{6} \partial_\rho \right)} e^{\rho L} = \frac{a}{1 - \frac{a}{6} L}.
\end{equation}
One can solve this 4th order ordinary differential equation for $\Sigma_C
(L)$ by integration. However, it is more instructive to solve it by Borel
transform. Using
\begin{equation}
  \frac{a}{1 - \frac{a}{6} L} = \int_0^{\infty} d t e^{- \frac{t}{a}}
  e^{\frac{t}{6} L},
\end{equation}
gives 
\begin{equation}
  \partial_L\Sigma_C = \int_0^{\infty} d t \mathcal{B} (t) e^{- \frac{t}{a}}, \quad
  \mathcal{B} (t) = \frac{1}{P \left(\frac{t}{6} \right) }
  e^{\frac{t}{6} L}.
  \label{chainsol}
\end{equation}
This says that the Borel transform of the chain graph self-energy sum\footnote{More precisely, $\cB(t)$ is the Borel transform of $\frac{1}{a} \partial_L \Sigma_C$,
but the leading asymptotic behavior of $f(a,L)$ and $\frac{1}{a} \partial_L f(a,L)$ are the same.}
 has poles at $t_i =6 \rho_i = -6, -12, -18$ where $\rho_i = -
1, - 2, - 3$ are the zeros of $P (\rho)$.
Based on the nature of the
singularities of $F (\rho)$, we can also conclude that the closest pole to the
origin, at $t = -6$ is of UV origin, while the $t = -12$ and $t=-18$ poles are
of IR origin.

\subsection{Rainbow Graphs \label{sec:rainbow2}}
The rainbow graphs consist of concentric self-energy bubbles:
\begin{equation}
  \Sigma_R^0 (p^2) =
  \bubble
  +\doublebubble
  +\rainbow
  +\cdots.
\end{equation}
In this case, the Schwinger-Dyson equation is 
\begin{equation}
  \Sigma_R^0 (p^2) = \frac{a}{\pi^3} \int d^6 k \frac{1}{p^2 k^2 (p + k)^2} [1 +
  \Sigma_R^0 (k^2)].
\end{equation}
In the momentum subtraction scheme this can be written as
\begin{equation}
  \Sigma_R (L) = a \lim_{\rho \rightarrow 0} [1 + \Sigma_R (\partial_\rho)]
  \frac{e^{\rho L} - 1}{\rho P (\rho)},
\end{equation}
which implies
\begin{equation}
  P (\partial_L) \partial_L \Sigma_R (L) = a [1 + \Sigma_R (L)].
  \label{RainbowSDE}
\end{equation}
This is a linear ODE and is solved by Laplace transform. Writing $\Sigma_R =
e^{-\gamma L} - 1$ gives
\begin{equation}
-  P (-\gamma)\gamma e^{-\gamma L} = a e^{-\gamma L}.
\end{equation}
So that $-P (-\gamma) \gamma = a$ which is a polynomial equation with roots 
$\gamma = \frac{1}{2} \left( 3 \pm\sqrt{5 \pm 4 \sqrt{1 + a}} \right)$. Picking the root with the correct boundary condition ($\Sigma = 0$ at $a=0$) gives
\begin{equation}
  \Sigma_R(L) = \exp \left[ \frac{L}{2} \left( - 3 + \sqrt{5 + 4 \sqrt{1 + a}}
  \right)  \right] - 1.
\end{equation}
This is an analytic function of $a$ which does not grow asymptotically at large orders in $a$.

\subsection{Hopf Graphs \label{sec:hopffirst}}
Next we consider the Hopf graphs. These include chain and rainbow graphs and
generally any self-energy bubble insertion anywhere into the $\phi$ leg 
\begin{equation}
  \Sigma_H^0 =
  \bubble + \doublebubble+ \chain + \rainbow + \cdots.
\end{equation}
In this case, the Schwinger-Dyson equation is 
\begin{equation}
  \Sigma_H^0 (p^2) = \frac{a}{\pi^3} \int d^6 k \frac{1}{p^2 k^2 (p + k)^2}
  \frac{1}{(1 - \Sigma_H^0 (k^2))}. \label{HopfSDE}
\end{equation}
So that the MOM-renormalized sum of Hopf graphs is
\begin{equation}
  \Sigma_H (L) = \lim_{\rho \rightarrow 0} \frac{a}{\left( 1 - \frac{a}{6}
  \Sigma_H (\partial_\rho) \right)} \frac{e^{\rho L} - 1}{\rho P (\rho)}.
\end{equation}
This can be solved iteratively:
\begin{equation}
  \Sigma_H (L) = \frac{a}{6} L + \frac{a^2}{36} \left( \frac{L^2}{2} -
  \frac{11}{6} L \right) + \frac{a^3}{6^3} \left( \frac{L^3}{2} - \frac{11}{3}
  L^2 + \frac{94}{9} L \right) + \cdots. \label{SigmaH3loop}
\end{equation}
We can also multiply by $P (\partial_L) \partial_L$ as with the chain graphs
to get
\begin{equation}
  P (\partial_L) \partial_L \Sigma_H (L) = \frac{a}{1 - \Sigma_H (L)}.
  \label{HopfSDEB}
\end{equation}
This is a much more complicated differential equation than the previous ones
and we are unable to solve it in closed form. But a closed form solution is not needed to find the asymptotic behavior. 

To find the asymptotic behavior for the Hopf series, we take inspiration from renormalon techniques (see~\cite{Beneke:1998ui} for a review). To find the leading renormalon
for, say, the Adler function in QED, one first sums the leading logarithmic
series of vacuum polarization graphs, then plugs these $\alpha^n L^n$ terms
back into the outer fermion loop  for example). To apply that approach here,
we first find the leading-logarithmic approximation, i.e. the terms $a^n L^n$
in $\Sigma_H$. Since acting with $\partial_L$ on such a series will change the
homogeneity in $a L$, we can drop any term in $P (\partial_L)$ other than a
constant. With this truncation the SDE reduces to
\begin{equation}
  6 \partial_L \Sigma^\LL_H (L) = \frac{a}{1 - \Sigma_H^\LL (L)}.
\end{equation}
This equation is homogeneous in $a L$ so it will give the leading log solution.
With boundary condition $\Sigma_H (0) = 0$ its solution is
\begin{equation}
  \Sigma^{\LL}_H(L) = 1 - \sqrt{1 -\frac{a}{3} L}.
\end{equation}
 Plugging this back into the full SDE gives
\begin{equation}
  P (\partial_L) \partial_L \Sigma_H^\LLL (L) = \frac{a}{\sqrt{1 - \frac{a}{3}
  L}} =\frac{\sqrt{a}}{\sqrt{\pi}} \int_0^{\infty} d t e^{- \frac{t}{a}}
  \frac{e^{\frac{ t}{3} L}}{\sqrt{t}},
\end{equation}
where the $\LLL$ superscript refers to the approximation we are using (inserting the leading log resummed result back into the SDE). Imposing the MOM renormalization conditions results in 
\begin{equation}
  \Sigma_H (L) \sim\Sigma_H^\LLL (L) =\sqrt{a} \int_0^{\infty} d t e^{- \frac{t}{a}} t^{-3/2} \cB_p(t) ,
  \qquad
  \cB_p(t) =\frac{3}{\sqrt{\pi}}  \frac{e^{\frac{t}{3} L}-1}{P \left( \frac{t}{3} \right)}.
\end{equation}
where $\sim$ means up to subleading asymptotic behavior. 
The integral is of the form of a Borel-Leroy transform. A Borel-Leroy transform is like a Borel transform with a different power of $a$ out front. Its singularites are at the same locations as for the Borel transform (see Appendix~\ref{app:Borel} for more details).   

The final result for the Hopf case is qualitatively similar to the chain case: three Borel poles. However, the Borel poles in the Hopf case are twice as close to the origin as in the chain case, at $t = 3 \rho_i = - 3, - 6, - 9$. We can verify that the correct asymptotic behavior is achieved with this approach by comparing to~\cite{Broadhurst:2000dq,Borinsky:2021hnd} or to the RGE-based method which we review next. 

\subsection{Renormalization group equations}\label{sec:RGEQFT}

In quantum field theory, local operators and Green's functions satisfy renormalization group equations (RGEs) determined by the scale dependence of the field strength
and coupling $Z$-factors. The form of the RGE is highly constrained by the locality of the theory. Therefeore, one has no reason to expect that a generic incomplete selection of Feynman diagrams and
counterterms contributing to a Green's function will also satisfy an RGE. However, if we hope to
associate the asymptotic growth of a selection of diagrams with the $\beta$
function, we must be able to associate some sort of RGE to the relevant
selection. Here we present two methods of deriving these RGEs.
Further discussion of the RGEs and their relation to local rormalization can be found in Section~\ref{sec:isHM}.

\subsubsection{Callan-Symanzik Equation \label{sec:CSE}}
The first way to derive the RGE is as in the Callan-Symanzik equation. The
Callan-Symanzik equation is derived from $\frac{d}{d \mu} G_{\phi}^0 = 0$ by rewriting the bare
Green's function $G_{\phi}^0$ in terms of renormalized
quantities. To renormalize, the bare fields are rescaled as $\phi_0 = \sqrt{Z}\phi$ to get $G_{\phi}^0 = \langle \phi_0 \phi_0 \rangle = Z \langle \phi \phi \rangle = Z G_{\phi}$ and the coupling is rescaled as $a_0 = a Z_a$. Both $Z$ and $Z_a$ are functions of the renormalized coupling $a$.
The general relation between the bare Green's function $G^0$ and the bare 1PI self-energy graphs $\Sigma^0$ is
\begin{equation}
  G_{\phi}^0 (a_0, p) = \frac{1}{p^2} \frac{1}{1 - \Sigma^0 (a_0, p)} = Z 
  \frac{1}{p^2} \frac{1}{1 - \Sigma (a Z_a, p)} = Z G_{\phi} (Z_a a, p)
  \label{hopfnoflip}.
\end{equation}
To implement these substitutions into SDEs, let us introduce the notation $1 - \Sigma^0_H = Y_H^0 = [p^2G_\phi^0]^{-1}$ and the bare coupling $a_0$. Then the bare Hopf SDE in Eq.~\eqref{HopfSDE} becomes
\begin{equation}
     Y^0_H (p^2) = 1- \frac{a_0}{\pi^3} \int d^d k \frac{1}{p^2 k^2 (p + k)^2}
  \frac{1}{Y^0_H(k^2)}.
  \label{bare1}
\end{equation}
After rescaling $\phi_0 =\sqrt{Z} \phi$ so that $Y_H^0 = \frac{1}{Z} Y_H$,
we then get
\begin{equation}
  Y_H (p^2) = Z- Z^2 \frac{a_0}{\pi^3} \int d^d k \frac{1}{p^2 k^2 (p + k)^2}
  \frac{1}{Y_H(k^2)}.
  \label{renorm}
\end{equation}
Now, assume that up to order in $a_0^n$, $Y_H(p^2)$ is finite. In the BPHZ proof of renormalizability~\cite{bogoliubov:1957,hepp:1966,zimmermann:1969},
a key lemma is if a divergent integral has no subdivergences then its divergences will be local.
Local divergences are ones that can have only
polynomial momentum dependence so they can be canceled by local counterterms. (An excellent
review of the BPHZ proof and this lemma can be found in~\cite{Collins:1984xc}.)
In this case, since $Y_H(p^2)$ is finite at order $a_0^n$, the integral will have no subdivergences up to order $a_0^n$ and therefore its divergences will be local. That means the integral can have divergences
$\eps^{- k}$ and non-analytic pieces $\ln^j(p^2)$
but not both at once (no terms like $\eps^{-k}\ln^j(p^2)$). Local divergences can be removed by $Z$ on the right-hand side. 
However, the $Z^2$ factor, which depends on $\eps^{- 1}$ can spoil this locality by multiplying $\ln^j(p^2)$ terms from the integral.
Conveniently, we can rescale the coupling to eliminate the $Z^2$ by writing
$a_0 = a \mu^{d-6} Z_a$
 with $Z_a = Z^{-2}$.
  The factor of $\mu^{d-6}$ is introduced conventionally to make the renormalized coupling $a$ dimensionless.
 After this rescaling, all the divergences in the integral are
removed with the remaining additive factor of $Z$, order-by-order in $a$.\footnote{If we also rescale the $\chi$ field, then $a_0 = a \mu^{d-6} \frac{Z_v^2}{Z^2 Z_\chi}.$ If we were to account for
 all vertex and propagator corrections,
the $Z_v$ term would be necessary to cancel extra $Z_v$ factors from the vertex corrections and $Z_\chi$ would be necessary to cancel $Z_\chi$ factors from the $\chi$ propagator. In our case there are no vertex or $\chi$ propagator counterterms so $Z_v = Z_\chi = 1$, and the net rescaling is $a_0 = a \mu^{d-6} 
 Z^{- 2}$.}  
 In other words, the additive renormalization used in the Hopf SDE does correspond to multiplicative renormalization $\phi \to \sqrt{Z} \phi$ as long as $a_0 \to \mu^{d-6} Z^{-2} a$ as well.  
 
 Then we can obtain an RGE for $\Sigma_H$ by writing 
\begin{equation}
    \mu\frac{ d}{d\mu}\big(1-\Sigma^0_H (a_0, p^2) \big) =  \mu\frac{ d}{d\mu} \frac{1}{Z}\big(1-\Sigma_H (\mu^{d-6} Z^{-2} a, L)\big) = \big(-2\partial_L + 2  \beta \partial_a +2 \gamma \big) (1-\Sigma_H), 
    \label{newrge0}
\end{equation}
where $\partial_L = -\frac{1}{2} \mu \partial_\mu$, $\gamma = -\frac{1}{2} \mu \frac{d}{d\mu} \ln Z$ 
with
\begin{equation}
    \beta =-\partial_L a =- \partial_L (a_0 Z^{2}\mu^{6-d})= \frac{1}{2}\mu \partial_\mu(a_0 Z^{2}\mu^{6-d}) \overset{d \rightarrow 6}{=}  -2 a\gamma. 
\end{equation}
A relation between $\beta$ and $\gamma$ like this is common in SDEs. Because $1-\Sigma^0_H$ is a bare quantity and therefore has no $\mu$ dependence the LHS of Eq.~\eqref{newrge0} is 0. We then find
\begin{equation}
    \bigl(\partial_L +2 \gamma a \partial_a - \gamma \bigr)(1- \Sigma_H) = 0,
    \label{newrge}
\end{equation}
which is an RGE for the renormalized Hopf graph self-energy function $\Sigma_H$.

We can follow a similar procedure for the rainbows. An important caveat is that for the rainbow graphs to be consistently treated with the Callan-Symanzik approach, we must only include 1PI graphs into the Green's function, rather than the entire geometric series of graphs. That is, instead of Eq.~\eqref{hopfnoflip} we take
\begin{equation}
  G_{\phi}^0 (a_0, p) = \frac{1}{p^2} \big(1 + \Sigma^0 (a_0, p)\big) = Z 
  \frac{1}{p^2} \big(1 + \Sigma(a Z_a, p)\big) = Z G_{\phi} (Z_a a, p).
\label{rainbowflip}
\end{equation}
We will justify this step shortly. 
Defining $ Y^0_R = 1 + \Sigma^0_R$, the bare rainbow graph SDE becomes
\begin{equation}
    Y^0_R (p^2) = 1+ \frac{a_0}{\pi^3} \int d^d k \frac{1}{p^2 k^2 (p + k)^2} Y^0_R(k^2).
    \label{rainbowrge}
\end{equation}
Because this equation is linear in $Y_R^0$ if we renormalize as $Y_H^0 = ZY_H$, we get
\begin{equation}
  Y_R (p^2) = \frac{1}{Z} + \frac{a_0}{\pi^3} \int d^d k \frac{1}{p^2 k^2 (p + k)^2}  Y_R(k^2).
\label{renrainbowrge}
\end{equation}
This already has the form of an additive subtraction required for local counterterms without renormalizing $a_0$. All we need in this case is then to write $a_0 = \mu^{d-6}a$ and the RGE is simply
\begin{equation}
     \bigr(\partial_L  + \gamma \bigl)(1+ \Sigma_R) = 0,
\end{equation}
with $\gamma = -\frac{1}{2} \mu \frac{d}{d\mu} \ln Z$ and  $\beta=0$.

Note that for both the Hopf and the rainbow cases, at leading order in $a$ we get $\partial_L \Sigma = - \gamma$. 
Now, to justify using Eq.~\eqref{rainbowflip}, we note that 
that if Eq.~\eqref{hopfnoflip} were used instead, the RGE for the rainbow that resulted would have been $(\partial_L - \gamma)(1+\Sigma_R)=0$ leading to $\partial_L\Sigma = +\gamma$ at leading order. So for the anomalous dimension the RGE to be consistent between Hopf and rainbow, which have the same graphs up to 2 loops, we should use Eq.~\eqref{rainbowflip} for rainbow.  
To be clear, the RGEs we derived for the 1PI graphs $\Sigma_H$ and $\Sigma_R$ hold regardless of how we define the Greens's function. We are not forced to have the anomalous dimension defined consistently for Hopf and rainbow if each one is studied in isolation. However, the way the Green's function is defined is important if we are to interpret $Z$ as a local field strength renormalization factor and $\beta$ as an approximation to the $\beta$ function of the theory. We return to these points in Section~\ref{sec:isHM}.

\subsubsection{Wilsonian rescaling}
An alternative approach to deriving the RGE is to search for a scaling relation between the renormalized coupling and renormalized Green's function without ever mentioning the bare quantities. We call this a Wilsonian rescaling, to contrast it with the continuum RGE approach of the Callan-Symanzik equation.
\cite{Schwartz:2014sze}.

Using $Y_H (a, L) \equiv 1 - \Sigma_H (a, L)$,  Eq.~\eqref{HopfSDE} becomes
\begin{equation}
  P (\partial_L) \partial_L Y_H = \frac{a}{Y_H}.
\end{equation}
Similarly, for the rainbow graphs, define $Y_R \equiv 1 + \Sigma_R$, so that Eq.~\eqref{RainbowSDE} becomes
\begin{equation}
  P (\partial_L) \partial_L Y_R = a Y_R.
\end{equation}
So both are of the form
\begin{equation}
  P (\partial_L) \partial_L Y  = a Y^s, \label{sDE}
\end{equation}
with $s = 1$ for Rainbow and $s = - 1$ for Hopf.

The differential equation Eq.~\eqref{sDE} has two invariances: if $Y (a,L)$ is a solution then so is $Y(a, L + \delta L)$ for any $\delta L$ as well as
$\lambda Y(\lambda^{s - 1} a, L)$ for any $\lambda$. This means in particular that we can compensate for a small $\delta L$ by $\lambda=\exp(\delta\lambda)$  with a small $\delta\lambda$ so that the
solution is identical. In equations, 
\begin{equation}
  Y (a, L) = e^{\delta \lambda} Y (e^{(s-1)\delta \lambda}
  a, L + \delta L).
\end{equation}
To leading order $\delta \lambda$ and $\delta L$ we get
\begin{equation}
Y= (1 + \delta \lambda) Y  + \delta \lambda a (s - 1) \partial_a Y + \delta L \partial_L Y,
\end{equation}
so that
\begin{equation}
  (-\partial_L + \gamma a (s - 1) \partial_a +\gamma) Y =0,
  \label{RGEform1}
\end{equation}
where $\gamma \equiv -\frac{\delta \lambda}{\delta L}$. The MOM renormalization condition $Y=1$ at $L=0$  then gives
\begin{equation}
  \gamma =  \partial_L Y (a, L)_{L = 0}.
\end{equation}
It is conventional to write Eq.~\eqref{RGEform1} as
\begin{equation}
  (-\partial_L + \beta_a \partial_a + \gamma) Y = 0, \label{RGEX}
\end{equation}
where $\beta_a = \gamma a (s - 1)$. 
Alternatively, we could write this equation as in Eq.~\eqref{gammadef}
\begin{equation}
  \frac{d}{d L} \ln Y (a (L), L) = \gamma,
\end{equation}
where $\beta = -\frac{d a}{d L}$. This final form allows us to map $\beta$ to
the $\beta$-function of the theory in the Hopf or rainbow approximation and
$\gamma$ to the anomalous dimension of $\Sigma$. Explicitly, for Hopf which has
$s = - 1$ and $Y_H = 1 - \Sigma_H$, 
we find $\beta = -2 \gamma a$ and 
\begin{equation}
  (\partial_L +2 \gamma a\partial_a) \ln (1 - \Sigma_H) = \gamma.
  \label{HopfRGE2}
\end{equation}
For rainbow which has $s = 1$ we get $\beta = 0$.

\subsubsection{Combining the RGE and the SDE}
For the Hopf case, we can now combine the RGE and the SDE to get an ordinary
differential equation for $\gamma (a)$. To do so, let us write
\begin{equation}
  \Sigma_H = \sum_{n = 1}^{\infty} \sigma_n (a) L^n. \label{SigmaHseries}
\end{equation}
Then evaluating
Hopf SDE in Eq.~\eqref{HopfSDE} at $L = 0$ gives
\begin{equation}
  \lim_{L \rightarrow 0} P (\partial_L) \partial_L \Sigma_H (L) = a,
\end{equation}
or, more explicitly,
\begin{equation}
  \sigma_1 (a) + \frac{11}{3} \sigma_2 (a) + 6 \sigma_3 (a) + 4 \sigma_4 (a) =
  \frac{a}{6}, \label{sdsol}
\end{equation}
which relates the lowest 4 logarithmic coefficients in $\Sigma_H$. We can
relate these coefficients another way, using the RGE. We first rewrite Eq.
\eqref{HopfRGE2} as 
\begin{equation}
  ( \partial_L + 2 a \gamma \partial_a) \Sigma_H =- \gamma (1 - \Sigma_H). 
\end{equation}
Plugging in Eq.~\eqref{SigmaHseries} gives at order $L^0$
\begin{equation}
   \sigma_1 =- \gamma,
\end{equation}
and at order $L^n$
\begin{equation}
  \sigma_{n + 1} = \frac{1}{n + 1} (  \gamma\sigma_n- 2 a \gamma \sigma_n').
  \label{gamman1}
\end{equation}
In this way $\sigma_1, \ldots, \sigma_4$ are expressible in terms of $\gamma$.
Plugging Eq.~\eqref{gamman1} into Eq.~\eqref{sdsol} gives
\begin{multline}
   6 \gamma + 11 \gamma^2 + 6 \gamma^3 + \gamma^4 - 22 a \gamma \gamma' -
  12 a \gamma^2 \gamma' - 4 a \gamma^3 \gamma'
  + 24 a^2 {\gamma \gamma'}^2 - 4 a^2 \gamma^2 {\gamma'}^2
  - 8 a^3 {\gamma  \gamma'}^3 \\+ 24 a^2 \gamma^2 \gamma'' - 8 a^2 \gamma^3 \gamma'' - 32 a^3
  \gamma^2 \gamma' \gamma'' - 8 a^3 \gamma^3 \gamma''' = -a. \label{biggamma}
\end{multline}
This differential equation appears in Refs.~\cite{Broadhurst:1999ys,Borinsky:2021hnd} where it was derived using
Hopf-algebraic techniques which we review in Section~\ref{sec:Hopf}. 

One can use Eq.~\eqref{biggamma} to confirm the asymptotic behavior of the
anomalous dimension at large $a$. If we take as an ansatz the form $\gamma
\sim \sum K A^n a^n n!$ at large $n$ for some $K$ and $A$, then because
$\gamma$ begins at order $a$ only a small number of the terms in this equation
will contribute at large $n$. For example, the dominant term in $\gamma^2$
takes the order $a$ term from one $\gamma$ and the order $n - 1$ from the
other so it is down by a factor of $n$ compared to $\gamma$, The surviving
terms at large $n$ are
\begin{equation}
  6 \gamma - 22 a \gamma \gamma' + 24 a^2 \gamma^2 \gamma'' - 8 a^3 \gamma^3
  \gamma''' = 0.
\end{equation}
For our ansatz to solve this equation at large $n$, we need
\begin{equation}
  6 - 22 \frac{\gamma'(0)}{A} + 24 \frac{[\gamma'(0)[^2}{A^2} - 8
  \frac{[\gamma'(0)]^3}{A^3} = 0,
\end{equation}
where $\gamma'(0) = \partial_a\gamma$ at $a=0$.
Using $\gamma'(0) = -\frac{1}{6}$ then gives
\begin{equation}
  A = - \frac{1}{9}, - \frac{1}{6}, - \frac{1}{3},
\end{equation}
which is exactly the leading asymptotic growth we found with the
leading-logarithms and Borel transform approach in Section~\ref{sec:hopffirst}.

For the chain graphs, we can define $Y_C=\partial_L \Sigma_C$. Then using the exact solution in Borel form in Eq.~\eqref{chainsol}, we see
\begin{equation}
    (\partial_L - \frac{a^2}{6} \partial_a) Y_C = 0. \label{chainRGE}
\end{equation}
We might then interpret the chain case as having $\beta = \frac{1}{6}a^2$, 
which can be compared to the Hopf case where $\beta = \frac{1}{3}a^2 - \frac{11 }{108} a^3 + \cdots$  and the rainbow case where $\beta = 0$. Writing $\beta = a^2 \beta_0 + \cO(a^3)$ in all three cases we then find the leading Borel pole is at $t=-\frac{1}{\beta_0}$ (the rainbow Borel pole is at $t=\infty$), consistent with a renormalon interpretation of these singularities. 

\subsection{Counting diagrams}
We saw that the behavior of the chain, rainbow and Hopf classes of diagrams in $\phi^2 \chi$ theory is parametrically $(-\beta_0)^n  a^n n!$ at large order in the coupling $a=\frac{\lambda^2}{(4\pi)^3}$ with $\beta_0$ given the interpretation of the leading $\beta$-function coefficient. This suggests that the corresponding poles in the Borel plane may be associated with renormalons,  rather than instantons or something entirely new. Another feature of renormalons that contrasts with instantons is that for renormalons the $n!$ can come from a single diagram at order $n$ that evaluates to $\cO(n!)$ (e.g $\int dx  \ln^n x \sim n!$) while for instantons the growth comes from an $n!$ number of Feynman diagrams each of which may be order $1$. In this section we count the number of diagrams in the various cases.

For the rainbow or chain graphs there is exactly one graph at each order $a^n$ so the number of graphs does not grow with $n$. To count Hopf graphs, we note that each graph has a string of non-overlapping loops. We can represent such graphs uniquely as strings of $\pm$ with $+$ for the start of a loop and $-$ for the end of a loop. Because the loops are non-overlapping, once the $\pm$ are assigned, the diagram is completely determined.
Hopf graphs can also be represented as rooted trees, where the root is the outer loop and $+$ means go down and $-$ means go up, as on the right. For example,
\begin{equation}
    \hopffour = \chaintreethree.
\end{equation}
 So the number of such diagrams is bounded by the number of 1D random walks with $2n$ steps, $4^n$, which is sub-factorial. 

To get a more precise counting, note that we always must start a loop before it ends. So the random walks start and end at 0 and never go negative. Such walks are known as Dyck paths. The number of Dyck paths is the Catalan number
\begin{equation}
    C_n  = \frac{1}{n+1} \binom{2n}{n}.
\end{equation}
The $\binom{2n}{n}$ arises because once we pick $n$ of the $2n$ internal vertices to be $+$, the other $n$ are fixed to be $-$. The $\frac{1}{n+1}$ arises from requiring the paths never to go negative and can be computed using the cycle lemma. 
 At large $n$, $C_n \sim \frac{4^n}{\sqrt{\pi} n^{3/2}}$ which is sub-factorial (and smaller than the number of 1D walks, as anticipated).

We can contrast this with the set of all self-energy corrections with no closed $\phi$ loops. These graphs can be drawn the same way but now the loops can start and end anywhere on the $\phi$ line. For example,
\begin{equation}
    \hopffive.
\end{equation}
Note that the boxed subdiagram is a vertex correction so such graphs are not in the Hopf class -- their Schwinger-Dyson equation is much more complicated. The number of such diagrams is the  number of ways to pair up the $2n$ internal vertices, $(2n-1)!! \sim \frac{n! 2^n}{\sqrt{\pi n}}$.\footnote{ 
In  $\frac{1}{6} \lambda \phi^3$ theory, the total number of diagrams for the 2-point function is $\frac{1}{n! 6^n}(3n+1)!!$~\cite{Cvitanovic:1978wc}. 
The number of 1PI diagrams scales like $(\frac{3}{2})^n n!$ at large $n$.~\cite{Borinsky:2017hkb}.
}
This $n!$ growth in the number of diagrams is characteristic of instantons, not renormalons.
 Although $\phi^3$ in $d=6$ has an instanton, as does the $\phi^2 \chi$ theory, other theories with the same diagram topologies, such as the Yukawa theory or QED do not. 
The asymptotic behavior of the sum of all self-energy graphs when vertex corrections are included is an open question.

\section{Asymptotic growth in other diagram classes}\label{sec:other}
Next we examine the asymptotic growth in other classes of diagrams and other theories, using the same approach.
First we review the previously-studied case of Yukawa theory in $d=4$. Then we discuss the two-arm Hopf series in $\phi^3$ theory in $d=6$. Finally, we present some examples in $\phi^4$ theory in $d=4$.

\subsection{Yukawa theory in \texorpdfstring{$d = 4$}{d = 4}}
Yukawa theory in $d = 4$ has  Euclidean action is given by
\begin{equation}
   S = \int d^4 x\left[ \frac{1}{2} (\partial_\mu\phi)^2 + \bar{\psi} \slashed{\partial} \psi - g \phi \bar{\psi}\psi\ \right] \, .
\end{equation}
In Euclidean space the Green's function is
\begin{equation}
    \langle 0| \psi(x) \bar{\psi}(0)|0\rangle =\int \frac{d^4 p}{(2\pi)^4} e^{i p x} \frac{-i\slashed{p}}{p^2(1-\Sigma_Y^0)},
\end{equation}
where the 1PI graphs in the Hopf approximation are
\begin{equation}
i \slashed{p} \Sigma_Y^0 =
  \bubbleA + \doublebubbleA + \chainA + \rainbowA + \cdots.
\end{equation}
In these diagrams, the black fermion line is $\psi$ and the plain red line is $\phi$. 
The SDE for the Hopf graphs is
\begin{equation}
i \slashed{p} \Sigma^0_Y(p^2)= \,g^2 \int\!\frac{d^4 k}{(2\pi)^4}\;
\frac{ -i\slashed{k} }{\,k^2\,[\,1-\Sigma^0_Y(k^2)\,]\,(p+k)^2 } .
\label{eq:SigmaStart}
\end{equation}
We can then get an equation for $\Sigma^0_Y(p^2)$ by multiplying with $\frac{-i \slashed{p}}{4p^2}$ and taking the trace: 
\begin{align}
\Sigma^0_Y(q^2)
&= \frac{-i g^2}{4p^2} \int\!\frac{d^4 k}{(2\pi)^4}\;
\frac{ \mathrm{Tr}[\slashed{p} i\slashed{k}]}{\,k^2\,[\,1-\Sigma^0_H(k^2)\,]\,(p+k)^2 } \\
&= \frac{a}{p^2 \pi^{2}}\int d^{4}k\,
\frac{\,p\!\cdot\!k\,}{k^{2}[1-\Sigma^0_Y(k^{2})]\,(p+k)^{2}}, \, 
\end{align}
where $a = (\frac{g}{4\pi})^2$.  We can rewrite the integral above by expressing $p \cdot k= \frac{1}{2}((p+k)^2 - p^2 -k^2)$. The term proportional to $(p+k)^2$ leads to scaleless integrals which we set to zero. The remaining terms give
\begin{equation}
\Sigma^0_Y(p^{2})
= -\lim_{\rho\rightarrow0}\frac{a}{2\pi^2} \frac{1}{1- \Sigma^0_Y(\partial_\rho)} \int  d^{d} k\, \left(
\frac{1}{k^2 (p+k)^{2}}  \, + \frac{1}{p^2 (p+k)^2} \right) \left(\frac{k^2}{\mu^2}\right)^\rho.  
\end{equation}
Computing the integrals and imposing MOM renormalization conditions leads to
\begin{equation}
\Sigma_Y(L) =  \lim_{\rho\rightarrow0} \frac{a}{1- \Sigma_Y(\partial_\rho)}\frac{e^{L\rho}-1}{(\rho+2)\rho}.   
\end{equation}
From here we can read off the ordinary differential equation satisfied by $\Sigma_Y(q^{2})$ as 
\begin{equation}
    P(\partial_L) \partial_L \Sigma_Y = \frac{a}{1-\Sigma_Y}, \label{ode}
\end{equation}
with $P(\rho) = \rho+2$. 
This differential equation is in agreement with~\cite{Broadhurst:1999ys}. 

From here on, the analysis is identical to the analysis of the Hopf graphs for $\phi^2 \chi$ theory in 
Section~\ref{sec:hopffirst}, but with a different
$P (\rho)$. 
For the leading-log solution we replace $P(\partial_L)$ by $P(0)$ giving
\begin{equation}
    2 \partial_L \Sigma^{\LL}_Y = \frac{a}{1-\Sigma^{\LL}_Y}
    \label{llyukawa},
\end{equation}
with solution 
\begin{equation}
    \Sigma^{\LL}_Y = 1- \sqrt{1-a  L}.
\end{equation}
Plugging this back into the SDE gives
\begin{equation}
  P (\partial_L) \partial_L \Sigma_Y^\LLL (L) = \frac{a}{\sqrt{1 - a  L}} =\sqrt{\frac{a}{\pi}}\int_0^{\infty} d t e^{- \frac{t}{a}}
  \frac{ e^{tL} }{ \sqrt{t}}.
\end{equation}
Imposing the MOM renormalization conditions results in 
\begin{equation}
  \Sigma_Y (L) \sim\Sigma_Y^\LLL (L) =\sqrt{\frac{a}{\pi}} \int_0^{\infty} d t e^{- \frac{t}{a}} t^{-3/2}   \frac{e^{t L}-1}{P(t)}.
\end{equation}
Since $P(\rho)$ has a single zero at $\rho = -2$ in in this case there is a single Borel pole at $t=-2$. This pole is an agreement with the asymptotic behavior $\gamma \sim (-\frac{1}{2})^n n!$ found in~\cite{Borinsky:2020HADSE}.

We can also analyze the rainbows and chains in Yukawa theory. For the rainbows we find, following the procedure in Section~\ref{sec:rainbow2},
\begin{equation}
    P(-\gamma)= a  \quad \rightarrow \quad \gamma = 1-\sqrt{1+a}.
\end{equation}
This then means that $\Sigma^C_{Y} = e^{-1+\sqrt{1+a}} -1$ which is analytic. 

For the chains we find, following the procedure of Section~\ref{sec:chains2},
\begin{equation}
\partial_L\Sigma_Y^C (L) 
= \lim_{\rho \rightarrow 0} \frac{a}{\left( 1 - \frac{a}{2}
  \partial_\rho \right)} \frac{e^{\rho L}}{\rho+2}
  = \int_0^\infty dt e^{-\frac{t}{a}}e^{- \frac{1}{2}t L}
  \frac{1}{\frac{t}{2}+2},
\end{equation}
which has a Borel poles at $t = -4$, twice as far away from the origin as the one in the Hopf approximation. 
The leading asymptotic behavior of the chain and rainbow Yukawa series we find  are in agreement with~\cite{Borinsky:2020HADSE}.

 \subsection{Two-arm Hopf series in $\phi^3$ theory}\label{sec:twoarm}
 Next we turn to 
$\phi^3$ theory in $d = 6$ with Euclidean action
\begin{equation}
 S = \int d^6 d\left[ \frac{1}{2} (\partial_{\mu} \phi)^2 - \frac{\lambda}{3!}  \phi^3 \right].
\end{equation}
In this theory we can consider the recursive insertion of the
self-energy bubble into the $\phi$ propagator. The SDE is pictorially 
\begin{equation}
  \Sigma_B (p^2) =
  \doublesigma,
\end{equation}
which corresponds to 
\begin{equation}
  \Sigma_B^0 (p^2) = \frac{\lambda^2}{2}\frac{1}{p^2} \int \frac{d^6 k}{(2\pi)^6} \frac{1}{(p + k)^2 (1 -
  \Sigma_B^0 ((p + k)^2))} \frac{1}{k^2 (1 - \Sigma_B^0 (k^2))}. \label{BSDE}
\end{equation}
The factor of $\frac{1}{2}$ for these graphs relative to those in $\phi^2\chi$ theory is a symmetry factor for the primitive 1-loop self energy kernel. 

The master integral in this case is again computable with Schwinger
parameters:
\begin{align}
  I_{\alpha, \beta} (q) &\equiv  \int \frac{d^d k}{(2 \pi)^d}
  \frac{1}{(k^2)^{\alpha}} \frac{1}{[(p + k)^2]^{\beta}} \\
  &= \left(p^2\right)^{\frac{d}{2} - \alpha - \beta}
  \frac{1}{(4\pi)^{\frac{d}{2}}} \frac{\Gamma \left( \alpha + \beta - \frac{d}{2} \right)
  \Gamma \left( \frac{d}{2} - \alpha \right) \Gamma \left( \frac{d}{2} - \beta
  \right)}{\Gamma (\alpha) \Gamma (\beta) \Gamma (d - \alpha - \beta)}.
  %
\end{align}
In $d = 6$ this reduces to 
\begin{equation}
  I_{\alpha, \beta} (p) = 
  (p^2)^{3- \alpha - \beta}
  \frac{1}{(4\pi)^3}
  \frac{\Gamma (3 - \alpha) \Gamma (3 - \beta) \Gamma (\alpha + \beta -
  3)}{\Gamma (\alpha) \Gamma (\beta) \Gamma (6 - \alpha - \beta)}.
\end{equation}
After renormalizing in MOM, we then find
\begin{equation}
  \Sigma_B (a,L) = \lim_{\rho, \sigma \rightarrow 0} 
  \frac{a}{2 [1 - \Sigma_B(a,\partial_{\rho})][1 - \Sigma_B (a,\partial_{\sigma})]} F (\rho,
  \sigma) [e^{L (\rho + \sigma)} - 1],
\end{equation}
where $L = \ln \frac{p^2}{\mu^2}$ and $a=\lambda^2/(4\pi)^3$ as usual and
\begin{equation}
  F (\rho, \sigma) =(4\pi)^3 (p^2)^{-1-\rho-\sigma}   
  I_{1 - \rho, 1 -
  \sigma} = \frac{\Gamma (2 + \rho) \Gamma (2 + \sigma) \Gamma (- 1 - \rho -
  \sigma)}{\Gamma (1 - \rho) \Gamma (1 - \sigma) \Gamma (4 + \rho + \sigma)}.
\end{equation}
It is then straightforward to expand $\Sigma_B$ perturbatively to as high
order as we want
\begin{equation}
  \Sigma_B (L) = \frac{a}{12} L + \left(\frac{a}{12}\right)^2 \left( - \frac{11}{3} L +L^2\right)
  + \left(\frac{a}{12}\right)^3 \left( \frac{685}{18} L -\frac{77}{6}L^2+\frac{5}{3}L^3\right)
  + \cdots.
\end{equation}
Moreover, $\Sigma_B$ satisfies the RGE
\begin{equation}
    (\partial_L +3\gamma a \partial_a -\gamma) (1-\Sigma_B) = 0,
\end{equation}
with
\begin{equation} 
    \gamma =  -[\partial_L\Sigma_B]_{L=0} =
    -\frac{1}{12}a+\frac{11}{432} a^2-\frac{685 }{31104}a^3 - {\left(\frac{\zeta_3}{1296}-\frac{16405}{559872}\right) a^4} + \cdots.
\end{equation}
This RGE can be derived as in Section~\ref{sec:CSE} or using Hopf algebraic techniques as we discuss below in Section~\ref{sec:multislot}. It can also be checked perturbatively. 

We are interested in the asymptotic behavior of $\Sigma_B$ at large order
in $a$. To find the asymptotic behavior, we follow the approach from the
$\phi^2 \chi$ theory: compute the LL approximation first, then insert into the
right hand side of the SDE.
To take the LL approximation we need the dominant behavior of $F(\rho,\sigma)$ near $\rho =
\sigma = 0$ which comes from $\Gamma (- 1 - \rho - \sigma) \sim
\frac{1}{\rho + \sigma}$. 
Including just this term, the SDE reduces to
\begin{equation}
  \Sigma_{B}^\LL (a, L) = \frac{a}{12} \lim_{\rho, \sigma \rightarrow
  0} \left[ \frac{1}{1 - \Sigma_B^\LL (a, \partial_{\rho})} \right]
  \left[ \frac{1}{1 - \Sigma_B^\LL (a, \partial_{\rho})} \right]
  \frac{e^{L (\rho + \sigma)} - 1}{\rho + \sigma},
\end{equation}
so that 
\begin{equation}
  \partial_L \Sigma_B^\LL (a, L) = \frac{a}{12} \lim_{\rho, \sigma \rightarrow
  0}  \left[ \frac{1}{1 - \Sigma_B^\LL (a, \partial_{\rho})} \right]
  \left[ \frac{1}{1 - \Sigma_B^\LL (a, \partial_{\rho})} \right]
  e^{L (\rho + \sigma)} 
  = \frac{a}{12} \left[ \frac{1}{1 - \Sigma_{\LL}
  (a, L)} \right]^2.
\end{equation}
Writing $Y = 1 - \Sigma_B^\LL$ we find once again a autonomous monomial ODE:
\begin{equation}
  \partial_L Y =-\frac{a}{12}Y^s, \label{phi4s}
\end{equation}
with $s = - 2$. 
The leading-log solution is 
\begin{equation}
  Y^\LL = \left( 1 - (1 - s) \frac{a}{12} L \right)^{\frac{1}{1 -   s}} 
  \xrightarrow{s=-2}{} \left( 1 - \frac{a}{4} L \right)^{1/3}.
\end{equation}
So
\begin{equation}
  \Sigma_B^\LL= 1 - \left( 1 - \frac{a}{4} L \right)^{\frac{1}{3}} =
  \frac{a}{12}  L + \left(\frac{a}{12} \right)^2 L^2 
  +\frac{5}{3}\left(\frac{a}{12} \right)^3 L^3  + \cdots,
\end{equation}
in agreement with the leading logs of the iterative solution.

Next, we plug the leading log solution into the SDE. This gives
\begin{equation}
  \partial_L \Sigma_B^\LLL = \frac{a}{2}\lim_{\rho, \sigma \rightarrow 0} {\left( 1 - \frac{ a}{4} \partial_{\rho} \right)^{- \frac{1}{3}}}  \left( 1 - \frac{a}{4}
  \partial_{\sigma} \right)^{- \frac{1}{3}} (\rho+\sigma)e^{L (\rho + \sigma)} 
  F(\rho,\sigma).
\end{equation}
To find the Borel transform we can use
\begin{equation}
  (1 - \kappa a \partial_{\rho})^{- \beta} = a^{- \beta} \int_0^{\infty} d t
  e^{- \frac{t}{a}} \frac{t^{\beta - 1}}{\Gamma (\beta)} e^{\kappa t
  \partial_{\rho}},
  \label{kappabetarel}
\end{equation}
where $\kappa = \frac{1}{4}$ and $\beta = \frac{1}{3}$. 
Then,
\begin{equation}
  (1 - \frac{1}{4} a \partial_{\rho})^{-\third} (1 - \frac{1}{4} a \partial_{\sigma})^{-\third} = a^{-\frac{2}{3}}\frac{1}{\Gamma (\third)^2} \int_0^\infty d u \int_0^\infty d v e^{- \frac{u  + v}{a}} (u v)^{-\frac{2}{3}} e^{\frac{1}{4} u \partial_{\rho} + \frac{1}{4} v
  \partial_{\sigma}},
\end{equation}
and
\begin{align}
  \partial_L \Sigma_B^\LLL &=
  {\frac{a^{1/3}}{2\Gamma (1/3)^2}} 
  \int_0^\infty d u \int_0^\infty dv e^{- \frac{u + v}{a}} (u v)^{-\frac{2}{3}} \lim_{\rho, \sigma \rightarrow 0}
  e^{\frac{1}{4} u \partial_{\rho} + \frac{1}{4} v \partial_{\sigma}} e^{L (\rho +  \sigma)} (\rho+\sigma)F (\rho, \sigma) \nonumber
\\
  &=  {\frac{a^{1/3}}{2\Gamma (1/3)^2}}
  \int_0^\infty d u \int_0^\infty d v e^{- \frac{u +  v}{a}} (u v)^{-\frac{2}{3}} e^{\frac{L}{4} (u + v)} \frac{u+v}{4} F (\frac{u}{4}, \frac{v}{4}).
\end{align}
Now change to $u + v = t, x = \frac{u}{t}$ with $d u d v = t d t d x$ 
to find 
\begin{equation}
 \partial_L \Sigma_B^\LLL  =
 {\frac{a^{1/3}}{8\Gamma (1/3)^2}}
  \int_0^\infty d t e^{- \frac{t}{a}} e^{\frac{L}{4} t} t^{2/3} \int_0^1 d x \big(x (1 -
  x)\big)^{-\frac{2}{3}} F \left(\frac{t x}{4}, \frac{t(1-x)}{4}\right).
\end{equation}
Noting that
\begin{equation}
    F(\rho,\sigma) = \Gamma(-3-\rho-\sigma)\frac{A(\rho)A(\sigma)}{\Gamma(2+\rho+\sigma)},
    \quad A (\rho) = \frac{\Gamma (2 + \rho)}{\Gamma (1 - \rho)},
\end{equation}
we can then write
\begin{equation}
 \partial_L \Sigma_B^\LLL  =
 {\frac{a^{1/3}}{8\Gamma (1/3)^2}}
 \int_0^\infty d t e^{- \frac{t}{a}} e^{L
  \frac{t}{4}}  \Gamma\left(-3-\frac{t}{4}\right)  t^{2/3} \int_0^1 d x \big(x (1 - x)\big)^{-\frac{2}{3}} \frac{A (\frac{t x}{4}) A (\frac{t(1-x)}{4}  )}{\Gamma(2+\frac{t}{4})}. 
\end{equation}
From here we can read off the Borel transform. More precisely, because of the $a^{1/3}$ factor, this gives the Borel-Leroy transform of index $p=-1/3$. 
Some properties of Borel-Leroy transforms are given in Appendix~\ref{app:Borel}. The key property 
relevant for our calculation is that the Borel-Leroy transform has singularities in the same places as the Borel transform although the order of the singularities is shifted by $p$.
In this case,  Borel-Leroy transform of $\partial_L \Sigma_B^\LLL$ is ${\mathcal B}_p(t/4)$ where 
\begin{equation} 
  {\mathcal B}_p(t) =  \frac{2 e^{L t}  t^2\Gamma(-3-t)}{\Gamma(1/3)^2} 
  \int_0^1 d x (x (1 - x))^{-\frac{2}{3}} \frac{A ( t x) A ( t (1 - x))}{\Gamma(2+  t)}.
  \label{FullB}
\end{equation}
We would next like to find the singularities of ${\mathcal B}_p(t)$.

The $\Gamma(-3- t)$ function out front leads to poles at $ t = -3,-2,-1,\ldots$.
For the terms involving $A$ we need to find the singularities of
\begin{equation}
  I_A ( t) = \int_0^1 d x (x (1 - x))^{- \frac{2}{3}} \frac{\Gamma (2 +
   t x)}{\Gamma (1 -  t x)} \frac{\Gamma (2 +  t (1 -
  x))}{\Gamma (1 -  t (1 - x))} \frac{1}{\Gamma(2+ t)}.
\end{equation}
Since $\Gamma(z)$ only has simple poles and no zeros for any $z\in \mathbb{C}$, the singularities in this integral can only come from singularities in the $\Gamma$ functions in the numerator. For the integral to be singular, the singularities of the integrand have to either be pinched or coincide with the endpoint of integration. For a pinch we would need $2+ t x = -n$ and $2+ t (1-x) = -m$ simultaneously for two non-negative integers $n$ and $m$. But then $ t = -4-m-n$ which is also a pole in the $\Gamma(2+ t)$ factor. Thus one pole is canceled and the contour is not actually pinched. Thus the only singularities can come from the endpoints $x=0$ and $x=1$. Since the integral is symmetric under $x\to 1-x$ we only need to consider $x=0$. 

Near $x=0$ the pole must come from the $\Gamma(2+ t(1-x))$ factor.  
Let us then write $ t = - 2 - n + y$  with $n = 0, 1,
2, \cdots$ and $|y|<1$ and examine the behavior of the integral over $0\le x \le \delta$ for small $\delta$. That is, we want to integrate over a small but fixed interval $0\le x \le \delta$ and take $|y|\to 0$ assuming $0 \ll |y| \ll \delta \ll 1$. Because $x$ is integrated over, we cannot drop $x$ with respect to $y$ or vice versa.
Keeping the leading terms at small $x$
and $y$ the expansion gives
\begin{equation}
 \frac{ \Gamma (2 +   t (1 - x))}{\Gamma(2+  t) }= \frac{\Gamma (- n + (2 + n) x + y)}{\Gamma(-n+y)} \approx  \frac{y}{(2 + n) x + y}.
\end{equation}
Then
\begin{equation}
  \int_0^{\delta} d x (x (1 - x))^{- \frac{2}{3}} \frac{\Gamma (2 +   t
  x)}{\Gamma (1 -   t x)} \frac{\Gamma (2 +   t (1 - x))}{\Gamma (1
  -   t (1 - x))} \frac{1}{\Gamma(2+  t)}
  \approx
  \int_0^{\delta} d x \frac{x^{- \frac{2}{3}}y}{(2 + n) x + y}
  \sim y^{1/3}.
\end{equation}
Thus $I_A(t)$ has branch points at $t = \ldots,-4,-3,-2$ of order $1/3$. The $\Gamma(-3-t)$ factor in Eq.~\eqref{FullB} has simple poles ($\sim y^{-1}$) at $t=-3,-2,-1,\ldots$. So the net effect is branch points of the form $(t-t_0)^s$ with 
 $s=1/3$ for integers $t_0 \le -4$, $s=-2/3$ for  $t_0=-3,-2$ and $s=-1$ branch points for $t=-1$ and $\ge 1$. There is no singularity at $t=0$ due to the $t^2$ factor. 
 
The locations of singularities are the same in the Borel-Leroy transfrom and the Borel transform,
although their strengths change. However, the 
strength of a singularity is also affected by subleading logarithmic effects which we have neglected. Thus, to our level of approximation we should be able to trust the location of the singularities, but not their strength. These singularities are at every nonzero integer.

 Recalling that the Borel transform of $\partial_L \Sigma_B^\LLL$ is proportional to ${\mathcal B}_p(t/4)$ we conclude that the Borel transform of $\Sigma_B^\LLL$ should have singularities at $t= 4 n$ for every integer $n \ne 0$.  The leading pole seems consistent with the results of~\cite{Bellon:2010ade,Bellon:2010sde} (after correcting for symmetry factors), which were derived using a different method.

\subsection{$\phi^4$ theory}\label{sec:phi4}
Next, let's consider some sets of graphs $\phi^4$ theory with Euclidean action
\begin{equation}
    S = \int d^d x \left[ \frac{1}{2} (\partial_\mu \phi)^2 + \frac{\lambda}{4!} \phi^4 \right],
\end{equation}
and expand in $d=4-2\eps$ dimensions. We consider first a set of graphs like the Hopf graphs but with 2-loop sunset rather than 1-loop self-energy insertions.  Then we consider a series of $s$ channel bubbles and finally diagrams involving $s,t$ and $u$ channel bubble insertions.

\subsubsection{Sunset graphs}
Let's consider first the Hopf sunset graphs 
\begin{equation}
\Sigma_S = \sunset+
\doublesun
+\threesuns
+\manysuns
+\cdots.
\end{equation}
The sunset itself is
\begin{equation}
  \Sigma_S^1 (p^2) = \frac{\lambda^2}{3!} \frac{1}{p^2} \int \frac{d^4 k_1}{(2 \pi)^4} 
  \frac{d^4 k_2}{(2 \pi)^4}  \frac{1}{k_1^2} \frac{1}{k_2^2} \frac{1}{(p - k_1
  - k_2)^2},
\end{equation}
with the $3!$ a symmetry factor. For the SDE we replace each
propagator by the 1PI sum
\begin{equation}
  \Sigma_S (p^2) = \frac{\lambda^2}{6} \int \frac{d^4 k_1}{(2 \pi)^4} 
  \frac{d^4 k_2}{(2 \pi)^4} \frac{1}{k_1^2} \frac{1}{k_2^2} \frac{1}{(p - k_1
  - k_2)^2}  \frac{1}{1 - \Sigma_S (k_1^2)} \frac{1}{1 - \Sigma_S (k_2^2)}
  \frac{1}{1 - \Sigma_S (p - k_1 - k_2)^2}.
\end{equation}
To determine $\Sigma_S$ we need the analytically regulated integral:
\begin{align}
  I (a, b, c) &= \frac{1}{p^2} \int \frac{d^d k_1}{(2 \pi)^4}  \frac{d^d
  k_2}{(2 \pi)^4}  \frac{1}{k_1^{2 (1 - a) }} \frac{1}{k_2^{2 (1 - b)}}
  \frac{1}{(p - k_1 - k_2)^{2 (1 - c)}}
  \nonumber
\\
  &= \frac{\pi^{d}}{(2 \pi)^{2d}}  (p^2)^{d - 4 + a + b + c} \frac{\Gamma \left(
  \frac{d}{2} - 1 + a \right) \Gamma \left( \frac{d}{2} - 1 + b \right) \Gamma
  \left( \frac{d}{2} - 1 + c \right) \Gamma (3 - d - a - b - c)}{\Gamma (1 -
  a) \Gamma (1 - b) \Gamma (1 - c) \Gamma \left( \frac{3}{2} d - 3 + a + b + c
  \right)}
  \nonumber
\\
&\xrightarrow[]{d \rightarrow 4}  \frac{1}{256 \pi^4} (p^2)^{a + b + c} F (a,
  b, c),
\end{align}
where
\begin{equation}
  F (a, b, c) = \frac{\Gamma (1 + a) \Gamma (1 + b) \Gamma (1 + c) \Gamma (- 1
  - a - b - c)}{\Gamma (1 - a) \Gamma (1 - b) \Gamma (1 - c) \Gamma (3 + a + b  + c)}. \label{Fabcdef}
\end{equation}
The SDE is then
\begin{equation}
  \Sigma_S (q^2) = \alpha \lim_{a, b, c \rightarrow 0} \frac{1}{1 - \Sigma
  (\partial_a)} \frac{1}{1 - \Sigma (\partial_b)} \frac{1}{1 - \Sigma
  (\partial_c)} F (a, b, c) (e^{L (a + b + c)} - 1),
\end{equation}
where
\begin{equation}
  \alpha = \frac{\lambda^2}{6 \times 256 \pi^4}.
\end{equation}

Before continuing, we observe that because the middle line in the sunset graph is shared between two loops, this graph has
overlapping subdivergences.  However,
because the remaining graph after contracting one of these loops to a point is a tadpole and therefore cannot generate a $\ln p^2/\eps$ divergence, there are actually no non-local divergences in the sunset itself. 
Thus we can renormalize the sunset using local counterterms, as if it were a primitive graph without overlapping divergences. 

To find the singularities in the Borel transform of $\Sigma_S$, we follow the procedure in Section~\ref{sec:twoarm}. First, we find the leading log solution by extracting the leading singular behavior as $a,b,c\to 0$ which comes from the pole in the $\Gamma(-1-a-b-c)$ term in Eq.~\eqref{Fabcdef}. Expanding around this pole leads to 
\begin{align}
  \partial_L \Sigma_S^\LL (q^2) &= \frac{\alpha}{2} 
  \lim_{a, b, c \rightarrow  0} 
  \frac{1}{1 - \Sigma_S^\LL (\partial_a)} 
  \frac{1}{1 - \Sigma_S^\LL (\partial_b)}
  \frac{1}{1 - \Sigma_S^\LL (\partial_c)} e^{L (a + b + c)} 
  &= \frac{\alpha}{2}\left( \frac{1}{1 - \Sigma_S^\LL} \right)^3,
\end{align}
with solution
\begin{equation}
  \Sigma_S^\LL = 1 - (1 - 2 \alpha L)^{1 / 4}.
\end{equation}
We then focus on solving
\begin{equation}
  \partial_L\Sigma_S^\LLL = \alpha \lim_{a, b, c \rightarrow 0} 
\left(  \frac{1}{1 - 2 \alpha \partial_a} 
  \frac{1}{1 - 2 \alpha \partial_b} 
  \frac{1}{1 - 2 \alpha \partial_c}\right)^{1/4}
  \frac{A(a)A(b)A(c)}{\Gamma(1+a+b+c)} B(a+b+c),
\end{equation}
where
\begin{equation}
     A(x) =\frac{\Gamma(1+x)}{\Gamma(1-x)},
     \quad
     B(x) = -e^{L x}\frac{x}{1+x}\Gamma(-2-x).
\end{equation}
Then we use again Eq.~\eqref{kappabetarel} but now with $\kappa = 2$ and $\beta = \frac{1}{4}$. This gives 
\begin{multline}
    \partial_L\Sigma_S^\LLL = \frac{\alpha^{-3/4} }{\Gamma(1/4)^3}
    \int_0^\infty du dv dw 
    e^{- \frac{u + v + w}{\alpha}} (u v  w)^{-3/4} 
    e^{2 u \partial_a + 2 v \partial_b + 2 w \partial_c}
\\
\times
\frac{A(a)A(b)A(c)}{\Gamma(1+a+b+c)} B(a+b+c).
\end{multline}
Going to Feynman parameters this reduces to
\begin{multline}
    \partial_L\Sigma_S^\LLL = \frac{\alpha^{-3/4} }{\Gamma(1/4)^3}
    \int_0^\infty d t e^{-\frac{t}{\alpha}}
    B(2t) t^{-\frac{1}{4}}\\
\times    \int_0^1 d^3x \delta(x_1+x_2+x_3-1) 
    (x_1 x_2 x_3)^{- \frac{3}{4}}
\frac{A(2 t x_1)A( 2 t x_2)A( 2 t x_3)}{\Gamma(1+2 t)},
\end{multline}
from which we can read off the Borel-Leroy transform. As before, the integral on the second line has no pinches. The endpoint singularities come from poles of the $A$ factors which are at $2t =-1,-2,-3,\ldots$. 
In addition,  the $\Gamma(-2-2t)$ in $B(2t)$ leads to singularities at  at $2t = -2,-1,\ldots $. So we find singularities at $t=n/2$ for any non-zero integer $n$.

\subsubsection{$s$-channel bubble chain}
The sunset insertions in $\phi^4$ are a series in $\lambda^2$. 
We can also consider sets of graphs which provide a series in $\lambda$. The first such series we consider are $s$-channel
chain graphs of the form
\begin{equation}
\bubblechain.
\end{equation}
These are insertions of a bubble chain into a primitive tadpole graph:
\begin{align}
  \Sigma_{s}^0 (p^2) &= -\frac{\lambda}{2} \int \frac{d^d k}{(2 \pi)^d}  \frac{1}{p^2 (p + k)^2}
  (1 -\lambda B (k^2) + \lambda^2 B (k^2)^2 + \cdots)
\\
  &= -\frac{\lambda}{2} \int \frac{d^d k}{(2 \pi)^d}  \frac{1}{p^2(p + k)^2} \frac{1}{1 +
  \lambda B(k^2)}. \label{sigmawithB}
\end{align}
The leading term is a tadpole
\begin{equation}
\tadpole 
  = -\frac{\lambda}{2} \int \frac{d^4 k}{(2 \pi)^4}  \frac{1}{p^2(p + k)^2}.
\end{equation}
which has no dependence on external momenta and can be consistently set to zero. The 1-loop scalar bubble $B(p^2)$ appearing in Eq.~\eqref{sigmawithB} is
\begin{align}
  B(p^2) &= \frac{1}{\lambda^2}  \snoblob = \frac{1}{2} \mu^{2\eps} \int \frac{d^d k}{(2 \pi)^d}  \frac{1}{k^2 (p + k)^2} \\
  &=\frac{1}{2} \frac{1}{(4\pi)^{2-\eps}}
  \frac{\Gamma (\eps) \Gamma (1 - \eps)^2}{\Gamma (2 - 2
  \eps)} \left( \frac{\mu^2}{p^2} \right)^{\eps}
\\
  &= \frac{1}{32 \pi^2} \left( \frac{1}{\eps} -\ln \frac{p^2}{\mu^2} + 2 \right). \label{sigma10}
\end{align}
The associated 1-loop counterterm graph in MOM is
\begin{equation}
\delta_\lambda=     \crossct = -\frac{\lambda^2}{32\pi^2}\left( \frac{1}{\eps} + 2\right).
\end{equation}
Note that this is not the same as the counterterm if the $t$ and $u$-channel 1-loop graphs were also included, but it is the appropriate counterterm of the $s$-channel bubble series. 
The analytically-regulated  master integral in $d=4$ dimensions is 
\begin{equation}
  \int \frac{d^4 k}{(2 \pi)^4}  \frac{1}{p^2 (p + k)^2} \left(
  \frac{k^2}{\mu^2} \right)^{\rho} = - \frac{1}{16 \pi^2} \frac{1}{(1+\rho)(2+\rho )} e^{\rho L}.
\end{equation}
This integral is UV divergent for $\text{Re}~\rho \ge -1$ and IR divergent for $\text{Re}~\rho \le -2$.

The SDE for the renormalized $\Sigma_{s}$ in MOM is then
\begin{equation}
    \Sigma_s(p^2) = -\frac{\lambda}{2} \int \frac{d^4 k}{(2 \pi)^4} 
    \frac{1}{p^2(p + k)^2} 
    \frac{1}{1 - a\ln\frac{k^2}{\mu^2}}
= \lim_{\rho \rightarrow 0} 
\frac{a}{1 - a  \partial_{\rho}} \frac{e^{\rho L}-1}{(1+\rho) (2+\rho )},
\label{SigmasSDE}
\end{equation}
where $a=\frac{\lambda}{32\pi^2}$ and so 
\begin{equation}
 (1 + \partial_L)(2 + \partial_L) \partial_L \Sigma_{s} 
 = \lim_{\rho \rightarrow 0} \frac{a}{1 - a  \partial_{\rho}} \rho e^{\rho L}
 =\partial_L \left[
     \lim_{\rho \rightarrow 0}
     \frac{a}{1 - a  \partial_{\rho}}  e^{\rho L}\right]
     =    \frac{a^2}{(1 - a L)^2}.
\end{equation}
This is similar to the chain in $\phi^3$.
Taking the Borel transform:
\begin{equation}
     \frac{a^2}{(1 - a L)^2} = \int_0^\infty dt e^{-\frac{t}{a}} t e^{tL},
\end{equation}
we can then write 
\begin{equation}
  \Sigma_{s} = \int_0^{\infty} d t 
  e^{-\frac{t}{a}} \frac{e^{t L} - 1}
  {(1+t)(2+t)},
  \label{SigmasB}
\end{equation}
which has poles at $t = -1$ and $t = -2$. The $t = -1$ pole is a UV renormalon and
the $t = -2$ is an IR renormalon.\footnote{Ref.~\cite{Loewe:2022aaw} computes a similar series with a mass regulator and finds Borel singularities at order $m^4$ and $m^6$ of the same magnitude but opposite sign (on the positive Borel axis).}

\subsubsection{ $stu$ bubble chain} 
As a generalization of the $s$ bubble chain, we next consider inserting the $s, t, u$ 1-loop bubbles into the 4-point vertex, extracting the leading logarithms, and inserting the sum into the 2-point function

At 1-loop there are three 4-point graphs
\begin{equation}
\midblob=
\cross 
+\snoblob
+\unoblob
+\tnoblob
+\cdots.
\end{equation}
So up to 1-loop
\begin{equation}
  \Gamma_4^0 (s, t, u) = -\lambda + \lambda^2 \left[B(s) + B(t) + B(u)\right]+\cdots,
\end{equation}
with $B(p^2)$ in Eq.~\eqref{sigma10}. Here, $s=(p_1+p_2)^2,t=(p_1+p_3)^2$ and $u=(p_1+p_4)^2$ with all momenta incoming.
In the MOM scheme the 1-loop counterterm is now
\begin{equation}
\delta_\lambda=    \crossct = -\frac{3 \lambda^2}{32\pi^2}\left( \frac{1}{\eps} + 2\right),
\end{equation}
which is three times bigger than when only $s$-channel is included and
\begin{equation}
   \Gamma_4 = - \lambda- \frac{\lambda^2}{32\pi^2}\left( \ln\frac{s}{\mu^2} + \ln\frac{t}{\mu^2} + \ln\frac{u}{\mu^2}\right) + \cdots.
\end{equation}
The anomalous dimension for the 4-point function starts at order $\lambda^2$, so the RGE to LL level is
\begin{equation}
    (\mu \partial_\mu + \beta_0 \lambda^2 \partial_\lambda )\Gamma_4(s,t,u,\mu) =0,\qquad \beta_0 = \frac{3}{16\pi^2}.
    \label{LLRGE1}
\end{equation}
Note that the RGE only dictates the $\mu$ dependence and not the dependence on $s,t,u$.

To determine what scales appear, consider next the 2-loop contributions to $\Gamma_4$. At the leading-log level, we only need to iteratively insert bubbles. Inserting an $s$-channel bubble into one of the vertices of the $s$-channel diagram gives
\begin{equation}
     \doubles = -\lambda^3 B(s)^2 = -
     \lambda\left(\frac{\lambda}{16\pi^2}\right)^2 \left[ \frac{1}{4
\varepsilon^2} -\frac{1}{2\eps}L_s+ \frac{1}{\eps} + \frac{1}{2} L_s^2 + \cdots \right],
\end{equation}
where $L_s = \ln\frac{s}{\mu^2}$.
There is only one other topology that shows up at 2-loops:
\begin{align}
    \icecream &=-\lambda^3 \mu^{2\eps}  \int \frac{d^dk}{(2\pi)^d} \frac{1}{k^2(p_1+p_2-k)^2} B[ (k+p_4)^2]\\[-15pt]
    &=  - \lambda^3 \frac{1}{(16 \pi)^{4 - 2 \varepsilon}} \left( \frac{\mu^2}{s}
\right)^{2 \varepsilon} \frac{\Gamma (\varepsilon) \Gamma (2 \varepsilon)
\Gamma (1 - \varepsilon)^2 \Gamma (1 - 2 \varepsilon)}{(1 - 2 \varepsilon)
\Gamma (2 - 3 \varepsilon)}\\
    &= - \lambda \left( \frac{\lambda}{16 \pi^2} \right)^2 \left[ \frac{1}{4
\varepsilon^2} -\frac{1}{2\eps}L_s+ \frac{1}{\eps} + \frac{1}{2} L_s^2 + \cdots \right].
\end{align}
So the leading logs in the ice-cream cone are the same as in the bubble. 
Note, importantly, that the ice-cream cone diagram depends only on $L_s$. This can be
seen simply because only $P = p_1 + p_2$ and $p_4$ enter the graph. So the only invariants we can have are $P^2 = s$, $P \cdot p_4 = -s/2$ and
$p_4^2 = 0$. 
There is another ice-cream cone where the bubble is inserted into the left vertex, which gives the same result. 
If we insert a
$t$-channel bubble instead of the $u$-channel one into the right vertex, the graph is topologically
identical and already included in the symmetry factor. The only other 2-loop graphs are crossings, starting with a $t$ or $u$-channel bubble.
There are two $s$-channel counterterm graphs, each of which give
\begin{equation}
\snoblobct =  -\lambda \delta_\lambda B(s) =
- \lambda \left( \frac{\lambda}{16 \pi^2} \right)^2 \left( -\frac{3}{4\varepsilon^2}
+\frac{3}{4\eps}L_s - \frac{3}{\eps} - \frac{3}{8} L_s^2 + \cdots \right).
\end{equation}
The factor of $\frac{3}{4}$ in the $\frac{L_s}{\eps}$ term is exactly what is needed to cancel the non-local divergences in the sum of the double bubble and ice-cream cone graphs. Adding everything up, result is that up to 2-loops, the leading-logs of the renormalized vertex function are
\begin{equation}
   \Gamma_4 = - \lambda - \lambda\left( \frac{\lambda}{16\pi^2}\right)\frac{1}{2}\left( L_s + L_t + L_u \right) -\lambda 
   \left(\frac{\lambda}{16\pi^2}\right)^2 \frac{3}{4}\left( L_s^2 + L_t^2 + L_u^2 \right) + \cdots.
\end{equation}
The $\mu$ dependence of this expression is  consistent with the LL RGE in Eq.~\eqref{LLRGE1}. 

At next order there are bubble-insertion graphs that can depend on multiple scales, but not at leading log. Consider for example, the baseball diagram
\begin{align}
    \baseball &= \lambda^4\int \frac{d^d k}{(2\pi)^d}\frac{1}{k^2(P-k)^2} B[(k-p_1)^2]B[(k+p_4)^2] \\
 &=\frac{\lambda^4}{\eps^2 (32\pi^2)^2} (\mu^2)^{3\eps} \int \frac{d^dk}{(2\pi)^d}
 \frac{[(k+p_1)^2]^{-\eps}[(k-p_4)^2]^{-\eps}}{k^2(P-k)^2}.
\end{align}
To see which $L^3$ terms appear, we can expand the various momenta at large $k$. For the inner bubbles
\begin{equation}    
    [(k+p)^2]^{-\eps}=(k^2+2 k\cdot p)^{-\eps}=(k^2)^{-\eps}(1-2 \eps \frac{k\cdot p}{k^2}+ \cdots). 
\end{equation}
To get $p_1\cdot p_4$ dependence, we would need the $\eps^2 (k\cdot p_1)(k\cdot p_4)/(k^2)$ cross term. This is down by a factor of $\eps^2$ so it cannot be leading log. For the outer bubble, there is no such $\eps$ suppression and we can get $L_s^3$ terms. Indeed,
\begin{equation}
\frac{1}{\eps^2} (\mu^2)^{3\eps}
\int d^dk \frac{(k^2)^{-2\eps}}{k^2(P-k)^2}= \frac{1}{4}\left(\frac{1}{\eps^3} +\cdots+ L_s^3 + \cdots \right).
\end{equation}

More generally, every graph we generate by sequentially inserting bubbles will be 2-particle-reducible in the channel $S=s,t,u$ determined by the first primitive graph. The leading logarithms of such graphs  depend only on $L_S$. That is, at LL level, $\Gamma_4$ is a sum over $L_s^n + L_t^n + L_u^n$ with the overall coefficient fixed by the LL RGE. 
Explicitly,
\begin{align}
    \Gamma_4^\LL(s,t,u) &=  
    -\lambda -\lambda \sum_{n=1}^\infty
    \frac{1}{3}\left(\frac{\beta_0 \lambda}{2}\right)^n
     (L_s^n + L_t^n + L_u^n)  \label{Gamma4LL}
      \\  
&= -\frac{\lambda}{3}  \left(
\frac{1}{1 - \frac{\beta_0}{2}  \lambda L_s} 
+ \frac{1}{1 - \frac{\beta_0}{2} \lambda L_t} 
+ \frac{1}{1 - \frac{\beta_0}{2} \lambda  L_u}
\right),
\end{align}
which is kind of an average of the LL resummation in each of the 3 channels separately.\footnote{Note that this result is different from the parquet approximation where the 4-point function is approximated as $\Gamma_4 = \lambda \left(1 +\frac{\lambda}{32\pi^2}(L_s + L_t + L_u)\right)^{-1}$ as in~\cite{Kugler:2018}.  The parquet approximation suggests that $L_s L_t$ cross terms appear at 2-loops which we do not find, and it will generate an infinite number of Borel singularities.}

For the asymptotic behavior of the two point function including these nested $s,t,u$ bubble insertions we need to calculate
\begin{equation}
  \Sigma_{stu}(p^2) =\tadpoleblob
   = \frac{1}{2p^2} \int \frac{d^4 k}{(2 \pi)^4} 
  \frac{1}{k^2} \Gamma_4 \left((p + k)^2, (p - k)^2, 0\right).
  \label{Sigstu}
\end{equation}
Since when the 4-point graphs are inserting into the tadpole vertex we automatically have $u=0$, all the $\ln u $ terms in $\Gamma_4$ gives scaleless integrals like the 1-loop tadpole; these vanish in dimensional regularization and can be dropped.

To get the asymptotic behavior, we insert Eq.~\eqref{Gamma4LL} into Eq.~\eqref{Sigstu}. Since $\Gamma_4^\LL$ is a sum of separate channels and symmetric in permutations of $s\leftrightarrow t$ we can write
\begin{equation}
  \Sigma_{stu}^\LLL 
  = 2\times \frac{\lambda}{2} \int \frac{d^d k}{(2 \pi)^d}  \frac{1}{p^2(p + k)^2} \frac{1}{1 -\frac{\beta_0}{2}\lambda \ln\frac{k^2}{\mu^2}}.
\end{equation}
This is the same as the $s$-channel SDE in Eq.~\eqref{SigmasSDE} up to an overall factor of $2$ and the replacement of $a=\frac{\lambda}{32\pi^2}$ with $\frac{\beta_0}{2}\lambda = \frac{3}{2} a$. The result is that
\begin{equation}
  \Sigma_{stu} \sim 2\int_0^{\infty} d t 
  e^{-\frac{2t}{\beta_0 \lambda}} \frac{e^{t L} - 1}{(1+t)(2+t)} 
=\beta_0 \int_0^{\infty} d t 
  e^{-\frac{t}{\lambda}} \frac{e^{\frac{\beta_0}{2} t L} - 1}{(1+\frac{\beta_0}{2}t)(2+\frac{\beta_0}{2} t)}.
\end{equation}
The singularities in the Borel plane are then identified as renormalons with $t = -\frac{2}{\beta_0}$ and $-\frac{4}{\beta_0}$.

\section{Hopf algebraic approach \label{sec:Hopf}}
We have derived the asymptotic expansion of a number of classes of diagrams in various quantum field theories. The asymptotic behavior of the Hopf series was first derived using Hopf algebraic techniques. In this section, we review the relevant Hopf algebra and attempt to summarize some of the key mathematical results in the
language of quantum field theory. There are many reviews of the Hopf algebra (e.g. ~\cite{Kreimer:2012rm,Manchon:2001bf,Manchon:2004qa,Borinsky:2018mdl,yeatsthesis,van_Suijlekom_2009,Weinzierl:2015nda,Balduf:2024pgk}) We do not attempt to be  comprehensive here; instead we attempt to isolate the key steps, notations and definitions needed to get efficiently between the formal mathematical properties of the Hopf algebra and some results relevant for physics.

\subsection{Connes-Kreimer Hopf algebra}
The Connes-Kreimer Hopf algebra is built on the observation  that the BPHZ forest formula perfectly matches the algebra of rooted trees. Recall that the Zimmerman forest formula~\cite{zimmermann:1969} expresses which counterterms are needed to cancel the non-local subdivergences of a divergent graph in QFT. For example, consider the 3-loop chain graph in $\phi^2\chi$ theory (cf. Eq.\eqref{chainexplicit})
\begin{equation}
    \chain   =
    -\frac{1}{3 \eps ^3}+\frac{\ln p^2 }{\eps^2}+\cdots.
\end{equation}
 Because of the $\ln p^2/\eps^2$ term, this diagram cannot be renormalized by rescaling a local term (polynomial in derivatives) in the Lagrangian. The BPHZ forest formula says that such non-local divergences are canceled by adding to a diagram other Feynman diagrams involving lower-order counterterms. In this case, we add
 \begin{equation}
 \bubbletwoct + \chainctR+ \chainctL \nonumber,
 \end{equation}
 and the sum will no longer have any divergences with non-analytic momentum dependence. Exactly which counterterm diagrams are needed is given by the forest formula: you contract every possible subgraph to a point and replace by the appropriate counterterm.  This forest formula has the same recursive structure as an algebra of rooted trees where the root is the main graph and the nodes represent divergent subdiagrams. 
 For example,
the 3-loop chain graph maps to a 3-node tree:
\begin{equation}
\chaintree[0.8]=\chain, 
\end{equation}
with the blue and green dots representing the two  1-loop subdivergences and the top dot the full graph.
The relevant divergent diagrams are 
\begin{equation}
    {\blue \bullet} ={\green \bullet} = \bubble,
    \quad
    \twostick[0.8]{black}{darkblue} = \doublebubble,
    \quad
    {\blue \bullet} {\green \bullet}
    = \left( \bubble \right)^2,
\end{equation}
with the first and the last being subdiagrams of the 3-loop chain. The middle diagram is formed by shrinking the subdiagram on the left or right to a point. We add color to the dots just for clarity; there is no difference between ${\green \bullet}$, ${\blue \bullet}$ and $\bullet$. 
The Connes-Kreimer approach to renormalization begins with the mapping between diagrams and trees and then uses algebraic machinery to establish properties and relations among counterterms and renormalized amplitudes.

In the Connes-Kreimer (CK) Hopf algebra $A$, 1PI diagrams are represented as rooted trees. The spin, momenta and other quantum numbers of the particles are left implicit in a {\bf decoration}.
Diagrams that are not 1PI are forests which are sets of trees. The Hopf algebra has a commutative
product $m : A \otimes A \rightarrow A$, given by the disjoint union of the
diagrams, and a coproduct $\Delta : A \rightarrow A \otimes A$
which encodes the forest formula.
We sometimes write its action as $\Delta \Gamma = \sum_\gamma \gamma \otimes \Gamma/\gamma$ where the sum is over divergent subgraphs.
The Hopf algebra is not co-commutative: the left and right entries of the
coproduct are distinguished. 
The left entries correspond to divergent subdiagrams $\gamma$ of the bigger diagram $\Gamma$. The corresponding right entries $\Gamma/\gamma$ are the remainder of the original diagram after the subdiagram is contracted to a point. 
For example,
\begin{equation}
  \Delta  \chaintree[0.8] = 
 \oneA \otimes  \chaintree[0.8] +   \chaintree[0.8] \otimes \oneA +
  {\blue  \bullet}\,
    {\green\bullet}
    \otimes
    \bullet
+
{\green \bullet}
\otimes
\twostick[0.8]{black}{darkblue}
 +
 {\blue \bullet}
 \otimes
 \twostick[0.8]{black}{darkgreen}.
 \label{examcop}
\end{equation}
The 5 entries on the right-hand side correspond to
\begin{equation}
 \chain  + \linectO[2pt] + \bubbletwoct + \chainctR+ \chainctL \nonumber,
\end{equation}
which are the original graph, the overall counterterm, and all the subgraphs with counterterms as dictated by BPHZ.
We denote the empty graph by $\oneA$, which is
the multiplicative identity of the algebra.

The coproduct of $\Gamma$ always has two special entries: $\oneA\otimes \Gamma$ 
for the graph itself, and $\Gamma \otimes \oneA$ for the overall counterterm. So the non-trivial information in the coproduct is contained in the
reduced coproduct $\Dtilde = \Delta - \id \otimes \oneA
- \oneA \otimes \id$. In the example in Eq.~\eqref{examcop},
\begin{equation}
\Dtilde  \chaintree[0.8] = 
  {\blue  \bullet}\,
    {\green\bullet}
    \otimes
    \bullet
+
{\green \bullet}
\otimes
\twostick[0.8]{black}{darkblue}
 +
 {\blue \bullet}
 \otimes
 \twostick[0.8]{black}{darkgreen}. \label{chaincoproduct}
\end{equation}
Algebra elements with $\Dtilde \Gamma = 0$ are called \textbf{primitive}.
They have no subdivergences. All 1-loop graphs are primitive. Any
finite graph at any-loop order is primitive. In addition, there are multi-loop
divergent primitive graphs, such as the 2-loop crossed triangle. As trees, any
primitive graph is often written as $\bullet$ with the distinguishing features left
implicit.

A Hopf algebra must have a counit $\epsco: A \rightarrow
\mathbb{C}$ which is a linear map satisfying, 
\begin{equation}
(\epsco \otimes
\id) \circ \Delta = \id = (\id \otimes \epsco)
\circ \Delta,
\end{equation} 
where $\id$ is the identity map. More intuitively, the
counit projects out the component of an algebra element along the
$\oneA$ direction. That is $x = \epsco (x)
\oneA + \tilde{x}$ where $\epsco (\tilde{x}) = 0$. In
the CK Hopf algebra, Feynman diagrams do not have any $\oneA$
component, so they satisfy $\epsco (x) = 0$. Such algebra
elements are said to belong to the \textbf{augmentation ideal}: $I = \{ x |
\epsco (x) = 0 \nobracket \}$. A Hopf algebra must also have a linear
endomorphism $S$ called the \textbf{antipode} which satisfies 
\begin{equation}
    m \circ (S \otimes \id) \circ \Delta (x) = m \circ (\id \otimes S) \circ \Delta (x) = \epsco (x) \oneA.
\end{equation}
For any $x \in I$ this means
\begin{equation}
  S (x) = - x - m \circ (S \otimes \id) \Dtilde x \label{Srec},
\end{equation}
which is a recursive formula for computing the antipode. On a primitive graph
$S (x) = - x$ which is the additive inverse. More generally, the antipode
allows for the construction of inverses in a sense that will be clear
momentarily.

The evaluation of a Feynman diagram in dimensional regularization is a
homomorphism called a
\textbf{character} from the algebra $A$ to functions of external momenta. Working at fixed external momenta and fixed spacetime dimension $d$ the target algebra is the complex numbers $\phi : A \rightarrow \mathbb{C}$.
Alternatively, if we expand around the critical dimension, e.g. $d = 6 - 2
\eps$, the target algebra can be thought of as  Laurent series in $\eps$ which we write as $\phi :
A \rightarrow L$ with $L=\mathbb{C} [\varepsilon, \varepsilon^{- 1}]$. All characters
should be algebra homomorphisms, meaning $\phi (x y) = \phi (x) \phi (y)$, and
{\bf{unital}}, meaning $\phi (\oneA) = 1$. 

An important feature of the
Hopf algebra is that characters form a Lie group with group multiplication given
by the \textbf{convolution product}, defined by
\begin{equation}
  \phi_1 \ast \phi_2 = m \circ \phi_1 \otimes \phi_1 \circ \Delta \label{convdef}.
\end{equation}
Using Sweedler
notation, $\Delta x = x^{(1)} \otimes x^{(2)}$ with the sum over entries
implicit, this means that for an algebra element $x$ that
\begin{equation}
  (\phi_1 \ast \phi_2) (x) = \phi_1 (x^{(1)}) \cdot \phi_2 (x^{(2)}),
\end{equation}
with $\cdot$ on the right side indicating multiplication in the target algebra $\mathbb{C}$ or
$\mathbb{C} [\varepsilon, \varepsilon^{- 1}]$. The identity element of the
group is the counit $\oneG = \epsco$. Group inverses are
constructed with the antipode. To see that, we compute
\begin{multline}
  \phi \ast (\phi \circ S) (x) = m \circ (\phi \otimes \phi \circ S) \circ
  \Delta (x)= m \circ (\phi \otimes \phi) \circ (\id \otimes S)\circ \Delta (x)
\\
= \phi \circ m \circ (\id \otimes S) \circ\Delta (x)
=\phi\big( \epsco(x) \oneA \big)
  = \epsco (x) \phi
  (\oneA) =  \epsco (x) = \oneG (x).
\end{multline}
So that $\phi \ast (\phi \circ S) = \oneG$ and therefore $\phi^{- 1}
= \phi \circ S$. Then a recursive formula for $\phi^{- 1}$ is
\begin{equation}
  \phi^{- 1} = - \phi - m \circ (\phi^{- 1} \otimes \phi) \circ \Dtilde,
  \label{invruleOld}
\end{equation}
which follows directly from Eq.~\eqref{Srec}.

Since the set of characters form a group for any target algebra,
the set of Hopf algebra endomorphisms $G_0 = \operatorname{Hom} (A, A)$ are a group. The identity map is in this group, $\id \in G_0$ but is not equal to the
identity element of the group which is $\oneGzero = \oneA
\epsco$. Because of the defining property of the antipode, $S
\ast \id = \id \ast S = \oneA \epsco =
\oneGzero$, the antipode can be thought of as the inverse of the
identity map in the character group of algebra endomorphisms. We can also
define $\pi = \id - \oneGzero$ as the projector onto the
augmentation ideal. Then, 
\begin{equation}
  S = \id^{- 1}  = (\oneGzero + \pi)^{- 1} = \sum_{k
  = 0} (- 1)^k \pi^{\ast k} = \oneGzero - \pi + \pi \ast \pi -
  \pi^{\ast 3} + \cdots. \label{Sformula}
\end{equation}
Thus $S$ is a sum over elements in the repeated coproduct with a sign
determined by the number of trees $k$ in each forest that appears. For
example,
\begin{equation}
S\left( \chaintree[0.6]\right) = -\left( \chaintree[0.6]\right)
+({\green \bullet}) \left(\twostick[0.6]{black}{darkblue}\right)
 +( {\blue \bullet}) \left(\twostick[0.6]{black}{darkgreen}\right)
- ( {\blue  \bullet})\, ( {\green\bullet})  ( \bullet).
\end{equation}
The signs are fixed by having 1, 2 or 3 trees. This perhaps gives some minimal
intuition for what $S$ actually does. More examples are in Appendix~\ref{app:examples}.

Let's return to the character group $G = \text{Hom} (A, L)$ where $L
=\mathbb{C} [\eps, \eps^{- 1}]$. Define the \textbf{bare
character} $\phi \in G$ to evaluate a diagram in $d$ dimensions in the
$\eps$ expansion. Any IR-finite $\ell$-loop diagram $x$ has $\phi (x) = \sum_{n = -
\ell}^{\infty} c_n \eps^n$ with a finite number of negative powers of
$\eps$ and possibly an infinite number of positive powers. From such a
character we can extract the UV divergences. In $\msbar$ this would
correspond to a map $R$ which acts as 
\begin{equation}
R [\phi (x)] = \sum_{n = - \ell}^{-1} c_n \varepsilon^n.   
\end{equation}
This combined map $R \circ \phi$ is not a
character, since it does not satisfy the homomorphism property $R [\phi (x) \phi (y)] = R [\phi (x)] R
[\phi (y)]$ (take for example $\phi (x) = \phi (y) = \frac{1}{\varepsilon} +
\varepsilon$). Remarkably, however, a map constructed recursively as
\begin{equation}
  \phi_- = - R [\phi + m \circ (\phi_- \otimes \phi) \Dtilde]
  \label{invrule}
\end{equation}
is a homomorphism and a character. 
We call this character $\phi_-$  the \textbf{counterterm character}. 
For a graph $x$ if the coproduct agrees with the BPHZ forest formula, then $\phi_-(x)$ will produce a local counterterm. 
That is $\phi_- (x) \in \mathbb{C} [\eps^{- 1}, p_i^{\mu}]$ is a polynomial in
$\eps^{-1}$ and the external momenta. This fact follows from the BPHZ  proof~\cite{bogoliubov:1957,hepp:1966,zimmermann:1969} {\it if} the coproduct is constructed correctly. 

We also can define the
\textbf{renormalized character} as
\begin{equation}
  \phi_+ = \phi_- \ast \phi. \label{BirkhoffF}
\end{equation}
The renormalized character will be finite at $\varepsilon = 0$. Note that
\begin{align}
  \phi_+ (x) &= \phi_- (\oneA) \phi (x) + \phi_- (x) \phi
  (\oneA) + m \circ (\phi_- \otimes \phi) \Dtilde x \nonumber\\
  &= (1 - R) [\phi + m \circ (\phi_- \otimes \phi) \Dtilde x], 
  \label{phiplusform}
\end{align}
which is valued in the complement of the image of $R$. That is, $\phi_+ \in
\mathbb{C} [\epsilon]$.

The decomposition of a character $\phi \in \mathbb{C}
[\varepsilon, \varepsilon^{- 1}]$ as $\phi = \phi_-^{- 1} \ast \phi_+$ into a
part $\phi_-^{- 1} \in \mathbb{C} [\varepsilon^{- 1}]$ and $\phi_+ \in
\mathbb{C} [\varepsilon]$ is an example of \textbf{Birkhoff factorization}~\cite{birkhoff1909}.
Birkhoff factorization characterizes a class of solutions to the
Riemann-Hilbert problem (\#21 of Hilbert's 23 problems) for group-valued meromorphic maps on the complex plane. This view of renormalization as factorization is a powerful perspective and  a foundational aspect of the Hopf algebra approach. The original Connes-Kreimer papers~\cite{Connes:1999yr,Connes:1999zw} entitled ``Renormalization in Quantum-Field Theory and the Reimann-Hilbert Problem'' explore this connection.

We emphasize that the abstract recursive relations in Eqs.~\eqref{invrule} and~\eqref{phiplusform} are in 1-to-1 correspondence with  renormalization using local counterterms and the BPHZ forest formula. For example, using the reduced coproduct in Eq.~\eqref{chaincoproduct} 
\begin{align}
    \phi_+\left( \chaintree[0.6]\right) &= \phi\left(\chaintree[0.6]\right)
    +\phi_-\left(\chaintree[0.6]\right)
    +\phi_-({\blue \bullet}\,{\green \bullet}) \phi(\bullet)
+
\phi_-({\blue \bullet})
\phi\left(\twostick[0.6]{black}{darkgreen}\right)
+
\phi\left(\twostick[0.6]{black}{darkblue}\right)
\phi_-({\green \bullet})
\\
&= \chain + \linectO[2pt] + \bubbletwoct + \chainctR+ \chainctL \nonumber.
\end{align}
This is exactly the sum of graphs needed to cancel all the divergences in the 3-loop chain graph. In this way, the coproduct encodes the divergent subgraphs, $\phi$ gives the bare amplitude of a graph or subgraph, 
$\phi_-$ gives the combination needed to produce a local counterterm, and $\phi_+$ gives the renormalized amplitude.

An important point is that Eq.~\eqref{invrule} will not produce a character for any $R$, but it will if
$R$ is a Rota-Baxter operator~\cite{
rota1969,baxter1960,
Connes:1998qv,
Manchon:2001bf,
EbrahimiFard:2004twg,
Bergbauer:2005fb,
EbrahimiFard:2006zqz}. A Rota-Baxter operator is an idempotent operator ($R^2 = R$) with
the property
\begin{equation}
  R (x) R (y) = R [R (x) y + \nobracket \nobracket x R (y) + \lambda x y],
\end{equation}
for some $\lambda$. In the case of $R$ for $\msbar$, $\lambda = - 1$. The proof that Rota-Baxter operators produce characters can be found in many places (e.g.~\cite{Manchon:2001bf}).

An alternative subtraction scheme to $\msbar$ is kinematic
renormalization, where the finite parts of the counterterms are determined by the
amplitude at a particular scale. An example of kinematic renormalization is
 on-shell renormalization in QED (see e.g.~\cite{Schwartz:2014sze}). In the particular case of a
single-scale function, such as the self energy graphs $\Sigma (L)$ which depend in dimensional regularization only on $L = \ln p^2/\mu^2,$ we can define the {\bf momentum subtraction} (MOM) operator as 
\begin{equation}
    R_p \big[f (L, \varepsilon)\big] = f(0, \varepsilon).
\end{equation}
In contrast to minimal subtraction, momentum subtraction
subtracts not just the poles but also any non-logarithmic finite parts. This
$R_p$ operator is an algebra homomorphism $R_p (x y) = R_p (x) R_p (y)$ and also a
Rota-Baxter operator with $\lambda = 0$. Therefore it also leads to Birkhoff
factorization $\phi_+ = \phi_- \ast \phi$ and $\phi_-$ in Eq.~\eqref{invrule} is also a
character with $R$ replaced with $R_p$. Because $R_p$ is a
homomorphism then
\begin{equation}
  R_p [\phi_- (x)] = R_p [\phi_+ \ast \phi^{- 1} (x)] = R_p [\phi_+ (x^{(1)})
  \phi^{- 1} (x^{(2)})] = R_p [\phi_+ (x^{(1)})] R_p [\phi^{- 1} (x^{(2)})],
\end{equation}
where the homomorphism property of $R_p$ was used in the last step (so this doesn't hold for $R$). Then since $R_p [\phi_+ (x^{(1)})] =
\epsco (x^{(1)})$ the only term from the coproduct that survives is the one with
$x^{(1)} = \oneA$ and $x^{(2)} = x$. So, using that $R_p$ is idempotent we get
\begin{equation}
  \phi_- = R_p \circ \phi^{- 1}.
\end{equation}
This feature makes momentum subtraction very easy to use: one can simply
compute $\phi^{-1}$ recursively using Eq.~\eqref{invrule} on the full bare
amplitudes and then act with $R_p$ only once at the end to get $\phi_-$, without ever expanding in $\eps$.
In fact, $\eps$ plays no role at all in
$R_p$. 
In contrast, with  $R$ one needs to recursively renormalize; one cannot construct $\phi_-(x)$ from $\phi^{-1}(x)$ alone.

Let us finally note that cancellation of non-local divergences requires the coproduct to encode the forest formula which is not guaranteed by the Hopf algebra structure itself.
If one designs the Hopf algebra correctly, then one can translate the BPHZ proof into Hopf algebra language. However, we are not aware of a proof of renormalizability using the Hopf algebra which is notably simpler than BPHZ. Rather than explore further in this direction, we will assume the coproduct aligns with the forest formula and examine next the way the Schwinger-Dyson equations and the renormalization group can be understood through the Hopf algebra.

\subsection{Combinatorial Schwinger Dyson equations}
We are now ready to review how the Schwinger-Dyson equations are expressed
using the CK Hopf algebra. We begin with SDE for the Hopf graphs in the $\phi^2 \chi$ theory in $d = 6$, as in Eq.~\eqref{HopfSDE}:
\begin{equation}
  \Sigma^0_H (p^2) = \frac{a}{\pi^3} \int d^6 k \frac{1}{p^2 k^2 (p + k)^2}
  \frac{1}{1 - \Sigma^0_H (k^2)} \label{SDEv2}.
\end{equation}
This SDE encodes the insertion of the bare Green's function $G_2^0 = [k^2 (1 -
\Sigma_H^0 (k^2))]^{- 1}$ into the $\phi$ leg of the primitive self-energy
graph. Each Feynman diagram summed into $\Sigma_H^0$ corresponds to a unique element of the
Hopf algebra and their sum $x$ is also in $A$.
 The SDE in Eq.~\eqref{SDEv2} can then be written as
\begin{equation}
  B_{\bullet} (x) =
  \Bx,
\end{equation}
where $B_{\bullet}: A \rightarrow A$ is an algebra map which inserts $\phi(x)$ into the internal $\phi$ propagator of the primitive self-energy graph. That is
\begin{equation}
  \phi\big(B_\bullet(x)\big)(p^2) = 
 \frac{a}{\pi^3 p^2} \int d^6 k 
  \frac{[\phi(x)(k^2)] }{k^2 (p + k)^2}.
  \label{Bbulletdef}
\end{equation}
Then, defining $X_H$ so that $1 - \Sigma_H^0 = \phi (X_H)$  the SDE can be written as
\begin{equation}
  X_H = \oneA - a B_{\bullet} (X_H^{- 1}).
\end{equation}
Here, $X_H^{-1} \in A$ is defined through its series expansion in $a$ using $X_H =\oneA + \mathcal{O} (a)$. This form of the SDE is called a \textbf{combinatorial SDE} or cSDE.

The operator $B_{\bullet}$ must be compatible with the coproduct, meaning that
the coproduct $\Delta B_{\bullet} (x)$ must contain all of the subdivergences
of $B_{\bullet} (x)$ just like $\Delta \Gamma = \sum_\gamma \gamma \otimes \Gamma/\gamma$ contains the divergent subgraphs $\gamma \subset \Gamma$ in the left entry.
To assure this we need $B_{\bullet}$ to satisfy
\begin{equation}
  \Delta B_{\bullet} = (\id \otimes B_{\bullet}) \Delta + B_{\bullet}
  \otimes \oneA. \label{corul}
\end{equation}
The first term on the right
accounts for the proper subdivergences, which are the subdivergences of the inserted graph $x$ with the new loop being appended to the right element of the coproduct (i.e. not contracted). The $B_{\bullet} \otimes \oneA$ term accounts for the new  (improper) subdivergence, namely the entire graph $B_\bullet(x)$. 
A linear map satisfying Eq.~\eqref{corul} is an example of a \textbf{Hochschild 1-cocycle}.


The Hochschild cocycle condition when combined with the cDSE implies~\cite{Foissy}
\begin{equation}
  \Delta X_H = \sum_{\ell = 0}^{\infty} X_H^{1 - 2 \ell} \otimes [X_H]_{\ell},
  \label{DeltaXH}
\end{equation}
where $[X]_{\ell}$ means the component of order $a^{\ell}$ of $X$, i.e the sum of all the
$\ell$-loop amplitudes in $X$. To understand this formula, it's helpful to think of the coproduct for a graph $\Gamma$ as
$\Delta \Gamma = \sum_\gamma \gamma \otimes \Gamma/ \gamma$
where $\gamma$ is a subgraph of $\Gamma$ while $\Gamma/\gamma$ is the remaining graph after the subgraph has been contracted to a point.
For a 1PI Hopf graph $\Gamma$, no matter what $\gamma$ is, the remainder $\Gamma/\gamma$ is always a 1PI Hopf graph. 
Therefore, the coproduct can be written as a sum of Hopf graphs $x$  as
$\Delta X_H = \sum_{x\in X_H} \gamma_x \otimes x$ for some algebra elements $\gamma_x$.
For a given $x$, $\gamma_x$ is the sum over all subgraphs that can possibly have been contracted out of any Hopf graph in $X_H$ to leave $x$. These subgraphs do not have to be 1PI, so they are not simply elements of $X_H$. However, each
$\ell$ loop Hopf graph $x$ has $2 \ell - 1$ $\phi$ propagators. Each such
propagator is a slot where we could have contracted any Hopf graph. So the
left entry is simply $\gamma_x = G_2^{(2 \ell - 1)}$ where $G_2 = \frac{1}{1 - \Sigma_H} =
\frac{1}{X_H}$ is the combinatorial connected-Green's-function. This explains
Eq.~\eqref{DeltaXH}.

For the rainbow graphs, we can define $X_R$ so that $\phi (X_R) = 1 +
\Sigma_R$. Then the cSDE has the form $X_R = \oneA + a B_{\bullet}
(X_R)$. So both the Hopf and Rainbow graphs lead to cSDEs of the form 
\begin{equation}
  X = \oneA \pm a B_{\bullet} (X^s),
  \label{cSDEs}
\end{equation}
where $s = 1$ and the plus sign used for rainbow and $s = - 1$ and the minus sign used for Hopf. The sign can be absorbed into $a$ or $B_\bullet$ if desired and does not affect the algebraic properties leading to Eq.~\eqref{DeltaXH}. The generalization of Eq.~\eqref{DeltaXH} when
Eq.~\eqref{cSDEs} is satisfied is
\begin{equation}
  \Delta X = \sum_{\ell = 0}^{\infty} X^{1 + (s - 1) \ell} \otimes [X]_{\ell}.
  \label{DeltaXgen}
\end{equation}
In the rainbow case this reduces to $\Delta X = X \otimes X$ which is
particularly simple. The fact that for any $s$ all the entries of the
coproduct are disjoint unions of the trees in $X$ means that these graphs form
a \textbf{closed Hopf subalgebra}~\cite{foissy2007faadibrunosubalgebras}.

Eq.~\eqref{DeltaXgen} can be combined with Birkhoff factorization, Eq.~\eqref{BirkhoffF} to generate the RGE almost immediately. We present this derivation in Section~\ref{sec:isHM}.
Before doing so, we discuss the abstract algebraic approach of Connes and Kreimer to the renormalization group.

\subsection{Anomalous dimension and the Lie Algebra}
To understand the Hopf algebraic approach to the renormalization group, we need to examine how  renormalized amplitudes depend on scale.
To begin, we can consider the \textbf{convolution anomalous
dimension} $\bgamma$ of the renormalized character, defined by
\begin{equation}
   \partial_L \phi_+ =  \bgamma \ast \phi_+ \label{anomdef},
\end{equation}
where $\partial_L = - \frac{1}{2} \mu \partial_\mu$ to be consistent with the convention $L = \ln \frac{p^2}{\mu^2}$. 
In dimensional regularization a bare $\ell$-loop
amplitude has the form
\begin{equation}
  A_{\ell} (p) = \int \mu^{2 \eps} \frac{d^d k_1}{(2 \pi)^d} \ldots
  \mu^{2 \eps} \frac{d^d k_{\ell}}{(2 \pi)^d} \mathcal{I} (k, p),
\end{equation}
for some integrand $\mathcal{I}$. Therefore, 
\begin{equation}
  \partial_L A_{\ell} =- \eps \ell A_{\ell}.
\end{equation}
We next introduce the \textbf{grading operator} $Y : A \rightarrow A$ which
acts on a $\ell$-loop algebra element $x$ as $Y(x) = \ell(x) x$ where $\ell(x)$ gives the loop order of $x$.
Then 
\begin{equation}
  \partial_L \phi (x) = - \phi (\eps Y \circ x).
  \label{dLphi}
\end{equation}
Note that $Y$ is a \textbf{derivation}:
\begin{equation}
  Y (a b) =  [\ell (a) + \ell (b)] a b = [\ell (a) a] b + a
  [\ell (b) b] = Y (a) b + a Y (b).
\end{equation}
In addition, since $\Delta$ preserves the grading, 
\begin{equation}
  \Delta Y  = (Y \otimes \id+ \id \otimes Y)  \Delta,
\end{equation}
so that $Y$ is also a \textbf{co-derivation} as well.

For the renormalized character things are more complicated since it involves summing over products of counterterms, which are $\mu$ independent,
and
subdiagrams which have lower loop order. Writing $\Dtilde x =
\tilde{x}^{(1)} \otimes \tilde{x}^{(2)}$ for the reduced coproduct, Eq.
\eqref{phiplusform} can be written as
\begin{equation}
  \phi_+ (x) = \phi (x) + \phi_- (x) + \phi_- (\tilde{x}^{(1)}) \phi
  (\tilde{x}^{(2)}).
\end{equation}
The counterterm character $\phi_-$ is
independent of $\mu$. So,
\begin{align}
  \partial_L \phi_+ (x) &= \partial_L \phi (x) + \phi_- (\tilde{x}^{(1)})
  \partial_L \phi (\tilde{x}^{(2)})
\\
  &=- \phi (\eps Y \cdot x) - \phi_- (\tilde{x}^{(1)}) \phi (\eps
  Y \tilde{x}^{(2)})
\\
  &= - m\circ(\phi_- \otimes \phi) (\id \otimes \eps Y) [x \otimes
  \oneA + \oneA \otimes x + \tilde{x}^{(1)} \otimes
  \tilde{x}^{(2)}]
\\
  &=- \phi_- \ast ( \phi \circ \varepsilon Y) (x).
\end{align}
Since $Y$ is a coderivation
\begin{equation}
  \phi \circ Y = (\phi_-^{- 1} \ast \phi_+) \Delta Y = (\phi_-^{- 1} \circ Y)
  \ast \phi_+ + \phi_-^{- 1} \ast (\phi_+ \circ Y).
\end{equation}
So
\begin{equation}
  \partial_L \phi_+ =- \phi_- \ast (\phi_-^{- 1} \circ \eps Y) \ast
  \phi_+ - (\phi_+ \circ \eps Y).
\end{equation}
Now we can take $\eps \rightarrow 0$. The $(\phi_+ \circ \eps
Y)$ term vanishes, since $\phi_+ \in \mathbb{C} [\eps]$ , but the other
term does not since $\phi_-$ is singular. Then
\begin{equation}
  \partial_L \phi_+ =- (\phi_-^{- 1} \circ S) \ast (\phi_-^{- 1} \circ
  \eps Y) \ast \phi_+ = -\phi_-^{- 1} (\eps S \ast Y) \ast \phi_+.
\end{equation}
Thus
\begin{equation}
  \bgamma = -\lim_{\eps \rightarrow 0}  \phi_-^{- 1} (\eps S \ast
  Y). \label{CKgamma}
\end{equation}
This is a main result of Connes and Kreimer~\cite{Connes:2000fe}.\footnote{Our Eq.~\eqref{CKgamma} is Eq. (30) in Section 2 of~\cite{Connes:1999zw} where it is written as $\beta = Y \operatorname{Res} \varphi_\eps$
where their
$\varphi_{\eps}$ is our $\phi_-^{- 1}$ and their $\beta$ is our $\bgamma$.}

In momentum subtraction things are even simpler. There, $\phi_- = R_p \circ \phi \circ S$ and therefore $\phi_-^{- 1} = \phi_- \circ S = R_p \circ \phi$ 
so that
\begin{equation}
  \bgammap = -\lim_{\eps \to 0}\phi_-^{- 1} (\eps S \ast Y) = 
  - \lim_{\eps \to 0}R_p \circ \phi (\eps
  S \ast Y) = 
  -\lim_{L,\eps \rightarrow 0} \phi (\eps S \ast Y).
\end{equation}
This is a somewhat surprising formula: it says that $\phi \circ S \ast Y$ acting on any algebra element will
only ever have $\frac{1}{\varepsilon}$ poles. 
Moreover, using Eq.~\eqref{dLphi} we have $\partial_L \phi(\eps S\ast Y) = -\eps^2\phi((S\ast Y)\circ Y )$ which then must vanish at $\eps\to 0$ and therefore $\bgammap$ cannot depend on $L$. So
we can simply drop the limit and write
\begin{equation}
  \bgammap = -\phi (\varepsilon S \ast Y), \label{gammap}
\end{equation}
which has a finite limit as $\varepsilon \rightarrow 0$. This means that if we
multiply the subgraphs by their loop order to compute $\phi \circ (S \ast Y)$
we will get some combination of the various loop integrals which has at
most a $\frac{1}{\eps}$ pole. The residue of that pole is the
convolution anomalous dimension of the renormalized character. In contrast
$\phi \circ S$, without the grading, has poles at arbitrarily high order.

Additionally, we can observe that in the momentum subtraction scheme there
are no constant terms in the renormalized amplitudes, so $\phi_+  = \oneG$ at
$L = 0$.
Then since $\bgammap \ast \oneG = \bgammap$, Eq.~\eqref{anomdef}
gives
\begin{equation}
  \bgammap = (\partial_L \phi_+)_{L = 0} \label{gammaPform}.
  \end{equation}
This is a convenient formula, since it allows us to anomalous dimension
$\bgammap$ directly from the renormalized amplitude.

The operator $D = S \ast Y$ appearing in Eq.~\eqref{gammap} called a
\textbf{Dynkin operator}. It is an element of the Lie algebra of the
character group. The \textbf{Lie algebra} $\mathfrak{g}$ is the tangent
space of the character group $G$ at the identity. Its elements are maps
$\bdelta : A \rightarrow \mathbb{C}$ which vanish on the identity $\bdelta
(\oneA) = 0$ and are \textbf{infinitesimal characters}, so they
satisfy
\begin{equation}
  \bdelta (x y) = \bdelta (x) \epsco (y) + \epsco (x)
  \bdelta (y),
\end{equation}
with $\epsco$ the counit. One can check that $D$ satisfies this
property.
It implies characters always vanish on decomposable elements $z = x y \in I^2$, since $\epsco (x) = \epsco (y) =
0$. The Lie bracket can be expressed with the convolution product as $[\bdelta_1,
\bdelta_2] = \bdelta_1 \ast \bdelta_2 - \bdelta_2 \ast \bdelta_1$.

Since every character must be the exponential of some element of the Lie
algebra, we know that $\ln^\ast \phi_+ \in \mathfrak{g}$
where $\ln^\ast = \sum \frac{1}{n} \ast^n$ is defined by its series. 
To identify this element, we can first formally solve Eq.~\eqref{anomdef} to get
\begin{equation}
     \phi_+ = \exp^\ast (L \bgamma) \ast \phi_+^0. \label{desol}
\end{equation}
Here $\phi_+^0=R_p \circ \phi_+$ means evaluate the renomalized character in whatever scheme we use then set $L=0$. In the momentum subtraction scheme, $\phi_+^0 = \oneG$ and therefore
\begin{equation}
  \phi_+ = \exp^\ast (L \bgammap), \label{phiplusdelta}
\end{equation}
where $\bgammap = -\phi (\eps D) 
= \frac{1}{L} \ln^\ast \phi_+
\in \mathfrak{g}$. 
MOM has the special property that $\phi_+(L_1+L_2) = \phi_+(L_1) \ast \phi_+(L_2)$,
which follows from Eq.~\eqref{phiplusdelta}.
This judicious property makes MOM a preferred scheme for the Hopf algebra approach as we will see in Section~\ref{sec:RGEHopf}. 

It is perhaps worth emphasizing that $\phi_+ = \exp^\ast (L \bgammap)$
is not a solution to the RG equations. Rather it is a formula that relates the
logarithms in any particular Feynman diagram after its subdivergences have
been removed. For any $\ell$-loop diagram $x$, $\bgammap^{\ast n} x$
vanishes for some $n \leqslant \ell$ so $\phi_+ (x)$ has terms at most $L^n$.
That is, if we write
\begin{equation}
  \phi_+ (x) = {\sum_{k = 0}^{\ell}}  c_k (x) L^k,
\end{equation}
then $c_k = \frac{1}{n!} \bgammap^{\ast k} \cdot x$. We verify this explicitly with some examples in Appendix~\ref{app:examples}.

\subsection{Renormalization group  evolution\label{sec:RGEHopf}}
Next, we will derive the RGE of the renormalized self-energy graphs directly from the combinatorial SDE, following the approach of~\cite{Balduf:2024pgk}.
To do so, we use Eq.~\eqref{phiplusdelta} which says that the in MOM, the renormalized character is the convolution exponential of its anomalous dimension.
So we can
write 
\begin{equation}
  \Pi (a, L) = \phi_+ (X) = \exp^\ast (L \bgammap) \circ X.
  \label{Piphidef}
\end{equation}
The anomalous dimension for $\Sigma(a,L) = \phi_+ (\oneA - X)$ is 
\begin{equation}
  \gamma =- \lim_{L \rightarrow 0} \partial_L \Sigma = \lim_{L \rightarrow
  0} \partial_L \phi_+ (X)
  =  \bgammap \circ X.
  \label{gammadefined}
\end{equation}
Here we distinguish the anomalous dimension $\gamma$ for the self energy graphs, which is a function of $a$, from the convolution anomalous dimension operator $\bgammaP$ which is element of the character Lie algebra. 
Our goal is to relate $\gamma$ to the $\beta$ function using Hopf algebra
techniques, as in Eq.~\eqref{RGEX}.

First observe that since \ $\phi_+$ is a homomorphism we get 
\begin{equation}
  \phi_+ (X^s) = \phi_+ (X)^s = [\exp \ast (L \bgammap) \circ X]^s =
  [\Pi (a, L)]^s.
\end{equation}
Then, since $\exp^\ast = \sum \frac{1}{n!} \ast^n$ has the same algebraic
properties as any other exponential, we can separate the exponent into  
\begin{equation}
  \Pi \mathcal{} (a, L_0 + L) = \exp^\ast ((L_0 + L) \bgammap) \circ
  X  = [\exp^\ast (L_0 \bgammap)] \ast [\exp^\ast (L
  \bgammap)] \circ X.
\end{equation}
Then, recalling the definition of a convolution product and using Eq.
\eqref{DeltaXgen} we get
\begin{align}
  \Pi  (a, L_0 + L) &= m \circ \sum_{\ell = 0}^{\infty} \exp^\ast (L_0
  \bgammap) \circ X^{1 + (s - 1) \ell} \otimes \exp^\ast (L  \bgammap) [X]_{\ell}\\
  &= \sum_{\ell = 0}^{\infty} [\Pi \mathcal{} (a, L_0)]^{1 + (s - 1) \ell} [\Pi(a, L)]_{\ell}.
\end{align}
Now $[\Pi \mathcal{} (a, L_0)]_{\ell}$ is proportional to $a^{\ell}$ by
definition. So we can write 
\begin{equation}
  \Pi (a, L_0 + L) = \Pi  (a, L_0) \Pi(\tilde{a} (L_0),  L),
\end{equation}
where 
\begin{equation}
  \tilde{a} (L_0) = a \Pi (a, L_0)^{s - 1}
  \label{atildedef}
\end{equation}
represents a running coupling. This suggests we can define
\begin{equation}
  \beta \equiv \lim_{L \rightarrow 0} \partial_L \tilde{a} (a, L) 
\end{equation}
as the $\beta$ function. Then from Eqs.~\eqref{atildedef} and~\eqref{Piphidef}, 
\begin{equation}
    \beta= a \lim_{L \rightarrow 0} \partial_L \Pi (a, L)^{s-1}  = a \lim_{L
  \rightarrow 0} \partial_L \phi_+ (X)^{s - 1} = a (s - 1) \lim_{L
  \rightarrow 0} \partial_L \phi_+ (X) = (s - 1) a\gamma, \label{betagamrel}
\end{equation}
and
\begin{align}
  \partial_L \Pi (a, L) &= \lim_{L_0 \rightarrow 0} \partial_{L_0} \Pi (a, L + L_0)
  = \lim_{L_0 \rightarrow 0} \partial_{L_0} \Pi (a, L_0) \Pi (\tilde{a} (L_0),  L) \\
  &= \gamma \Pi (a, L) + (s - 1) a\gamma \partial_a \Pi (a, L).
\end{align}
Since $\Pi = 1-\Sigma$, this implies
\begin{equation}
  (\partial_L - \gamma - (s - 1)a \gamma \partial_a) (1- \Sigma) = 0, \label{RGEgammas}
\end{equation}
which is our RGE from Eqs. \eqref{newrge}, \eqref{RGEform1} or \eqref{RGEX}. Thus we have shown that the Hopf algebraic approach to the RGE is equivalent to the approach using more traditional QFT methods.

Note that although we defined $\gamma =- (\partial_L \Sigma)_{L = 0}$ in Eq.~\eqref{gammadefined}, 
holds at any $L$. It can be rewritten in a more standard form as
\begin{equation}
  \gamma =\frac{d}{d L} \ln (1-\Sigma), 
\end{equation}
where $\frac{d}{d L} = \partial_L -\beta \partial_a$ with $\beta =  (s - 1)
a \gamma$.

\subsection{Multi-slot cocycles \label{sec:multislot}}
So far we have seen that the Hopf algebra is useful for determining the RGE,
in particular the relationship between the anomalous dimension and the beta
function, directly from the combinatorial SDE. The RGE was not necessary in
Sections \ref{sec:phichi} or \ref{sec:other} to determine the asymptotic
behavior. However, it is necessary if we are to attempt to interpret
asymptotic growth as associated with a renormalon and the $\beta$ function.
Now we proceed to examine the cSDEs for other cases studied in Section
\ref{sec:other} using the Hopf algebra.

Section~\ref{sec:twoarm} considered $\phi^3$ theory in $d = 6$ with
self-energy insertions into both $\phi$ legs of the primitive 1-loop graph.
The SDE was in Eq.~\eqref{BSDE}:
\begin{equation}
  \Sigma_B^0 (p^2) = \frac{a}{2\pi^3} \int d^6 k \frac{1}{p^2 (p + k)^2 (1 -
  \Sigma_B^0 ((p + k)^2))} \frac{1}{k^2 (1 - \Sigma_B^0 (k^2))}.
\end{equation}
Writing $1 - \Sigma_B^0 = \phi (X_B)$ as with the $\phi^2 \chi$ theory we can
write the cSDE as
\begin{equation}
  X_B = \oneA + a B_{\bullet} \left( \frac{1}{X_B} \otimes
  \frac{1}{X_B} \right).
\end{equation}
In this case $B_{\bullet} : A \otimes A \rightarrow A$ is a generalization of
the Hochschild cocycle with the first argument inserted into the top leg of
the bubble and the second argument inserted into the bottom leg
\begin{equation}
  B_{\bullet} (x \otimes y) =
  \Bxy,
\end{equation}
which corresponds to 
\begin{equation}
  \phi(B_\bullet(x\otimes y))(p^2) = 
  \frac{a}{2\pi^3} \int d^6 k \frac{1}{p^2}\frac{[\phi(x)(p+k)^2]}{ (p + k)^2}\frac{[\phi(y)(k)^2]}{  k^2}.
\end{equation}
The coproduct of $B_{\bullet} (x \otimes y)$ must contain all of the
subdivergences of the new graph. This includes not only the subdivergences of
$x$ and $y$ but also all of the cross terms. In Sweedler notation, we want
\begin{equation}
  \Delta B_{\bullet} (x \otimes y) = x^{(1)} y^{(1)} \otimes B_{\bullet}
  (x^{(2)} \otimes y^{(2)}) + B_{\bullet} (x \otimes y) \otimes \oneA,
\end{equation}
which we can write more algebraically as
\begin{equation}
\Delta   B_{\bullet} = (m \otimes B_{\bullet}) \circ (\Delta \otimes \Delta) +
  B_{\bullet} \otimes \oneA.
\end{equation}
Here $m$ means we multiply the subdivergences, generating a decomposable
disjoint-union algebra element. The tensor product of coproducts employs a
(standard) convention:
\begin{equation}
  (\Delta \otimes \Delta) (x \otimes y) = (x^{(1)} \otimes x^{(2)}) \otimes
  (y^{(1)} \otimes y^{(2)}) \equiv x^{(1)} \otimes y^{(1)} \otimes x^{(2)}
  \otimes y^{(2)}.
\end{equation}
In this way $m$ in $m \otimes B_{\bullet}$ will act on $x^{(1)} \otimes
y^{(1)}$ and $B_{\bullet}$ on $x^{(2)} \otimes y^{(2)}$.

Next, suppose we have an SDE involving the 2-slot cocycle of the form
\begin{equation}
  X = \oneA + a B_{\bullet} (X^s \otimes X^t),
\end{equation}
then the generalization of Eq.~\eqref{DeltaXgen} is
\begin{equation}
  \Delta X = \sum_{\ell = 0}^{\infty} X^{1 + (s + t - 1)
  \ell} \otimes [X]_{\ell}.
\end{equation}
The derivation is identical to the 1-slot case. It then follows that
\begin{equation}
  \beta = (s + t - 1) \gamma a,
\end{equation}
and the RGE is
\begin{equation}
    (\partial_L - \gamma - (s +t- 1)a \gamma \partial_a) (1- \Sigma) = 0.
  \label{RGEst}
\end{equation}
For the special case of the SDE from Eq.~\eqref{BSDE} we have $s = t = - 1$ so
$\beta = - 3 \gamma a$. 

The other examples in Section~\ref{sec:other} can be analyzed the same way,
and in all cases we find that the pole locations are all proportional to the
$\beta$ function coefficient. This suggests that we can identify all of these
Borel singularities as renormalons.

Finally, let us comment on the natural generalization, where we would define
\begin{equation}
  B_{\bullet} (x \otimes y \otimes z \otimes w) = 
 \Bxyzw.
 \label{Bprob}
\end{equation}
The picture here means we construct a new algebra element by inserting the
subdiagrams $x, y, z, w$ into the 1-loop primitive bubble in the places as
indicated. One can certainly define $B_\bullet$ this way, but it is challenging to
interpret it as a cocycle. The problem is that inserting a triangle graph into
$z$ or $w$ will give a two-loop graph with overlapping divergences~\cite{Kreimer:1997dp,Connes:1998qv, Kreimer:1998ood,Krajewski:1998xi,Manchon:2004qa,OlsonHarris:2025dse}.
This significantly complicates the analysis and is beyond the scope of the current study.

\section{Is Hopf renormalization multiplicative? \label{sec:isHM}}
The previous section described how the CK Hopf algebra can be used to describe
renormalization, Schwinger-Dyson equations and the renormalization group. In
this section we attempt to compare and contrast some elements of the Hopf
approach to the use of more traditional QFT methods.

In the Hopf approach, renormalized amplitudes are constructed algebraically. The renormalized character is as in Eq.~\eqref{phiplusform}:
\begin{equation}
  \phi_+ = (1 - R) [\phi + m \circ (\phi_- \otimes \phi) \Dtilde ]. 
\end{equation}
This equation encodes the recursive BPHZ  prescription for removing
subdivergences so that the resulting counterterms are local. We call this
procedure \textbf{Hopf renormalization}. Hopf renormalization does not refer to bare or renormalized
fields or couplings at all; indeed for any graph $x$, $\phi_+(x)$ is a sum of
terms homogeneous in the coupling and therefore indifferent to any rescaling
or redefinition of the coupling. In Hopf renormalization, divergences are
removed via subtractions rather than rescalings. Since every graph is
renormalized on its own, relations among the counterterms for different graphs
are not transparent.

In contrast, in \textbf{multiplicative renormalization}, we write (for the
$\phi^2 \chi$ theory for concreteness)
\begin{equation}
  \mathcal{L} = \frac{1}{2} (\partial_{\mu} \phi_0)^2 + \frac{1}{2}
  (\partial_{\mu} \chi_0)^2 -\frac{1}{2} \lambda_0 \phi_0^2 \chi_0 =
  \frac{1}{2} Z_{\phi} (\partial_{\mu} \phi^2) + \frac{1}{2} Z_{\phi}
  (\partial_{\mu} \chi^2) - \frac{\lambda}{2} Z_v \phi^2 \chi,
\end{equation}
where $\lambda_0 = Z_{\lambda} \lambda$, $\phi_0 = \sqrt{Z_{\phi}} \phi$,
$\chi_0 = \sqrt{Z_{\chi}} \chi$ and the vertex counterterm is
\begin{equation}
  Z_v = Z_{\lambda} Z_{\phi} \sqrt{Z_{\chi}} \label{Zvform}.
\end{equation}
Then the $Z$ factors are expanded as $Z_j = 1  +\delta_j$ with $\delta_j$ the counterterms.
 These counterterms are determined order-by-order in $\lambda$ and the same counterterms are used for any graph.

We know that Hopf renormalization and multiplicative renormalization must be equivalent when all graphs are included. This follows because the Hopf algebra encodes the relevant subdivergences exactly as required by BPHZ. Then the equivalence is a corollary of the textbook demonstration that the BPHZ prescription is multiplicative (see Section 5.6 of~\cite{Collins:1984xc} for example). What we are concerned with in this section is 1) on an abstract level, understanding {\textit{how}} the multiplicative nature of renormalization is encoded in the Hopf algebra and 2) on a practical level, whether the equivalence between multiplicative and Hopf renormalization still holds if only a subset of diagrams are included, such as those represented by a SDE. The second point is particularly relevant if we hope to identify singularities in the Borel plane with renormalons, since renormalons are associated with the rescaling of the coupling. 

With multiplicative renormalization, Green's functions rescale according to the
powers of the fields included. For example, $\langle \phi_0 \phi_0 \rangle =
Z_{\phi} \langle \phi \phi \rangle$. With Hopf renormalization, the diagrams
contributing to $\langle \phi \phi \rangle$ are represented by some algebra
element $X$, and then the bare Green's function is $\langle \phi_0 \phi_0
\rangle = \phi (X)$, where we are unfortunately using the same symbol $\phi$ for the field and the bare character. The renormalized Green's function is $\langle \phi \phi
\rangle = \phi_+ (X)$ and the field strength renormalization factor is $Z = \phi_- (X)$. These are
related by Birkhoff factorization
\begin{equation}
  \phi_+ = \phi_- \ast \phi, \label{BirkhoffFB}
\end{equation}
which looks the same as $\langle \phi_0 \phi_0 \rangle = Z_{\phi} \langle \phi
\phi \rangle$ except for the replacement of a product with a convolution
product. In the special case when $\Delta X = X \otimes X$, which holds for $X$ satisfying a linear cSDE (as with the rainbow graphs in $\phi^2 \chi$ theory) the two are the same; that is, for a linear cSDE, Eq.~\eqref{BirkhoffFB} implies $\phi_+
(X) = \phi_- (X) \phi (X)$. 
In the general case, however, the recursive
subtractions implicit in the coproduct account for the coupling renormalization in addition to field-strength renormalization. 
When $X$ contains elements of a closed Hopf subalgebra and Eq.~\eqref{DeltaXgen} holds, we can then act with $\phi_+$ on Eq.~\eqref{DeltaXgen} and use Birkhoff factorization to get
\begin{equation}
    \phi_+ = \sum_{\ell = 0}^{\infty} \phi_-(X)^{1 + (s - 1)\ell} \phi([X]_\ell).
\end{equation}
Recalling that $[X]_\ell$ means the $\cO(a^\ell)$ part of $X$, then 
$\phi([X]_\ell)$ is proportional to $a^\ell$. So writing $Z=\phi_-(X)$ and $Z_a = Z^{s-1}$ this becomes 
\begin{equation}
    \phi_+ (X) = Z \phi(\widehat{X}),\quad \widehat{X} =\sum_{\ell = 0}^{\infty} (a Z_a)^\ell [X]_\ell.
\end{equation}
In words, the renormalized character is produced by rescaling the bare character and rescaling the bare coupling. 
The relation $Z_a=Z^{s-1}$ then leads immediately to the RGE in Eq.~\eqref{RGEgammas}.
This connection between Birkhoff factorization and multiplicative renormalization is actually quite elegant, and, in our opinion, significantly simpler than the Lie-algebra based derivation of Eq.~\eqref{RGEgammas} we reviewed in Section~\ref{sec:RGEHopf}.

It is also instructive to explore whether Hopf renormalization and multiplicative renormalization correspond to the computation of the same Feynman diagrams. In the Hopf algebraic approach, counterterms come from contracting subdiagrams to a point. In multiplicative renormalization, counterterms come from interaction vertices generated in renormalized perturbation theory.
Now, the diagrams we consider in this paper are not all the possible diagrams contributing to a Green's function; we have restricted to diagrams arising from insertions into propagators only. By inserting into propagators, one might expect that the only relevant counterterms are those from the field strength renormalization, not vertex corrections. That is how the counterterms appear in the Hopf approach. However, in multiplicative renormalization, if we just rescale the fields and not the coupling, then $Z_\lambda=1$ in Eq.~\eqref{Zvform} so $Z_v = Z_\phi \sqrt{Z_\chi} \ne 1$. That is,
vertex counterterms {\textit{are}} generated, generically. For them not to be generated, we must have $Z_v=1$, which leads to a relation between $Z_\lambda$ and $Z_\phi$, or equivalently between $\beta$ and $\gamma$. 

To be more explicit, consider again the Hopf and rainbow, graphs in $\phi^2 \chi$ theory to 2 loops. We take $Z_\chi=1$ as $Z_\chi$ plays no role in either approach for these graphs. 
In the Hopf algebra approach, the  computation amounts to summing the following character values
\begin{equation}
    \phi_+(x_1+x_2) =
    \left(
    \looseoverset[8pt]{\bubble}{\phi(x_1)}
    +
    \looseoverset[2pt]{\linect[1pt]}{\phi_-(x_1)}
    \right)
    +
    \left(
    \looseoverset[8pt]{\doublebubble}{\phi(x_2)}
    +
    \looseoverset[8pt]{\bubblect}{\phi_-(x_1)\phi(x_2)}
     +
    \looseoverset[2pt]{\linectO[1pt]}{\phi_-(x_2)}
    \right),
    \label{SigmacompleteA}
\end{equation}
where $x_1$ is the primitive element for the 1-loop graph, with $\Sigma_1^0 = \phi(x_1)$ and  $x_2$ is the 2-loop algebra element with $\Sigma_2^0 = \phi(x_2)$.
In renormalized perturbation theory, we  get instead:
\begin{multline}
    \Sigma_1 + \Sigma_2 = 
    \looseoverset[8pt]{\bubble}{\Sigma_1^0}
    +
    \looseoverset[2pt]{\linect[1pt]}{-\delta_\phi^1}
    \\
+\looseoverset[8pt]{\doublebubble}{\Sigma_2^0}
+\looseoverset[8pt]{\bubblect}{-\delta_\phi^1\Sigma_1^0}
+\looseoverset[2pt]{\bubblectL}{\delta_v^1 \Sigma_1^0} 
+\looseoverset[2pt]{\bubblectR}{\delta_v^1 \Sigma_1^0}
+\looseoverset[2pt]{\linectO[1pt]}{-\delta_\phi^2},
\label{SigmamultA}
\end{multline}
where $\delta_j^\ell$ is the order $a^\ell$ part of $\delta_j$.
In order for Eqs.~\eqref{SigmacompleteA} and~\eqref{SigmamultA} to agree, we need $\delta_\phi^1 = - \phi_-(x_1)$, $\delta_\phi^2 = - \phi_-(x_2)$ and $\delta_v=0$. Then we have 
\begin{align}
    \delta_\phi^1 &= R[\Sigma_1^0],
    \\
    \delta_\phi^2 &= R\Big[\Sigma_2^0 - \delta_\phi^1 \Sigma_1^0\Big]
    = R\Big[\Sigma_2^0 - R\big[\Sigma_1^0\big] \Sigma_1^0\Big].
    \label{localcombo}
\end{align}
Exactly these combinations are needed for the counterterms to be local (cf. Eq.\eqref{phiminus2}).
The condition $\delta_v=0$ is very natural, since the SDE was designed to avoid vertex insertions. It implies $Z_v=1$ so that $Z_{\lambda} = Z_{\phi}^{-
1}$ or, recalling $a=\lambda^2/(4\pi)^3$, that 
$Z_a = Z_{\phi}^{-2}$.
In multiplicative renormalization, the beta function for $a$ and anomalous dimension for $\phi$ are 
\begin{equation}
\beta_a = - \frac{a}{2}  \mu \frac{d}{d\mu} \ln Z_a,
\qquad 
\gamma = - \frac{1}{2}\mu \frac{d}{d\mu} \ln Z_\phi.
\end{equation}
So that setting $Z_a = Z_{\phi}^{-2}$ leads automatically to  $\beta = - 2 a \gamma$. 
This is 
the same relationship between $\beta$ and $\gamma$ we found the Hopf graphs (see Eq.~\eqref{betagamrel} with $s=-1$). 

Although we can map Hopf renormalization to multiplicative renormalization cleanly for the Hopf graphs, the rainbow graphs are not quite as clean. The rainbow graphs are identical to the Hopf graphs at this order, so they should also have $\beta = - 2 a \gamma$ too. But this is not true. The rainbow graphs satisfy an RGE with $\beta=0$. Thus how can Hopf and multiplicative renormalization also be equivalent for the rainbow graphs?

To reconcile $\beta = 0$ for the rainbow graphs with multiplicative renormalization, we recall the observation from Section~\ref{sec:CSE}
that for the Callan-Symanzik equation to apply for the rainbow graphs with the conventional definition of the anomalous dimension, the Green's function must include only 1PI graphs, not the geometric sum of 1PI propagator insertions. That is, while for the Hopf graphs, the
bare and renormalized Green's function are related as in Eq.\eqref{hopfnoflip}: $\frac{1}{1-\Sigma_0}= Z_\phi\frac{1}{1-\Sigma}$, for the rainbow graphs $1+\Sigma_0 = Z_\phi(1+\Sigma_R)$ as in Eq.~\eqref{rainbowflip}.
This can be effected in renormalized perturbation theory by flipping the sign of all the bare and renormalized graphs ($\Sigma_\ell^0 \to -\Sigma_\ell^0$ and $\Sigma_\ell \to -\Sigma_\ell$) and taking $Z_\phi \to 1/Z_\phi$. 
Let us write $1/Z_\phi =1+ \sum_{\ell\ge 1} a^\ell \hat{\delta}_\phi^\ell$. Then at two loops with these modifications we get
\begin{multline}
    -\Sigma_1  -\Sigma_2 = 
    \looseoverset[8pt]{\bubble}{-\Sigma_1^0}
    +
    \looseoverset[2pt]{\linect[1pt]}{-\hat\delta_\phi^1}
    \\
+\looseoverset[8pt]{\doublebubble}{-\Sigma_2^0}
+\looseoverset[8pt]{\bubblect}{\hat\delta_\phi^1\Sigma_1^0}
+\looseoverset[2pt]{\bubblectL}{-\delta_v^1 \Sigma_1^0} 
+\looseoverset[2pt]{\bubblectR}{-\delta_v^1 \Sigma_1^0}
+\looseoverset[2pt]{\linectO[1pt]}{-\hat\delta_\phi^2}.
\label{SigmamultB}
\end{multline}
So 
\begin{align}
    \hat\delta_\phi^1 &= -R[\Sigma_1^0] = \delta_\phi^1,
    \\
    \hat\delta_\phi^2 &= R\big[-\Sigma_2^0 + \hat\delta_\phi^1 \Sigma_1^0 - 2 \delta_v^1 \Sigma_1^0\big]
    = -R\Big[\Sigma_2^0 + (R[\Sigma_1^0] - 2 \delta_v^1)\Sigma_1^0\Big].
\end{align}
We already know the combination in Eq.~\eqref{localcombo} is local, so for $\hat\delta_\phi^2$ also to be local
we must have $\delta_v^1 = R[\Sigma_1^0] = -\hat\delta^1 = \delta_\phi^1$. That is, $Z_v=Z_\phi$ to this order, as expected. 

So we can conclude that the rainbow graphs can be multiplicatively renormalized as long as only 1PI graphs are included in the Green's function and if $Z_v = Z_\phi$ so that $\beta=0$. This does not in itself mean that Hopf renormalization is equivalent to multiplicative renormalization for the rainbows. The issue remains that the Hopf approach is predicated on counterterm graphs arising from the contraction of subgraphs to a point, but there are no vertex graphs in the rainbow approximation whose contraction matches onto the $\delta_v$ counterterms.
In other words, if the requirement for Hopf renormalization being equivalent to multiplicative renormalization is  only that the renormalization of a class of graphs can somehow be achieved with multiplicative field strength and coupling constant renormalization, then both Hopf and rainbow graphs  qualify. However, if a stronger requirement is imposed that the elements of the Hopf coproduct match directly onto counterterm diagrams arising from renormalized perturbation theory, then Hopf qualifies but rainbow does not.

Going to diagram classes that include vertex insertions in addition to propagator insertions adds complexity to both the SDEs and the Hopf algebra. As mentioned at the end of Section~\ref{sec:multislot}, with vertex insertions we generically have overlapping divergences and the insertions are not naturally described with Hochschild cocycles. In the Hopf literature~\cite{Kreimer:2006va,Kreimer:2009iy,OlsonHarris:2025dse}, these cases are often discussed using an object called the {\textbf{invariant charge}}.
The invariant charge refers to a ratio such as, in $\phi^2 \chi$ theory,  
\begin{equation}
  Q \equiv \frac{G_{2, 1} (p_{i j}, \mu)}{\sqrt{G_{2, 0} (p_1, \mu)}
  \sqrt{G_{2, 0} (p_2, \mu)} \sqrt{G_{0, 2} (p_2, \mu)}},
\end{equation}
where $G_{n,m} \sim \langle \phi^n \chi^m \rangle$ are renormalized connected Green's functions. The Callan-Symanzik equation then implies $\frac{d}{d\mu} Q=0$, which is why $Q$ is called
invariant. At leading order $Q = \lambda + \cO (\lambda^3)$ which is
why $Q$ is called a charge. 
In the Hopf algebra, $Q$ has a combinatorial
avatar $\widehat{ Q}$ so that $Q = \lambda \phi_+ ( \widehat{Q})$, where 
\begin{equation}
 \widehat{Q} = 
  \frac{ X^V } {X^\phi\sqrt{X^\chi}},
\end{equation}
and the $X^j$ are $\oneA$ plus the sum of appropriate 1PI graphs: $X^V$ include 1PI vertex corrections, $X^\chi$ include 1PI contributions to the $\chi$ propagator 
and  $X^\phi$ include 1PI $\phi$ propagator corrections. Using $\widehat{Q}$ is often the starting point for an analysis of the RGE using the Hopf approach. However, again, due to complications associated with overlapping divergences, the full form of $\widehat{Q}$ is rarely used. Nevertheless, the invariant charge seems to play a prominent role in attempts to solve SDEs using Hopf algebras~\cite{Kreimer:2006va,OlsonHarris:2025dse,yeatsthesis,Balduf:2024pgk}.

\section{Conclusions \label{sec:conclusions}}
In this paper we have examined the asymptotic behavior of various sums of
Feynman diagrams at large order in the coupling. These asymptotic expansions
can be efficiently computed using Schwinger-Dyson equations and Borel
transforms. Although some expansions, such as insertions into one arm of the
self-energy graph in $\phi^3$ in $d = 6$ or Yukawa theory in $d = 4$ were
originally studied using Hopf algebraic techniques, we find these results can
be derived efficiently using more conventional quantum field theory
methodology. We include analysis of the Borel singularities for other classes of diagrams in $\phi^3$ and $\phi^4$ theory as well.

One question we attempted to answer is whether the various diagram sums we
consider should be thought of as renormalons, instantons, or some other source
of asymptotic growth. More precisely, if all the singularities in the Borel
plane were known for a theory, could there be singularities in addition to
renormalons and instantons? 
This question was partly motivated by
the work of~\cite{Broadhurst:1998ij,Broadhurst:2000dq,Borinsky:2020HADSE,Borinsky:2021hnd}
which observed that a class of ``chain'' diagrams, similar to the vacuum-polarization diagrams generating the renormalon in gauge theories, has the leading asymptotic behavior $ (-\frac{1}{6})^n n! a^n$ while another class of ``Hopf'' graphs which are not in clear correspondence with vacuum polarization graphs, has the leading behavior $ (-\frac{1}{3})^n n! a^n$.
To determine whether these are renormalons we looked for a connection between the location of the associated Borel singularity and the leading-order $\beta$ function coefficient $\beta_0$. The modification of the $\beta$ function when only a limited set of graphs are included can be determined by the renormalization group equation. By examining the RGEs for various diagram types, we find that in fact both the chain and Hopf graph asymptotic behavior can be associated with renormalons. We draw the same conclusion for the other classes of graphs in other theories we considered.

We also provided a review of some relevant aspects of the Connes-Kreimer Hopf algebra. 
The Hopf algebraic machinery is quite elegant in some respects and can produce some results which appear, to us at least, surprising from QFT. For example, a classic result of Connes-Kreimer is that $ \varepsilon \phi (S \ast Y)$ is finite as
$\varepsilon \rightarrow 0$. In words, the bare character $\phi$ evaluated on
the convolution product of the antipode and the grading operator for any
Feynman diagram should have at most a $\eps^{-1}$ pole. This means that if you sum over the divergent subgraphs weighting the uncontracted
subdiagram by the loop order, the result will have at most a
$\eps^{-1}$ pole and no non-local subdivergences. For example, the 3-loop chain graph in $\phi^2 \chi$ theory is
\begin{equation}
 \chain = \left(\frac{a}{6}\right)^3 \left(\frac{p^2}{\mu^2}\right)^{-3 \eps }\frac{ \Gamma (2-3 \eps ) \Gamma (2-\eps )^5 \Gamma (\eps -1)^2 \Gamma (3 \eps -1)}{\Gamma (4-4 \epsilon ) \Gamma (4-2 \eps )^2 \Gamma (2 \eps +1)},
\end{equation}
which has a $\eps^{-3}$ pole, non-local divergences of the form $\eps^{-2}\ln p^2$ and so on (see Eq.~\eqref{chainexplicit}). However, if we multiply it by 3 and add the product of some of its subdiagrams weighted by their loop order we get only a single, local, $\eps^{-1}$ pole:
\begin{equation}
    3\chain - 4 \bubble \times \doublebubble +\left(\bubble\right)^3
= -\left(\frac{a}{6}\right)^3\frac{85}{18\eps} + \cO(\eps^0).
\end{equation}
Regardless of whether this construction has practical applications, it at least demonstrates that the Hopf algebra techniques can reveal hidden structures in QFT.

Another application of the Hopf algebraic approach is that it allows the
Schwinger-Dyson equations to be written as combinatorial Schwinger-Dyson
equations, as relationships between the algebra elements themselves. For
example, the Hopf graph cSDE is $X = 1 - a B_{\bullet} (X^{- 1})$ with
$B_{\bullet}$ a Hochschild 1-cocycle for inserting a subgraph into the internal
$\phi$ propagator of the primitive self-energy graph in $\phi^2 \chi$ theory.
Diagrams are evaluated with a character $\phi : A \rightarrow
\mathbb{C}$ and the set of characters form a Lie group. This Lie group $G$ has a Lie algebra $\mathfrak{g}$ and one can write $\phi = \exp^\ast (L
\bgammap)$ with $\phi \in G$ and $\bgammap \in
\mathfrak{g}$. When combined with the cSDE the Lie algebra element can be
interpreted as the anomalous dimension of the field. Then the renormalization
group equations can be derived in cases where the SDEs are harder to solve.

The Hopf algebraic approach runs into difficulties for SDEs which involve
inserting subdiagrams into vertices of other diagrams. Such insertions lead to
overlapping divergences. Although the Hopf algebra can handle overlapping
divergences it is at the expense of a significant increase in complexity.
Studying SDEs with vertex insertions is an active area of research~\cite{Panzer:2012gp,foissy2015mulitgradeddysonschwingersystems,OlsonHarris:2025dse,Balduf:2025fjp}. An interesting open question is whether when vertex
corrections are included one will still find all the Borel singularities to be
associated with the $\beta$ function of the theory or if new sources of
asymptotic behavior might appear. For example, in QED, one cannot consider only Hopf-type graphs with self-energy insertions alone because these are gauge-dependent. If these QED Hopf graphs are supplemented by the minimal set of vertex corrections need to restore gauge-invariance, what is the nature of the resulting series? In general, there are many interesting open questions related to asymptotic growth that may be approached with a combination of QFT and Hopf-algebraic techniques.

\section{Acknowledgments}
The authors would like thank Gerald Dunne, Paul Balduf, Arindam Bhattacharya and Aurélien Dersy for valuable conversations. This work has been supported by the U.S. Department of Energy Office of High-energy Physics
under contract DE-SC0013607. LC was supported by the NSF - Graduate Research Fellowship Program. 

\appendix
\section{Explicit calculation of characters up to three loops
\label{app:examples}}
To understand the Hopf algebra techniques, it can be helpful to have explicit equations to check. We provide some examples for that purpose in this appendix. We work here in the $\phi^2 \chi$ theory in $d=6-2\eps$ dimensions.  In the following we give the first few terms in the expansion of various quantities in $a/6$, e.g.
\begin{equation}
    \Sigma = \sum \left(\frac{a}{6} \right)^n \Sigma_n(L).
\end{equation}
Equivalently, for an $\ell$ loop amplitude, we strip off a factor of $(a/6)^\ell$.

The master integral we need is the analytically-regulated $d$ dimensional bubble (the $d$-dimensional version of Eq.~\eqref{Fd6}):
\begin{align}
  F (\rho) &=  \mu^{6 - d} \int \frac{d^d k}{\pi^{d/2}} 
  \frac{1}{p^2 k^2 (p + k)^2} \left( \frac{k^2}{\mu^2} \right)^{\rho}
\\
  &=e^{L \left( \frac{d-6}{2} + \rho
  \right)} \Gamma \left( \frac{d}{2} - 1 \right) \frac{\Gamma \left( 2 -
  \frac{d}{2} - \rho \right)}{\Gamma (1 - \rho)} \frac{\Gamma \left(
  \frac{d}{2} - 1 + \rho \right)}{\Gamma (d - 2 + \rho)},
\end{align}
where $L = \ln \frac{p^2}{\mu^2}$. From here we compute the 1-loop self-energy
\begin{align}
    \Sigma_1(p^2) &=\phi(\bullet)= \bubble = F(0) = 6 e^{L \left( \frac{d-6}{2}
  \right)} 
    \frac{ \Gamma \left(2-\frac{d}{2}\right) \Gamma \left(\frac{d}{2}-1\right)^2}{\Gamma (d-2)}
    \nonumber
\\
&=-\frac{1}{\eps }+L-\frac{8}{3} + \cOe.
\end{align}
Since $\Sigma_1(L) \propto (p^2)^{(d-6)/2}$ we can insert $\Sigma_1$ back into the integral and use the same master integral for the two-loop self energy. This gives
\begin{align}
    \Sigma_2(L) &=  \phi\left(\twostick[0.6]{black}{black}\right)
    =\doublebubble 
        = 6^2 e^{2L \left( \frac{d}{2}  - 3
  \right)} 
\frac{\Gamma (5-d) \Gamma \left(2-\frac{d}{2}\right) \Gamma \left(\frac{d}{2}-1\right)^3 \Gamma (d-4)}{\Gamma \left(4-\frac{d}{2}\right) \Gamma (d-2) \Gamma \left(\frac{3 d}{2}-5\right)} \nonumber\\
&=
\frac{1}{2 \eps ^2}-\frac{L}{\eps }+\frac{43}{12 \eps }-\frac{\pi ^2}{12}+\frac{1207}{72}-\frac{43}{6}L+L^2 +\cOe.
\end{align}
At 3-loops the rainbow is
\begin{align}
\phi\left(\threestick[0.6]{black}{black}{black}\right)
&=\rainbow  = 6^3
e^{3L(\frac{d-6}{2})}
\frac{ \Gamma \left(8-\frac{3 d}{2}\right) \Gamma (5-d) \Gamma \left(2-\frac{d}{2}\right) \Gamma \left(\frac{d}{2}-1\right)^4 \Gamma (d-4) \Gamma \left(\frac{3 d}{2}-7\right)}{\Gamma (7-d) \Gamma \left(4-\frac{d}{2}\right) \Gamma (d-2) \Gamma \left(\frac{3 d}{2}-5\right) \Gamma (2 d-8)}
\nonumber\\
&=
-\frac{1}{6 \eps ^3}-\frac{9}{4 \eps ^2}-\frac{3 L^2}{4 \eps }+\frac{L}{2 \eps ^2}+\frac{27 L}{4 \eps }+\frac{\pi ^2}{24 \eps }-\frac{3937}{216 \eps }+\frac{29 \zeta_3}{6}\\
&\hspace{2cm}+\frac{9 \pi ^2}{16}-\frac{49939}{432}-\frac{\pi ^2}{8}L
+\frac{3937}{72}L-\frac{81}{8} L^2+\frac{3}{4} L^3
+\cOe,
\end{align}
and the chain
\begin{align}
\phi\left(\chaintree[0.6]\right)      &=
    \chain=  6^3 e^{3L(\frac{d-6}{2})} \frac{\Gamma \left(8-\frac{3 d}{2}\right) \Gamma \left(2-\frac{d}{2}\right)^2 \Gamma \left(\frac{d}{2}-1\right)^5 \Gamma \left(\frac{3 d}{2}-7\right)}{\Gamma (7-d) \Gamma (d-2)^2 \Gamma (2 d-8)} \nonumber\\
    &=-\frac{1}{3 \eps ^3}+\frac{L}{\eps^2}-\frac{35}{9 \eps ^2}-\frac{3 L^2}{2 \eps }+\frac{35 L}{3 \eps }+\frac{\pi ^2}{12 \eps }-\frac{746}{27 \eps } +\frac{35 \pi ^2}{36}\nonumber\\
    &\hspace{2cm}
   -\frac{12628}{81} + \frac{23 \zeta_3}{3}-\frac{\pi ^2}{4}L+\frac{746}{9}L -\frac{35}{2} L^2+\frac{3}{2} L^3+ \cOe.
   \label{chainexplicit}
\end{align}

The coproducts are
\begin{align}
    \Delta (\bullet) &= \oneA \otimes (\bullet) + (\bullet) \otimes \oneA, \\
      \Delta \left( \twostick[0.6]{black}{darkgreen}\right) &= \oneA \otimes  \left(\twostick[0.6]{black}{darkgreen}\right) +  \left(\twostick[0.6]{black}{darkgreen}\right)\otimes \oneA  + ({\green \bullet}) \otimes (\bullet), \\
    \Delta\left(\threestick[0.6]{black}{darkgreen}{darkblue}\right)
    &= \oneA \otimes \left(\threestick[0.6]{black}{darkgreen}{darkblue}\right)
    +\left(\threestick[0.6]{black}{darkgreen}{darkblue}\right)  \otimes \oneA
    + ({\blue \bullet}) \otimes \left(\twostick[0.6]{black}{darkgreen}\right)
    +\left(\twostick[0.6]{darkgreen}{darkblue} \right)   \otimes(\bullet),\\
    \nonumber
      \Delta \left(\chaintree[0.6]\right) &= 
\left(  \chaintree[0.6]\right) \otimes \oneA + \oneA \otimes \left( \chaintree[0.6]\right) 
+({\green \bullet})\otimes \left(\twostick[0.6]{black}{darkblue}\right)
 + ( {\blue \bullet}) \otimes\left(\twostick[0.6]{black}{darkgreen}\right)
+  ( {\blue  \bullet})\, ( {\green\bullet}) \otimes   ( \bullet).
\end{align}
Note that $({\green \bullet})\otimes \left(\twostick[0.6]{black}{darkblue}\right)
 =
( {\blue \bullet}) \otimes\left(\twostick[0.6]{black}{darkgreen}\right)$; colors are only added as a reminder of the origin of these two terms.

The antipodes are
\begin{align}
S(\bullet) &= -( \bullet),
\qquad S\left[(\bullet)({\green \bullet})\right] = ( \bullet)({\green \bullet}),
\qquad S\left[(\bullet)^k\right] =(-1)^{k+1} (\bullet)^k,
\\
 S\left(\twostick[0.6]{black}{darkgreen}\right) &= 
 -\left(\twostick[0.6]{black}{darkgreen}\right)
 +({\green \bullet} )({\black \bullet} ), \\
    S\left(\threestick[0.6]{black}{darkgreen}{darkblue}   \right)
    &= -\left(\threestick[0.6]{black}{darkgreen}{darkblue}\right)
    + ({\blue \bullet} )\left( \twostick[0.6]{black}{darkgreen} \right)
        +({\black \bullet} ) \left( \twostick[0.6]{darkgreen}{darkblue} \right)
- ( {\blue  \bullet})\, ( {\green\bullet})  ( \bullet),\\
 S\left( \chaintree[0.6]\right) &= -\left( \chaintree[0.6]\right)
+({\green \bullet}) \left(\twostick[0.6]{black}{darkblue}\right)
 +( {\blue \bullet}) \left(\twostick[0.6]{black}{darkgreen}\right)
- ( {\blue  \bullet})\, ( {\green\bullet})  ( \bullet).
\end{align}
consistent with the M{\"o}bius inversion formula $S = \id^{\ast - 1} = \sum_{k= 0} (- 1)^k \pi^{\ast k}$. 

The inverse characters mirror the antipode  
\begin{align}
\phi^{-1}(\bullet) &= -\phi( \bullet), \label{phiinv1}
\\
 \phi^{-1}\left(\twostick[0.6]{black}{black}\right) &= 
 -\phi\left(\twostick[0.6]{black}{black}\right)
 +\phi(\bullet)^2, \\
   \phi^{-1}\left(\threestick[0.6]{black}{darkgreen}{darkblue}   \right)
    &= -\phi\left(\threestick[0.6]{black}{darkgreen}{darkblue}   \right)
    + 2\phi(\bullet)\, \phi\!\left( \twostick[0.6]{black}{black} \right)
-\phi ( {\bullet})^3,\\
\phi^{-1}\left( \chaintree[0.6]\right) &= -\phi\left( \chaintree[0.6]\right)
+ 2\phi(\bullet) \, \phi\!\left( \twostick[0.6]{black}{black} \right)
-\phi ( {\bullet})^3. \label{phiinv4}
\end{align}

The counterterm characters in $\msbar$ are 
\begin{align}
  \phi_-(\bullet) &= -R\left[\phi(\bullet)\right] = \frac{1}{\eps},  \\
 \phi_-\left(\twostick[0.6]{black}{black}\right) &= 
 -R\left[ \phi\left(\twostick[0.6]{black}{black}\right)
 +\phi_-(\bullet) \phi(\bullet)\right]=\frac{1}{\eps^2}-\frac{11}{12\eps} \label{phiminus2},\\
   \phi_-\left(\threestick[0.6]{black}{darkgreen}{darkblue}   \right)
    &=  -R\left[\phi\left(\threestick[0.6]{black}{darkgreen}{darkblue}   \right)
    + \phi_-\left(\twostick[0.6]{black}{black}\right)\phi(\bullet)
    + \phi_-(\bullet)\phi\left( \twostick[0.6]{black}{black} \right)
\right]
=\frac{1}{6\eps^3}-\frac{11}{12\eps^2}+\frac{103}{54\eps},\\
\phi_-\left( \chaintree[0.6]\right) &=
 -R\left[\phi\left( \chaintree[0.6]\right)
    +2 \phi_-(\bullet)\phi\left( \twostick[0.6]{black}{black}\right)
    +\phi_-(\bullet)^2 \phi(\bullet)
\right]
=\frac{1}{3\eps^3}-\frac{11}{18\eps^2}-\frac{13}{108\eps}.
\end{align}
The renormalized characters in $\msbar$ are given by the same formulas with $\Rbar$ replacing $R$, resulting in
\begin{align}
  \phi_+(\bullet) &=   -\frac{8}{3}+L  \label{phiplus1}, \\
 \phi_+\left(\twostick[0.6]{black}{black}\right) &=  \frac{791}{72}-\frac{9}{2}L+\frac{1}{2}L^2,  \label{phiplus2}
  \\
   \phi_+\left(\threestick[0.6]{black}{darkgreen}{darkblue}   \right)
    &= 
-\frac{5507}{108}+\frac{2 \zeta_3}{3}+\frac{1555}{72} L -\frac{19}{6}L^2+\frac{1}{6}L^3,\\
\phi_+\left( \chaintree[0.6]\right) &=
-\frac{24155}{648}-\frac{2 \zeta_3}{3} +\frac{389}{18}L-\frac{9}{2}L^2+\frac{1}{3}L^3.
\label{phi+3}
\end{align}
In momentum subtraction, the formulas are again the same using $R_p$ and $1-R_p$ instead of $R$. The reuslts are
\begin{align}
\phi^P_-(\bullet) &= -\Sigma_1(0)= 6
    \frac{ \Gamma^2(2-\eps) \Gamma (\eps - 1)}{\Gamma (4-2\eps)} = \frac{1}{\eps} + \frac{8}{3} + \cOe,\\
 \phi^P_- \left(\twostick[0.6]{black}{black}\right) &=
 \frac{1}{2\eps^2}
+ \frac{7}{4\eps}
+ \left(\frac{137}{72} - \frac{\pi^2}{12}\right) + \cOe,\\
 \phi_-^P\left(\threestick[0.6]{black}{darkgreen}{darkblue}\right) &=
 \frac{1}{6\eps^3}
+ \frac{5}{12\eps^2}
- \frac{329}{216\eps}
- \frac{\pi^2}{24\eps}
+ \left(-\frac{4571}{432} - \frac{5\pi^2}{48} + \frac{7}{6}\zeta_3\right)
+\cOe, \\
 \phi_-^P\left( \chaintree[0.6]\right) &=
 \frac{1}{3\eps^3}
+ \frac{37}{18\eps^2}
+ \frac{851}{108\eps}
- \frac{\pi^2}{12\eps}
+ \left(\frac{19259}{648} - \frac{37\pi^2}{72} - \frac{5}{3}\zeta_3\right) + \cOe. 
\end{align}
Note that in momentum subtraction, there the $\cOe$ terms are not dropped, only the $L$ dependence. Although the counterterms are more complicated, the renormalized amplitudes are simpler (at least after setting $\eps=0$):
\begin{align}
\phi^P_+(\bullet) &= 6 \left(e^{\eps L } - 1\right)
    \frac{ \Gamma^2(2-\eps) \Gamma (\eps - 1)}{\Gamma (4-2\eps)}= L + \cOe,
    \label{phiP1}
\\
 \phi^P_+ \left(\twostick[0.6]{black}{black}\right) &=-\frac{11}{6} L + \frac{1}{2} L^2+ \cOe,\\
 \phi_+^P\left(\threestick[0.6]{black}{darkgreen}{darkblue}\right) &=
 \frac{103}{18} L - \frac{11}{6}L^2 + \frac{1}{6}L^3+ \cOe, \\
 \phi_+^P\left( \chaintree[0.6]\right) &=
 \frac{1}{3}L^3 - \frac{11}{6} L^2+\frac{85}{18}L + \cOe.
 \label{phiP4}
\end{align}
We emphasize the coefficients of the various powers of $L$ are not the same in the renormalized $\msbar$ and momentum-subtraction amplitudes. We also emphasize that one cannot set $\eps=0$ even in the renormalized characters or else the Birkhoff factorization $\phi_+ = \phi_- \ast \phi$ will fail.

The convolution anomalous dimension $\bgamma = -\lim_{\varepsilon \rightarrow 0} \phi_-^{-1} (\varepsilon S \ast  Y)$ from  Eq.~\eqref{CKgamma}  
requires $\phi_-^{-1}(x)$ which is computed using Eq.~\eqref{invrule} with $\phi_-$ replacing $\phi$. From here, we compute
$ -\phi_-^{-1}(\eps S \ast Y x) =-\eps \phi_-(x^{(1)}) \phi_-^{-1}(Y x^{(2)})$. Then, in $\msbar$
\begin{equation}
\bgamma(\oneA) = 0,\quad
\bgamma(\bullet) = 1 ,
\quad  \bgamma\left(\twostick[0.6]{black}{black}\right) =-\frac{11}{6},
\quad
 \bgamma\left(\threestick[0.6]{black}{darkgreen}{darkblue}\right) = \frac{103}{18},
\quad  \bgamma\left( \chaintree[0.6]\right) = -\frac{13}{36}
\end{equation}
We can then verify that $\partial_L \phi_+ =\bgamma \ast \phi_+$. For example, 
\begin{equation}
    (\bgamma \ast \phi_+)\left(\twostick[0.6]{black}{black}\right)  =  
    \bgamma\left(\twostick[0.6]{black}{black}\right) +\bgamma(\bullet)\phi_+(\bullet) = -\frac{9}{2} + L=
    \partial_L \phi_+\left(\twostick[0.6]{black}{black}\right) .
\end{equation}
In momentum subtraction, using  Eq.~\eqref{CKgamma} gives
\begin{equation}
\bgammaP(\bullet) = 1,
\quad  
\bgammaP\left(\twostick[0.6]{black}{black}\right) =-\frac{11}{6}, 
\quad
 \bgammaP\left(\threestick[0.6]{black}{darkgreen}{darkblue}\right) = \frac{103}{18}  ,
\quad
\bgammaP\left( \chaintree[0.6]\right) = \frac{85}{18}, 
\end{equation}
which satisfy $\partial_L \phi_+^P =\bgammaP \ast \phi_+^P$. In momentum subtraction, these coefficients can be computed more simply using Eq.~\eqref{gammaPform}, i.e. by reading off the coefficients of the single logs in Eqs.~\eqref{phiP1}-\eqref{phiP4}.

Finally, we verify that $\phi_+ = \exp^\ast( L \bgammap)$ in Eq.~\eqref{phiplusdelta} in the momentum subtraction scheme. We have already verified this relation at order $L$, in that $\bgammap$ agrees with the coefficient of the $L$ term in $\phi_+(x)$. Next since $\bgammaP$ is an infinitesimal character $\bgammaP \ast \bgammaP$ vanishes on primitive elements, 
\begin{align}
  \frac{1}{2}  \bgammaP \ast\bgammaP \left(\twostick[0.6]{black}{black}\right)  &= \frac{1}{2} [\bgammaP(\bullet)]^2=\frac{1}{2},\\
  \frac{1}{2}  \bgammaP\ast\bgammaP  \left(\threestick[0.6]{black}{darkgreen}{darkblue}\right) &=
  \frac{1}{2} 2 \bgammaP (\bullet) \bgammaP\left(\twostick[0.6]{black}{black}\right) = - \frac{11}{6},\\
  \frac{1}{2}  \bgammaP \ast \bgammaP \left( \chaintree[0.6]\right)&=\frac{1}{2} \left[2 \bgammaP(\bullet) \bgammaP \left(\twostick[0.6]{black}{black}\right)  + \bgammaP(\bullet \bullet)\bgammaP(\bullet) \right] = -\frac{11}{6}.
\end{align}
These are in agreement with the coefficients of the double logs in Eqs.~\eqref{phiP1}-\eqref{phiP4}. Finally,
\begin{align}
    \frac{1}{6}  \bgammaP \ast\bgammaP\ast\bgammaP  \left(\threestick[0.6]{black}{darkgreen}{darkblue}\right) &=
  \frac{1}{6} \big[\bgammaP\ast \bgammaP \left(\twostick[0.6]{black}{black}\right) \big]\bgammaP(\bullet) = [\bgammaP(\bullet)]^3 =\frac{1}{6},\\
  \frac{1}{6}  \bgammaP \ast \bgammaP\ast\bgammaP \left( \chaintree[0.6]\right)&=
  \frac{1}{6} \left\{2\big[\bgammaP\ast\bgammaP (\bullet) \big]\bgammaP\left(\twostick[0.6]{black}{black}\right)  + \big[\bgammaP\ast \bgammaP(\bullet^2)\big]\bgammaP(\bullet) \right\} = \frac{1}{3},
\end{align}
in agreement with the $L^3$ terms. 

We can also verify Eq.~\eqref{desol} in \msbar. Using the notation $\phi_+^0 = R_P \circ \phi_+$ for $\phi_+$ eveluated at $L=0$, we have
\begin{align}
 \bgamma\ast  \phi_+^0 (\bullet) &= \bgamma(\oneA) \phi_+^0(\bullet) + 
 \bgamma(\bullet) \phi_+^0(\oneA) =1,
 \\
 \bgamma\ast  \phi_+^0\left(\twostick[0.6]{black}{black}\right)
 &=\bgamma \left(\twostick[0.6]{black}{black}\right) \phi_+^0(\oneA)
 +\bgamma(\bullet) \phi_+^0(\bullet)
 = -\frac{9}{2},
 \\
    \bgamma\ast  \phi_+^0\left(\threestick[0.6]{black}{darkgreen}{darkblue}\right)
&= \bgamma\left(\threestick[0.6]{black}{darkgreen}{darkblue}\right) \phi_+^0 (\oneA)+\bgamma(\bullet) \phi_+^0 \left( \twostick[0.6]{black}{black}\right) + \bgamma\left(\twostick[0.6]{black}{black}\right)\phi_+^0(\bullet) 
= \frac{1555}{72},
\\
    \bgamma\ast  \phi_+^0\left(\chaintree[0.6]\right)
 &= \bgamma\left(\chaintree[0.6]\right) \phi_+^0 (\oneA)+2\bgamma(\bullet) \phi_+^0 \left( \twostick[0.6]{black}{black}\right) 
 = \frac{389}{18}.
\end{align}
These expressions match the coefficients of the single logs in Eqs.~\eqref{phiplus1}-\eqref{phi+3}.

\input{BorelAppendix}

\bibliographystyle{utphys}
\bibliography{references}
\end{document}

%% file: tikzdiagrams.tex
\newcommand{\bubble}{%
  \vcenter{\hbox{%
   \begin{tikzpicture}[baseline=(current bounding box.center), scale=0.6, line width=1.2pt]
  \begin{feynman}
   \vertex (L) at (-1.5,0);
    \vertex (A) at (-0.7,0);
    \vertex (B) at (0.7,0);
    \vertex (R) at (1.5,0);
    \diagram*{
      (L) -- [plain] (A) -- [plain] (B) -- [plain] (R),
      (A) -- [plain, half left,  looseness=1.5,draw=red] (B),
    };
  \end{feynman}
\end{tikzpicture}
  }}%
}

\newcommand{\bubblediagramlabel}{%
  \vcenter{\hbox{%
   \begin{tikzpicture}[baseline=(current bounding box.center), scale=0.9, line width=1.2pt]
  \begin{feynman}
   \vertex (L) at (-1.5,0);
    \vertex (A) at (-0.7,0);
    \vertex (B) at (0.7,0);
    \vertex (R) at (1.5,0);
    \diagram*{
      (L) -- [plain] (A) -- [plain, edge label'=\(-k\)] (B) -- [plain] (R),
      (A) -- [plain, half left,  looseness=1.5, draw=red, edge label=\(q{+}k\)] (B),
    };
  \end{feynman}
\end{tikzpicture}
  }}%
}

\newcommand{\doublebubble}{%
  \vcenter{\hbox{%
   \begin{tikzpicture}[baseline=(current bounding box.center), scale=0.6, line width=1.2pt]
  \begin{feynman}
   \vertex (L) at (-1.5,0);
    \vertex (A) at (-0.7,0);
    \vertex (A2) at (-0.3,0);
    \vertex (B2) at (0.3,0);
    \vertex (B) at (0.7,0);
    \vertex (R) at (1.5,0);
    \diagram*{
      (L) -- [plain] (A) -- [plain] (A2) -- [plain] (B2) --[plain] (B) -- [plain] (R),
      (A) -- [plain, half left, looseness=1.5, draw=red] (B),
      (A2) -- [plain, half left, draw=red] (B2),
    };
  \end{feynman}
\end{tikzpicture}
  }}%
}

\newcommand{\chain}{%
  \vcenter{\hbox{%
   \begin{tikzpicture}[baseline=(current bounding box.center), scale=0.6, line width=1.2pt]
  \begin{feynman}
    \vertex (L) at (-1.5,0);
    \vertex (A1) at (-0.9,0);
    \vertex (A2) at (-0.6,0);
    \vertex (A3) at (-0.15,0);
    \vertex (B3) at (0.15,0);
    \vertex (B2) at (0.6,0);
    \vertex (B1) at (0.9,0);
    \vertex (R) at (1.5,0);
    \diagram*{
      (L) -- [plain] (A1) -- [plain] (A2) -- [plain] (A3) -- [plain] (B3) -- [plain] (B2) --[plain] (B1) -- [plain] (R),
      (A1) -- [plain,  half left,  looseness=1.5, draw=red] (B1),
      (A2) -- [plain, half left,  looseness=2.0,  draw=red] (A3),
      (B3) -- [plain, half left,  looseness=2.0, draw=red] (B2),
    };
  \end{feynman}
\end{tikzpicture}
  }}%
}

\newcommand{\chainwide}{%
  \vcenter{\hbox{%
   \begin{tikzpicture}[baseline={([yshift=-3pt]current bounding box.center)}, scale=0.6, line width=1.2pt]
  \begin{feynman}
    \vertex (L) at (-1.5,0);
    \vertex (A1) at (-0.9,0);
    \vertex (A2) at (-0.6,0);
    \vertex (A3) at (-0.15,0);
    \vertex (A4) at (0.15,0);
    \vertex (A5) at (0.6,0);
    \vertex (A6) at (0.9,0);
    \vertex (A7) at (1.35,0);
    \vertex (A8) at (1.65,0);
    \vertex (R) at (2.25,0);
    \diagram*{
      (L) -- [plain] (A1) -- [plain] (A2) -- [plain] (A3) -- [plain] (A4) -- [plain] (A5) --[plain] (A6)--[plain] (A7)--[plain] (A8) -- [plain] (R),
      (A1) -- [plain,  half left,  looseness=1.5, draw=red] (A8),
      (A2) -- [plain, half left,  looseness=2.0,  draw=red] (A3),
      (A4) -- [plain, half left,  looseness=2.0, draw=red] (A5),
      (A6) -- [plain, half left,  looseness=2.0, draw=red] (A7),
    };
  \end{feynman}
\end{tikzpicture}
  }}%
}

\newcommand{\rainbow}{%
  \vcenter{\hbox{%
   \begin{tikzpicture}[baseline=(current bounding box.center), scale=0.6, line width=1.2pt]
  \begin{feynman}
    \vertex (L) at (-1.5,0);
    \vertex (A1) at (-1.0,0);
    \vertex (A2) at (-0.6,0);
    \vertex (A3) at (-0.2,0);
    \vertex (B3) at (0.2,0);
    \vertex (B2) at (0.6,0);
    \vertex (B1) at (1.0,0);
    \vertex (R) at (1.5,0);
    \diagram*{
      (L) -- [plain] (A1) -- [plain] (A2) -- [plain] (A3) -- [plain] (B3) -- [plain] (B2) --[plain] (B1) -- [plain] (R),
      (A1) -- [plain,  half left,  looseness=1.5, draw=red] (B1),
      (A2) -- [plain, half left,  looseness=1.5,  draw=red] (B2),
      (A3) -- [plain, half left,  looseness=2.0, draw=red] (B3),
    };
  \end{feynman}
\end{tikzpicture}
  }}%
}

\newcommand{\doublesigma}{%
  \vcenter{\hbox{%
   \begin{tikzpicture}[baseline=(current bounding box.center), scale=1.2, line width=1.2pt]
  \begin{feynman}
   \vertex (L) at (-1.5,0);
    \vertex (A) at (-0.7,0);
    \vertex (B) at (0.7,0);
    \vertex (R) at (1.5,0);
    \diagram*{
      (L) -- [plain] (A),
      (B) -- [plain] (R),
      (A) -- [plain, half left, looseness=1.5] (B),
       (A) -- [plain, half right, looseness=1.5] (B),
    };
  \end{feynman}
\node[fill=white, draw=green!50!black, rounded corners=2pt, inner sep=3pt]
       at ($ (A)!0.5!(B) + (0, 0.60) $) {$\Sigma_B$};
\node[fill=white, draw=green!50!black, rounded corners=2pt, inner sep=3pt]
       at ($ (A)!0.5!(B) + (0,-0.60) $) {$\Sigma_B$};
\end{tikzpicture}
  }}%
}

\newcommand{\sunset}{%
  \vcenter{\hbox{%
   \begin{tikzpicture}[baseline=(current bounding box.center), scale=0.6, line width=1.2pt]
  \begin{feynman}
   \vertex (L) at (-1.5,0);
    \vertex (A) at (-0.7,0);
    \vertex (B) at (0.7,0);
    \vertex (R) at (1.5,0);
    \diagram*{
      (L) -- [plain] (A) -- [plain] (B) -- [plain] (R),
      (A) -- [plain, half left,  looseness=1.5,draw=black] (B),
      (A) -- [plain, half right,  looseness=1.5,draw=black] (B),
    };
  \end{feynman}
\end{tikzpicture}
  }}%
}

\newcommand{\doublesun}{%
  \vcenter{\hbox{%
   \begin{tikzpicture}[baseline=(current bounding box.center), scale=0.6, line width=1.2pt]
  \begin{feynman}
   \vertex (L) at (-1.5,0);
    \vertex (A1) at (-1.0,0);
    \vertex (A2) at (-0.5,0);
    \vertex (B2) at (0.5,0);
    \vertex (B1) at (1.0,0);
    \vertex (R) at (1.5,0);
    \diagram*{
      (L) -- [plain] (A1) -- [plain] (A2) -- [plain](B2)-- [plain] (B1) -- [plain] (R),
      (A1) -- [plain, half left,  looseness=1.5,draw=black] (B1),
      (A2) -- [plain, half left,  looseness=1.5,draw=black] (B2),
      (A2) -- [plain, half right,  looseness=1.5,draw=black] (B2),
      (A1) -- [plain, half right,  looseness=1.5,draw=black] (B1),
    };
  \end{feynman}
\end{tikzpicture}
  }}%
}

\newcommand{\threesuns}{%
  \vcenter{\hbox{%
  \begin{tikzpicture}
  [baseline=(current bounding box.center), scale=0.6, line width=1.2pt, line cap=round, line join=round]
  \def\cr{5pt}
  \tikzset{
    circle-mid/.style={
      postaction={
        decorate,
        decoration={
          markings,
          mark=at position .5 with {\draw (0,0) circle (\cr);}
        }
      }
    }
  }
\begin{feynman}
  \vertex (L) at (-1.5,0);
    \vertex (A1) at (-1.0,0);
    \vertex (A2) at (-0.5,0);
    \vertex (B2) at (0.5,0);
    \vertex (B1) at (1.0,0);
    \vertex (R) at (1.5,0);
    \diagram*{
      (L) -- [plain] (A1) -- [plain] (A2) -- [plain](B2)-- [plain] (B1) -- [plain] (R),
      (A1) -- [plain, bend left=90,looseness=1.5] (B1),
      (A2) -- [plain, bend left=90,looseness=1.5] (B2),
      (A2) -- [plain, bend right=90,looseness=1.5] (B2),
      (A1) -- [plain, bend right=90,looseness=1.5] (B1),
    };
  \end{feynman}
   \draw[circle-mid] (A1) to[bend left=90,looseness=1.5] (B1);                 
\end{tikzpicture}
  }}%
}

\newcommand{\manysuns}{%
  \vcenter{\hbox{%
  \begin{tikzpicture}
  [baseline=(current bounding box.center), scale=0.6, line width=1.2pt, line cap=round, line join=round]
  \def\cr{5pt}
  \tikzset{
    circle-mid/.style={
      postaction={
        decorate,
        decoration={
          markings,
          mark=at position .5 with {\draw (0,0) circle (\cr);}
        }
      }
    }
  }
  \tikzset{
    circle-quart/.style={
      postaction={
        decorate,
        decoration={
          markings,
          mark=at position .25 with {\draw (0,0) circle (\cr);}
        }
      }
    }
  }
\begin{feynman}
   \vertex (L1) at (-2,0);
   \vertex (L) at (-1.5,0);
    \vertex (A1) at (-1.0,0);
    \vertex (A2) at (-0.5,0);
    \vertex (B2) at (0.5,0);
    \vertex (B1) at (1.0,0);
    \vertex (R) at (1.5,0);
    \vertex (R1) at (2,0);
    \diagram*{
     (L1) -- [plain] (L) -- [plain] (A1) -- [plain] (A2) -- [plain](B2)-- [plain] (B1) -- [plain] (R) -- [plain] (R1),
      (A1) -- [plain, bend left=90,looseness=1.5] (B1),
      (A2) -- [plain, bend left=90,looseness=1.5] (B2),
      (A2) -- [plain, bend right=90,looseness=1.5] (B2),
      (A1) -- [plain, bend right=90,looseness=1.5] (B1),
       (L) -- [plain, bend left=90,looseness=1.5] (R),
      (L) -- [plain, bend right=90,looseness=1.5] (R),

    };
  \end{feynman}
   \draw[circle-mid] (A1) to[bend left=90,looseness=1.5] (B1);                 
   \draw[circle-quart] (A1) to[bend left=90,looseness=1.5] (B1);                 
   \draw[circle-mid] (A2) to[bend right=90,looseness=1.5] (B2);                  
   \draw[circle-quart] (B1) to[bend left=90,looseness=1.5] (A1);                    \draw[circle-quart] (R) to[bend right=90,looseness=1.5] (L);  
\end{tikzpicture}
  }}%
}

\newcommand{\bubblechain}{%
  \vcenter{\hbox{%
   \begin{tikzpicture}[baseline=(current bounding box.center), scale=1.2, line width=1.2pt]
     \usetikzlibrary{calc}
     \begin{feynman}
       \vertex (L) at (-1.5,0);
       \vertex (A) at (-0.7,0);
       \vertex (B) at ( 0.7,0);
       \vertex (R) at ( 1.5,0);
       \diagram*{
         (L) -- [plain] (A),
         (B) -- [plain] (R),
         (A) -- [plain, half left, looseness=1.5 ] (B), 
       };
     \end{feynman}
     \pgfmathsetmacro{\r}{1.4/8} 
     \foreach \i in {0,1,2,3}{
       \draw ($ (A) + (2*\i*\r+\r ,0) $) circle (\r);
     }
   \end{tikzpicture}
  }}%
}

\newcommand{\tadpole}{%
  \vcenter{\hbox{%
   \begin{tikzpicture}[baseline=(current bounding box.center), scale=0.6, line width=1.2pt]
  \begin{feynman}
   \vertex (L) at (-1.0,0);
    \vertex (A) at (0,0);
    \vertex (R) at (1.0,0);
    \diagram*{
      (L) -- [plain] (A)  -- [plain] (R),
    };
  \end{feynman}
   \draw ($  (A)+(0,0.5) $) circle (0.5);
\end{tikzpicture}
  }}%
}

\newcommand{\midblob}{%
  \vcenter{\hbox{%
   \begin{tikzpicture}[baseline=(current bounding box.center), scale=0.6, line width=1.2pt]
  \begin{feynman}
    \vertex (Ldown) at (-0.8, 0.6);
    \vertex (Lup) at (-0.8,-0.6);
    \vertex (A) at (0,0);
    \vertex (Rdown) at (0.8, 0.6);
    \vertex (Rup) at (0.8,-0.6);
    \diagram*{
      (Lup) -- [plain] (A) -- [plain] (Ldown),
      (Rup) -- [plain] (A) -- [plain] (Rdown),
    };
  \end{feynman}
  \node[circle, draw=black, fill=green!70!black, minimum size=10pt] at (A) {};
  \node at (Ldown) [left=0.1,scale=0.8] {$p_2$};
  \node at (Lup) [left=0.1,scale=0.8] {$p_1$};
  \node at (Rup) [right=0.1,scale=0.8] {$p_4$};
  \node at (Rdown) [right=0.1,scale=0.8] {$p_3$};
\end{tikzpicture}
  }}%
}

\newcommand{\cross}{%
  \vcenter{\hbox{%
   \begin{tikzpicture}[baseline=(current bounding box.center), scale=0.6, line width=1.2pt]
  \begin{feynman}
    \vertex (Ldown) at (-0.7, 0.6);
    \vertex (Lup) at (-0.7,-0.6);
    \vertex (A) at (0,0);
    \vertex (Rdown) at (0.7, 0.6);
    \vertex (Rup) at (0.7,-0.6);
    \diagram*{
      (Lup) -- [plain] (A) -- [plain] (Ldown),
      (Rup) -- [plain] (A) -- [plain] (Rdown),
    };
  \end{feynman}
\end{tikzpicture}
  }}%
}

\newcommand{\crossct}{%
  \vcenter{\hbox{%
   \begin{tikzpicture}[baseline=(current bounding box.center), scale=0.6, line width=1.2pt]
  \begin{feynman}
    \vertex (Ldown) at (-0.7, 0.6);
    \vertex (Lup) at (-0.7,-0.6);
    \vertex (A) at (0,0);
    \vertex (Rdown) at (0.7, 0.6);
    \vertex (Rup) at (0.7,-0.6);
    \diagram*{
      (Lup) -- [plain] (A) -- [plain] (Ldown),
      (Rup) -- [plain] (A) -- [plain] (Rdown),
    };
  \end{feynman}
    \node[star, star points=5, star point ratio=2.25,
          draw, fill=green, minimum size=6pt, inner sep=0pt] at (A) {};
  \end{tikzpicture}
  }}%
}

\newcommand{\tnoblob}{%
  \vcenter{\hbox{%
   \begin{tikzpicture}[baseline=(current bounding box.center), scale=0.6, line width=1.2pt]
  \begin{feynman}
    \vertex (Ldown) at (0.6,-1.5);
    \vertex (Lup) at (-0.6,-1.5);
    \vertex (A) at (0,-0.8);
    \vertex (B) at (0,0.8);
    \vertex (Rdown) at (0.6,1.5);
    \vertex (Rup) at (-0.6,1.5);
    \diagram*{
      (Lup) -- [plain] (A),
      (Ldown) -- [plain] (A),
      (A) -- [plain, half left, looseness=1.2] (B),
      (A) -- [plain, half right, looseness=1.2] (B),
      (B) -- [plain] (Rup),
      (B) -- [plain] (Rdown),
    };
  \end{feynman}
\end{tikzpicture}
  }}%
}

\newcommand{\unoblob}{%
  \vcenter{\hbox{%
   \begin{tikzpicture}[baseline=(current bounding box.center), scale=0.6, line width=1.2pt]
  \begin{feynman}
    \vertex (Ldown) at (0.6,-1.5);
    \vertex (Lup) at (-0.6,-1.5);
    \vertex (A) at (0,-0.8);
    \vertex (B) at (0,0.8);
    \vertex (Rdown) at (0.6,1.5);
    \vertex (Rup) at (-0.6,1.5);
    \diagram*{
      (Lup) -- [plain] (A),
      (Ldown) -- [plain,bend right=0, looseness=0.5] (B),
      (A) -- [plain, half left, looseness=1.2] (B),
      (A) -- [plain, half right, looseness=1.2] (B),
      (B) -- [plain] (Rup),
      (A) -- [plain,bend right=0, looseness=0.5] (Rdown),
    };
  \end{feynman}
\end{tikzpicture}
  }}%
}

\newcommand{\snoblob}{%
  \vcenter{\hbox{%
   \begin{tikzpicture}[baseline=(current bounding box.center), scale=0.6, line width=1.2pt]
  \begin{feynman}
    \vertex (Ldown) at (-1.5, 0.6);
    \vertex (Lup) at (-1.5,-0.6);
    \vertex (A) at (-0.8,0);
    \vertex (B) at ( 0.8,0);
    \vertex (Rdown) at (1.5, 0.6);
    \vertex (Rup) at (1.5,-0.6);
    \diagram*{
      (Lup) -- [plain] (A),
      (Ldown) -- [plain] (A),
      (A) -- [plain, half left, looseness=1.2] (B),
      (A) -- [plain, half right, looseness=1.2] (B),
      (B) -- [plain] (Rup),
      (B) -- [plain] (Rdown),
    };
  \end{feynman}
\end{tikzpicture}
  }}%
}

\newcommand{\snoblobct}{%
  \vcenter{\hbox{%
   \begin{tikzpicture}[baseline=(current bounding box.center), scale=0.6, line width=1.2pt]
  \begin{feynman}
    \vertex (Ldown) at (-1.5, 0.6);
    \vertex (Lup) at (-1.5,-0.6);
    \vertex (A) at (-0.8,0);
    \vertex (B) at ( 0.8,0);
    \vertex (Rdown) at (1.5, 0.6);
    \vertex (Rup) at (1.5,-0.6);
    \diagram*{
      (Lup) -- [plain] (A),
      (Ldown) -- [plain] (A),
      (A) -- [plain, half left, looseness=1.2] (B),
      (A) -- [plain, half right, looseness=1.2] (B),
      (B) -- [plain] (Rup),
      (B) -- [plain] (Rdown),
    };
  \end{feynman}
      \node[star, star points=5, star point ratio=2.25,
          draw, fill=green, minimum size=6pt, inner sep=0pt] at (B) {};
\end{tikzpicture}
  }}%
}

\newcommand{\doubles}{%
  \vcenter{\hbox{%
   \begin{tikzpicture}[baseline=(current bounding box.center), scale=0.6, line width=1.2pt]
  \begin{feynman}
    \vertex (Ldown) at (-1.5, 0.6);
    \vertex (Lup) at (-1.5,-0.6);
    \vertex (A) at (-0.8,0);
    \vertex (B) at ( 0.8,0);
    \vertex (C) at ( 2.4,0);
    \vertex (Rdown) at (3.1, 0.6);
    \vertex (Rup) at (3.1,-0.6);
    \diagram*{
      (Lup) -- [plain] (A),
      (Ldown) -- [plain] (A),
      (A) -- [plain, half left, looseness=1.2] (B),
      (A) -- [plain, half right, looseness=1.2] (B),
      (B) -- [plain, half left, looseness=1.2] (C),
      (B) -- [plain, half right, looseness=1.2] (C),
      (C) -- [plain] (Rup),
      (C) -- [plain] (Rdown),
    };
  \end{feynman}
\end{tikzpicture}
  }}%
}

\newcommand{\icecream}{%
  \vcenter{\hbox{%
   \begin{tikzpicture}[baseline=(current bounding box.center), scale=1.0, line width=1.2pt]
  \begin{feynman}
    \vertex (Ldown) at (-1.5, 0.6);
    \vertex (Lup) at (-1.5,-0.6);
    \vertex (A) at (-0.8,0);
    \vertex (B) at ( 0.8,0);
    \vertex (C) at ( 0.6,0.4);
    \vertex (D) at ( 0.6,-0.4);
    \vertex (Rdown) at (1.5, 0.6);
    \vertex (Rup) at (1.5,-0.6);
    \diagram*{
      (Lup) -- [plain] (A),
      (Ldown) -- [plain] (A),
      (A) -- [plain, half left, looseness=1.2] (B),
      (A) -- [plain, half right, looseness=1.2] (B),
      (C) -- [plain, half right, looseness=1.2] (D),
      (D) -- [plain] (Rup),
      (C) -- [plain] (Rdown),
    };
  \end{feynman}
  \draw[->,thin] (-0.3,-0.7) to[bend right=10] (0.3,-0.7) node[midway, below=0.8,scale=0.8] {$k$};
  \draw[->,thin] (-0.3,0.7) to[bend left=10] (0.3,0.7) node[midway, above=0.8,scale=0.8] {$p_1+p_2 - k$};
  \draw[->,thin] (1.3, -0.7) -- (0.8, -0.6) node[midway, below=0.1,scale=0.8] {$p_4$};
\end{tikzpicture}
  }}%
}

\newcommand{\baseball}{%
  \vcenter{\hbox{%
   \begin{tikzpicture}[baseline=(current bounding box.center), scale=1.0, line width=1.2pt]
  \begin{feynman}
    \vertex (Ldown) at (-1.5, 0.6);
    \vertex (Lup) at (-1.5,-0.6);
    \vertex (A) at (-0.8,0);
    \vertex (B) at ( 0.8,0);
    \vertex (C) at ( 0.6,0.4);
    \vertex (D) at ( 0.6,-0.4);
    \vertex (E) at ( -0.6,0.4);
    \vertex (F) at ( -0.6,-0.4);
    \vertex (Rdown) at (1.5, 0.6);
    \vertex (Rup) at (1.5,-0.6);
    \diagram*{
      (Lup) -- [plain] (F),
      (Ldown) -- [plain] (E),
      (E) -- [plain, half left, looseness=1.2] (F),
      (A) -- [plain, half left, looseness=1.2] (B),
      (A) -- [plain, half right, looseness=1.2] (B),
      (C) -- [plain, half right, looseness=1.2] (D),
      (D) -- [plain] (Rup),
      (C) -- [plain] (Rdown),
    };
  \end{feynman}
  \draw[->,thin] (-1.3, -0.7) -- (-0.8, -0.6) node[midway, below=0.1,scale=0.8] {$p_1$};
  \draw[->,thin] (1.3, -0.7) -- (0.8, -0.6) node[midway, below=0.1,scale=0.8] {$p_4$};
  \draw[->,thin] (-0.3,-0.7) to[bend right=10] (0.3,-0.7) node[midway, below=0.8,scale=0.8] {$k$};
  \draw[->,thin] (-0.3,0.7) to[bend left=10] (0.3,0.7) node[midway, above=0.8,scale=0.8] {$P - k$};
\end{tikzpicture}
  }}%
}

\newcommand{\tadpoleblob}{%
  \vcenter{\hbox{%
   \begin{tikzpicture}[baseline=(current bounding box.center), scale=0.6, line width=1.2pt]
  \begin{feynman}
   \vertex (L) at (-2.0,0);
    \vertex (A) at (0,0);
    \vertex (R) at (2.0,0);
    \diagram*{
      (L) -- [plain] (A)  -- [plain] (R),
    };
  \end{feynman}
 \draw (0,1.0) ellipse [x radius=0.5, y radius=1.0];
  \node[circle, draw=black, fill=green!70!black, minimum size=10pt] at (A) {};
  \draw[->,thin] (0.5,-0.2) -- (1.0,-0.2) node[midway, below=2pt] {$p$};
  \draw[->,thin] (-1.0,-0.2) -- (-0.5,-0.2) node[midway, below=2pt] {$p$};
  \draw[->,thin] (0.6,0.3) to[bend right=30] (0.8,1.2) node[right] {$k$}; \draw[->,thin] (-0.8,1.2) to[bend right=30] (-0.6,0.3) node[above left = 2pt] {$k$};
\end{tikzpicture}
  }}%
}

\newcommand{\chaintree}[1][1.0]{%
  \begin{tikzpicture}[baseline={([yshift=-3pt]current bounding box.center)}, line width=1pt, scale=#1]
    \coordinate (T) at (0,0.4);
    \coordinate (L) at (-0.4,-0.3);
    \coordinate (R) at ( 0.4,-0.3);
    \draw[red] (T) -- (L);
    \draw[red] (T) -- (R);
    \fill[black] (T) circle (3pt);
    \fill[blue]  (L) circle (3pt);
    \fill[darkgreen] (R) circle (3pt);
  \end{tikzpicture}%
}

\newcommand{\twostick}[3][1.0]{%
  \begin{tikzpicture}[baseline={([yshift=-3pt]current bounding box.center)}, line width=1pt, scale=#1]
    \coordinate (T) at (0,0.4);
    \coordinate (B) at (0,-0.4);
    \draw[red] (T) -- (B);
    \fill[#2] (T) circle (3pt);
    \fill[#3] (B) circle (3pt);
  \end{tikzpicture}%
}

\newcommand{\threestick}[4][1.0]{%
  \begin{tikzpicture}[baseline={([yshift=-3pt]current bounding box.center)}, line width=1pt, scale=#1]
    \coordinate (T) at (0,0.6);
    \coordinate (M) at (0,0);
    \coordinate (B) at (0,-0.6);
    \draw[red] (T) -- (M) -- (B);
    \fill[#2] (T) circle (3pt);
    \fill[#3] (M) circle (3pt);
    \fill[#4] (B) circle (3pt);
  \end{tikzpicture}%
}

\newcommand{\Bx}{%
  \vcenter{\hbox{%
   \begin{tikzpicture}[baseline=(current bounding box.center), scale=1.0, line width=1.2pt]
  \begin{feynman}
   \vertex (L) at (-1.5,0);
    \vertex (A) at (-1,0);
    \vertex (B) at (1,0);
    \vertex (R) at (1.5,0);
    \diagram*{
      (L) -- [plain] (A) -- [plain] (B) -- [plain] (R),
      (A) -- [plain, half left,  looseness=1.5,draw=red] (B),
    };
  \end{feynman}
  \node[fill=white, draw=green!50!black, rounded corners=2pt, inner sep=4pt]
       at ($ (A)!0.5!(B)$) {$x$};
\end{tikzpicture}
  }}%
}

\newcommand{\Bxy}{%
  \vcenter{\hbox{%
   \begin{tikzpicture}[baseline=(current bounding box.center), scale=1.0, line width=1.2pt]
  \begin{feynman}
   \vertex (L) at (-1.5,0);
    \vertex (A) at (-0.7,0);
    \vertex (B) at (0.7,0);
    \vertex (R) at (1.5,0);
    \diagram*{
      (L) -- [plain] (A),
      (B) -- [plain] (R),
      (A) -- [plain, half left, looseness=1.5] (B),
       (A) -- [plain, half right, looseness=1.5] (B),
    };
  \end{feynman}
\node[fill=white, draw=green!50!black, rounded corners=2pt, inner sep=4pt]
       at ($ (A)!0.5!(B) + (0, 0.60) $) {$x$};
\node[fill=white, draw=green!50!black, rounded corners=2pt, inner sep=4pt]
       at ($ (A)!0.5!(B) + (0,-0.60) $) {$y$};
\end{tikzpicture}
  }}%
}

\newcommand{\Bxyzw}{%
  \vcenter{\hbox{%
   \begin{tikzpicture}[baseline=(current bounding box.center), scale=1.0, line width=1.2pt]
  \begin{feynman}
   \vertex (L) at (-1.8,0);
    \vertex (A) at (-1,0);
    \vertex (B) at (1,0);
    \vertex (R) at (1.8,0);
    \diagram*{
      (L) -- [plain] (A) -- [plain] (B) -- [plain] (R),
      (A) -- [plain, half left,  looseness=1.5,draw=darkred] (B),
    };
  \end{feynman}
  \node[fill=white, draw=green!50!black, rounded corners=2pt, inner sep=4pt]
       at ($ (A)!0.5!(B)$) {$x$};
  \node[fill=white, draw=green!50!black, rounded corners=2pt, inner sep=4pt]
       at ($(A)$) {$z$};
  \node[fill=white, draw=green!50!black, rounded corners=2pt, inner sep=4pt]
       at ($(B)$) {$w$};
  \node[fill=white, draw=green!50!black, rounded corners=2pt, inner sep=4pt]
       at ($ (A)!0.5!(B) + (0,0.9)$) {$y$};
\end{tikzpicture}
  }}%
}

\newcommand{\linect}[1][-3pt]{
  \begin{tikzpicture}[baseline={([yshift=#1]current bounding box.center)}, line width=1pt, scale=1.0]
    \coordinate (L) at (-0.5,0);
    \coordinate (M) at (0,0);
    \coordinate (R) at (0.5,0);
    \draw[black] (L) -- (R);
    \node[star, star points=5, star point ratio=2.25,
          draw, fill=green, minimum size=6pt, inner sep=0pt] at (M) {};
 \end{tikzpicture}%
}

\newcommand{\linectO}[1][-3pt]{
  \begin{tikzpicture}[baseline={([yshift=#1]current bounding box.center)}, line width=1pt, scale=1.0]
    \coordinate (L) at (-0.5,0);
    \coordinate (M) at (0,0);
    \coordinate (R) at (0.5,0);
    \draw[black] (L) -- (R);
    \node[star, star points=5, star point ratio=2.25,
          draw, fill=orange, minimum size=7pt, inner sep=0pt] at (M) {};
 \end{tikzpicture}%
}

\newcommand{\bubblect}{%
  \vcenter{\hbox{%
   \begin{tikzpicture}[baseline=(current bounding box.center), scale=0.6, line width=1.2pt]
  \begin{feynman}
   \vertex (L) at (-1.5,0);
    \vertex (A) at (-0.7,0);
    \vertex (M) at (0,0);
    \vertex (B) at (0.7,0);
    \vertex (R) at (1.5,0);
    \diagram*{
      (L) -- [plain] (A) -- [plain] (B) -- [plain] (R),
      (A) -- [plain, half left,  looseness=1.5,draw=red] (B),
    };
  \end{feynman}
  \node[star, star points=5, star point ratio=2.25,
          draw, fill=green, minimum size=6pt, inner sep=0pt] at (M) {};
\end{tikzpicture}
  }}%
}

\newcommand{\bubblectL}{%
  \vcenter{\hbox{%
   \begin{tikzpicture}[baseline=(current bounding box.center), scale=0.6, line width=1.2pt]
  \begin{feynman}
   \vertex (L) at (-1.5,0);
    \vertex (A) at (-0.7,0);
    \vertex (M) at (0,0);
    \vertex (B) at (0.7,0);
    \vertex (R) at (1.5,0);
    \diagram*{
      (L) -- [plain] (A) -- [plain] (B) -- [plain] (R),
      (A) -- [plain, half left,  looseness=1.5,draw=red] (B),
    };
  \end{feynman}
  \node[star, star points=5, star point ratio=2.25,
          draw, fill=darkblue, minimum size=6pt, inner sep=0pt] at (A) {};
\end{tikzpicture}
  }}%
}

\newcommand{\bubblectR}{%
  \vcenter{\hbox{%
   \begin{tikzpicture}[baseline=(current bounding box.center), scale=0.6, line width=1.2pt]
  \begin{feynman}
   \vertex (L) at (-1.5,0);
    \vertex (A) at (-0.7,0);
    \vertex (M) at (0,0);
    \vertex (B) at (0.7,0);
    \vertex (R) at (1.5,0);
    \diagram*{
      (L) -- [plain] (A) -- [plain] (B) -- [plain] (R),
      (A) -- [plain, half left,  looseness=1.5,draw=red] (B),
    };
  \end{feynman}
  \node[star, star points=5, star point ratio=2.25,
          draw, fill=darkblue, minimum size=6pt, inner sep=0pt] at (B) {};
\end{tikzpicture}
  }}%
}

\newcommand{\bubbletwoct}{%
  \vcenter{\hbox{%
   \begin{tikzpicture}[baseline=(current bounding box.center), scale=0.6, line width=1.2pt]
  \begin{feynman}
   \vertex (L) at (-1.5,0);
    \vertex (A) at (-0.7,0);
    \vertex (M) at (0,0);
    \vertex (B) at (0.7,0);
    \vertex (R) at (1.5,0);
    \diagram*{
      (L) -- [plain] (A) -- [plain] (B) -- [plain] (R),
      (A) -- [plain, half left,  looseness=1.5,draw=red] (B),
    };
  \end{feynman}
  \node[star, star points=5, star point ratio=2.25,
          draw, fill=green, minimum size=6pt, inner sep=0pt] at (-0.4,0) {};
  \node[star, star points=5, star point ratio=2.25,
          draw, fill=green, minimum size=6pt, inner sep=0pt] at (0.4,0) {};
\end{tikzpicture}
  }}%
}

\newcommand{\chainctL}{%
  \vcenter{\hbox{%
   \begin{tikzpicture}[baseline=(current bounding box.center), scale=0.6, line width=1.2pt]
  \begin{feynman}
    \vertex (L) at (-1.5,0);
    \vertex (A1) at (-0.9,0);
    \vertex (A2) at (-0.6,0);
    \vertex (A3) at (-0.15,0);
    \vertex (B3) at (0.15,0);
    \vertex (B2) at (0.6,0);
    \vertex (B1) at (0.9,0);
    \vertex (R) at (1.5,0);
    \diagram*{
      (L) -- [plain] (A1) -- [plain] (A2) -- [plain] (A3) -- [plain] (B3) -- [plain] (B2) --[plain] (B1) -- [plain] (R),
      (A1) -- [plain,  half left,  looseness=1.5, draw=red] (B1),
      (A2) -- [plain, half left,  looseness=2.0,  draw=red] (A3),
    };
  \end{feynman}
 \node[star, star points=5, star point ratio=2.25,
          draw, fill=green, minimum size=6pt, inner sep=0pt] at (0.4,0) {};
\end{tikzpicture}
  }}%
}

\newcommand{\chainctR}{%
  \vcenter{\hbox{%
   \begin{tikzpicture}[baseline=(current bounding box.center), scale=0.6, line width=1.2pt]
  \begin{feynman}
    \vertex (L) at (-1.5,0);
    \vertex (A1) at (-0.9,0);
    \vertex (A2) at (-0.6,0);
    \vertex (A3) at (-0.15,0);
    \vertex (B3) at (0.15,0);
    \vertex (B2) at (0.6,0);
    \vertex (B1) at (0.9,0);
    \vertex (R) at (1.5,0);
    \diagram*{
      (L) -- [plain] (A1) -- [plain] (A2) -- [plain] (A3) -- [plain] (B3) -- [plain] (B2) --[plain] (B1) -- [plain] (R),
      (A1) -- [plain,  half left,  looseness=1.5, draw=red] (B1),
      (B3) -- [plain, half left,  looseness=2.0,  draw=red] (B2),
    };
  \end{feynman}
 \node[star, star points=5, star point ratio=2.25,
          draw, fill=green, minimum size=6pt, inner sep=0pt] at (-0.4,0) {};
\end{tikzpicture}
  }}%
}

\newcommand{\hopffour}{%
  \vcenter{\hbox{%
   \begin{tikzpicture}[baseline=(current bounding box.center), scale=0.8, line width=1.2pt]
  \begin{feynman}
    \vertex (L) at (-1.9,0);
    \vertex (A1) at (-1.4,0);
    \vertex (A2) at (-1.0,0);
    \vertex (A3) at (-0.6,0);
    \vertex (A4) at (-0.2,0);
    \vertex (A5) at (0.2,0);
    \vertex (A6) at (0.7,0);
    \vertex (A7) at (1.3,0);
    \vertex (A8) at (1.7,0);
    \vertex (R) at (1.9,0);
    \diagram*{
      (L) -- [plain] (A1) -- [plain] (A2) -- [plain] (A3) -- [plain] (A4) -- [plain] (A5) -- [plain] (A6) -- [plain] (A7) -- [plain] (A8) -- [plain] (R),
      (A1) -- [plain,  half left,  looseness=1.5, draw=red] (A8),
      (A2) -- [plain, half left,  looseness=1.5,  draw=red] (A5),
      (A3) -- [plain, half left,  looseness=2.0, draw=red] (A4),
      (A6) -- [plain, half left,  looseness=2.0, draw=red] (A7),
    };
  \end{feynman}
  \node[scale=1.0, transform shape, darkblue] at ($(A2) - (0, 0.2)$) {$+$};
  \node[scale=1.0, transform shape, darkblue] at ($(A3) - (0, 0.2)$) {$+$};
  \node[scale=1.0, transform shape, darkgreen] at ($(A4) - (0, 0.2)$) {$-$};
  \node[scale=1.0, transform shape, darkgreen] at ($(A5) - (0, 0.2)$) {$-$};
  \node[scale=1.0, transform shape, darkblue] at ($(A6) - (0, 0.2)$) {$+$};
  \node[scale=1.0, transform shape, darkgreen] at ($(A7) - (0, 0.2)$) {$-$};
\end{tikzpicture}
  }}%
}

\newcommand{\hopffive}{%
  \vcenter{\hbox{%
   \begin{tikzpicture}[baseline=(current bounding box.center), scale=0.8, line width=1.2pt]
  \begin{feynman}
    \vertex (L) at (-1.9,0);
    \vertex (A1) at (-1.4,0);
    \vertex (A2) at (-1.0,0);
    \vertex (A3) at (-0.6,0);
    \vertex (A4) at (-0.2,0);
    \vertex (A5) at (0.2,0);
    \vertex (A6) at (0.7,0);
    \vertex (A7) at (1.3,0);
    \vertex (A8) at (1.7,0);
    \vertex (R) at (1.9,0);
    \diagram*{
      (L) -- [plain] (A1) -- [plain] (A2) -- [plain] (A3) -- [plain] (A4) -- [plain] (A5) -- [plain] (A6) -- [plain] (A7) -- [plain] (A8) -- [plain] (R),
      (A1) -- [plain,  half left,  looseness=1.5, draw=red] (A8),
      (A2) -- [plain, half left,  looseness=1.5,  draw=red] (A6),
      (A3) -- [plain, half left,  looseness=2.0, draw=red] (A4),
      (A5) -- [plain, half right,  looseness=1.5, draw=red] (A7),
    };
  \end{feynman}
  \draw[darkgreen, dashed] ($(0, -0.6)$) rectangle (1.5,0.4);
\end{tikzpicture}
  }}%
}

\newcommand{\FSEone}[1]{
  \pgfmathsetmacro{\L}{1.80*#1}   
  \pgfmathsetmacro{\A}{1.50*#1}   
  \pgfmathsetmacro{\Cx}{0.55*#1}  
  \pgfmathsetmacro{\Hy}{0.90*#1}  
  \pgfmathsetmacro{\By}{0.82*#1}  
  \pgfmathsetmacro{\R}{0.34*#1}   

  \draw (-\L,0) -- (\L,0);

  \draw[decorate,decoration={snake,amplitude=1.1pt,segment length=3pt}]
       (-\A,0) .. controls (-\Cx,\Hy) and (\Cx,\Hy) .. (\A,0);

  \filldraw[fill=white] (0,\By) circle (\R);
}

\newcommand{\FSErec}[2]{
  \FSEone{#2}%
  \ifnum#1>1
    \pgfmathtruncatemacro{\nextd}{#1-1}
    \pgfmathsetmacro{\childs}{0.58*#2}   
    \pgfmathsetmacro{\shiftx}{0.85*#2}   
    \begin{scope}[shift={(-\shiftx,0)}]
      \FSErec{\nextd}{\childs}
    \end{scope}
    \begin{scope}[shift={( \shiftx,0)}]
      \FSErec{\nextd}{\childs}
    \end{scope}
  \fi
}

\newcommand{\chaintreethree}{%
  \begin{tikzpicture}[baseline={([yshift=-3pt]current bounding box.center)}, line width=1pt, scale=1.0]
    \coordinate (T) at (0,0.4);
    \coordinate (L) at (-0.4,-0.3);
    \coordinate (R) at ( 0.4,-0.3);
    \coordinate (L1) at ( -0.8,-1.0);
    \coordinate (L2) at ( 0,-1.0);
    \coordinate (shift) at (0.2,0);
    \coordinate (shifty) at (0,0.1);
    \draw[red] (T) -- (L);
    \draw[red] (T) -- (R);
    \draw[red] (L) -- (L1);
    \draw[red] (L) -- (L2);
    \fill[black] (T) circle (3pt);
    \fill[black]  (L) circle (3pt);
    \fill[black] (R) circle (3pt);
    \fill[black] (L1) circle (3pt);
    \fill[black] (L2) circle (3pt);

    \draw[->, thin,darkblue]  ($(T)!0.2!(L) - (shift)$) -- ($(T)!0.8!(L) - (shift)$);   
    \node[scale=0.8, darkblue] at ($(T)!0.4!(L) - 2*(shift)$) {$+$};
    
    \draw[->, thin,darkblue]  ($(L)!0.2!(L1) - (shift)$) -- ($(L)!0.8!(L1) - (shift)$); 
    \node[scale=0.8, darkblue] at ($(L)!0.4!(L1) - 2*(shift)$) {$+$};

    \draw[->, thin, darkgreen]  ($(L)!0.9!(L1) + (shift)$) -- ($(L)!0.6!(L1) + (shift)$); 
    \node[scale=0.6, darkgreen] at ($(L)!0.9!(L1) + (shift) - (shifty)$) {$-$};

    \draw[->, thin, darkblue]  ($(L)!0.6!(L2) - (shift)$) -- ($(L)!0.9!(L2) - (shift)$); 
    \node[scale=0.6,darkblue ] at ($(L)!0.9!(L2) - (shift) - (shifty)$) {$+$};

    \draw[->, thin, darkgreen]  ($(L)!0.8!(L2) + (shift)$) -- ($(L)!0.2!(L2) + (shift)$); 
    \node[scale=0.8,darkgreen ] at ($(L)!0.6!(L2) + 2*(shift)$) {$-$};

    \draw[->, thin, darkgreen]  ($(T)!0.9!(L) + (shift)$) -- ($(T)!0.6!(L) + (shift)$); 
    \node[scale=0.6, darkgreen] at ($(T)!0.9!(L) + (shift) - (shifty)$) {$-$};

    \draw[->, thin, darkblue]  ($(T)!0.6!(R) - (shift)$) -- ($(T)!0.9!(R) - (shift)$); 
    \node[scale=0.6,darkblue ] at ($(T)!0.9!(R) - (shift) - (shifty)$) {$+$};

    \draw[->, thin,darkgreen]  ($(T)!0.8!(R) + (shift)$) -- ($(T)!0.2!(R) + (shift)$);   
    \node[scale=0.8, darkgreen] at ($(T)!0.4!(R) + 2*(shift)$) {$-$};

  \end{tikzpicture}%
}
\usepackage{tikz}
\usetikzlibrary{arrows.meta}
\tikzset{
  fermion/.style={draw=black, line width=0.9pt, -{Stealth[length=2pt,width=2pt]}},
  scalar/.style={draw=red,   line width=0.9pt, line cap=round}
}

\usepackage{tikz}
\usetikzlibrary{arrows.meta,decorations.markings}

\tikzset{
  fline/.style={
    draw=black, line width=1pt,
    postaction={decorate},
    decoration={markings,
      mark=at position 0.17 with {\arrow{Stealth[length=4pt,width=4.2pt]}},
      mark=at position 0.50 with {\arrow{Stealth[length=4pt,width=4.2pt]}},
      mark=at position 0.83 with {\arrow{Stealth[length=4pt,width=4.2pt]}}}},
  sline/.style={draw=red, line width=1pt, line cap=round}
}

\usetikzlibrary{arrows.meta,decorations.markings}

\usepackage{tikz}
\usetikzlibrary{arrows.meta,decorations.markings}

\tikzset{
  fseg/.style={draw=black, line width=1.2pt, line cap=round,
    postaction={decorate},
    decoration={markings,
      mark=at position 0.5 with {\arrow{Stealth[length=5.5pt,width=5.8pt]}}}},
  sarch/.style={draw=red, line width=1.2pt, line cap=round}
}

\tikzset{
  fplain/.style={draw=black, line width=1.2pt, line cap=round},
  farrow/.style={draw=black, line width=1.2pt, line cap=round,
                 -{Stealth[length=5.5pt,width=5.8pt]}},
  sarch/.style={draw=red, line width=1.2pt, line cap=round}
}

\tikzfeynmanset{
    midarrow/.style={
      /tikz/postaction={decorate},
      /tikz/decoration={
        markings,
        mark=at position #1 with {\arrow{Latex[length=3pt, width=5pt]}}
      }
    },
    midarrow/.default=0.6
}
  
\newcommand{\bubbleA}{%
  \vcenter{\hbox{%
\begin{tikzpicture}[baseline=(current bounding box.center), scale=0.6, line width=1.2pt]
  \begin{feynman}
    \vertex (L) at (-1.5,0);
    \vertex (A) at (-0.7,0);
    \vertex (B) at (0.7,0);
    \vertex (R) at (1.5,0);
    \diagram*{
      (L) -- [plain, midarrow=0.8] (A)
          -- [plain, midarrow] (B)
          -- [plain, midarrow=0.8] (R),
      (A) -- [plain, half left, looseness=1.5, draw=red] (B),
    };
  \end{feynman}
\end{tikzpicture}
  }}%
}

\newcommand{\doublebubbleA}{%
  \vcenter{\hbox{%
   \begin{tikzpicture}[baseline=(current bounding box.center), scale=0.6, line width=1.2pt]
  \begin{feynman}
   \vertex (L) at (-1.5,0);
    \vertex (A) at (-0.7,0);
    \vertex (A2) at (-0.3,0);
    \vertex (B2) at (0.3,0);
    \vertex (B) at (0.7,0);
    \vertex (R) at (1.5,0);
    \diagram*{
      (L) 
      -- [plain, midarrow=0.8] (A)
      -- [plain, midarrow=0.8] (A2) 
      -- [plain, midarrow=0.8] (B2) 
      -- [plain, midarrow=0.8] (B) 
      -- [plain, midarrow=0.8] (R),
      (A) -- [plain, half left, looseness=1.5, draw=red] (B),
      (A2) -- [plain, half left, draw=red] (B2),
    };
  \end{feynman}
\end{tikzpicture}
  }}%
}

\newcommand{\chainA}{%
  \vcenter{\hbox{%
   \begin{tikzpicture}[baseline=(current bounding box.center), scale=0.6, line width=1.2pt]
  \begin{feynman}
    \vertex (L) at (-1.5,0);
    \vertex (A1) at (-0.9,0);
    \vertex (A2) at (-0.6,0);
    \vertex (A3) at (-0.15,0);
    \vertex (B3) at (0.15,0);
    \vertex (B2) at (0.6,0);
    \vertex (B1) at (0.9,0);
    \vertex (R) at (1.5,0);
    \diagram*{
      (L) 
      -- [plain, midarrow=0.8] (A1)
      -- [plain, midarrow=0.9] (A2) 
      -- [plain, midarrow=0.8] (A3) 
      -- [plain, midarrow=0.8] (B3) 
      -- [plain, midarrow=0.8] (B2) 
      -- [plain, midarrow=0.9] (B1) 
      -- [plain, midarrow=0.7] (R),
      (A1) -- [plain,  half left,  looseness=1.5, draw=red] (B1),
      (A2) -- [plain, half left,  looseness=2.0,  draw=red] (A3),
      (B3) -- [plain, half left,  looseness=2.0, draw=red] (B2),
    };
  \end{feynman}
\end{tikzpicture}
  }}%
}
\newcommand{\rainbowA}{%
  \vcenter{\hbox{%
   \begin{tikzpicture}[baseline=(current bounding box.center), scale=0.6, line width=1.2pt]
  \begin{feynman}
    \vertex (L) at (-1.5,0);
    \vertex (A1) at (-1.0,0);
    \vertex (A2) at (-0.6,0);
    \vertex (A3) at (-0.2,0);
    \vertex (B3) at (0.2,0);
    \vertex (B2) at (0.6,0);
    \vertex (B1) at (1.0,0);
    \vertex (R) at (1.5,0);
    \diagram*{
      (L) 
      -- [plain, midarrow=0.8] (A1) 
      -- [plain, midarrow=0.8] (A2) 
      -- [plain, midarrow=0.8] (A3) 
      -- [plain, midarrow=0.8] (B3) 
      -- [plain, midarrow=0.8] (B2) 
      -- [plain, midarrow=0.8] (B1) 
      -- [plain, midarrow=0.8] (R),
      (A1) -- [plain,  half left,  looseness=1.5, draw=red] (B1),
      (A2) -- [plain, half left,  looseness=1.5,  draw=red] (B2),
      (A3) -- [plain, half left,  looseness=2.0, draw=red] (B3),
    };
  \end{feynman}
\end{tikzpicture}
  }}%
}

%% file: BorelAppendix.tex
\section{Borel and Borel-Leroy transforms \label{app:Borel}}
A formal series $f(a)$
is related to its Borel transform $\cB(t)$ by
\begin{equation}
f(a) = \sum_{n=0}^\infty c_n a^n = \frac{1}{a}\int_0^\infty dt e^{-t/a}\,\cB(t),
\qquad
\cB(t)=\sum_{n\ge0}\frac{c_n}{n!}\,t^n.
\end{equation}
If the coefficients of the series in $f(a)$ behave asymptotically at large $n$ as
\begin{equation}
    c_n \sim C\,n^{-r} A^{\,n}\,n!.
\end{equation}
then the Borel transform behaves asymptotically like
\begin{equation}
\cB(t)\sim \sum_{n=0}^\infty \frac{c_n}{n!} t^n = C \sum_{n =0}^\infty (A t)^n n^{-r}
= C\,\Li_{r}(A t).\label{Borelexample}
\end{equation}
The polylogarithm $\Li_r(At)$ 
has a singularity at $t_0=1/A$ for any $r \in \mathbb{R}$. It is a branch point for all $r \in \mathbb{R}$ except for non-positive integers $r=0,-1,-2,\cdots$ where the singularity is a pole.
To be explicit, the form of the expansion near the singularity 
as $\delta=1-At\to 0^+$ depends on $r$:
\begin{equation}
\Li_{r}(1-\delta)\sim
\begin{cases}
\displaystyle
\Gamma(1-r)\,\delta^{\,r-1}\;+\;\text{(less singular and/or analytic in $\delta$)}, 
& r\notin \mathbb{Z},\\[8pt]
\displaystyle
\frac{(-1)^{r}}{(r-1)!}\,\delta^{\,r-1}\log\delta\;+\;\text{(analytic in $\delta$)}, 
& r= 1,2,3,\dots,\\[10pt]
\displaystyle
\frac{(-r)!}{\delta^{\,1-r}}\;+\;\text{(less singular terms)}, 
& r = 0,-1,-2,\dots.
\end{cases}
\end{equation}
We will generally call this behavior a Borel singularity of strength $r$.
Thus the Borel transform of an asymptotic series as in Eq.~\eqref{Borelexample} always has a singularity at $t=1/A$ but its strength depends on subleading effects (the $n^{-r} \ll A^n$ term in the series).
Note that if $r>1$ then $\Li_r(1-\delta)$ is finite as $\delta\to 0$ but the singularity shows up after some number of derivatives.

A Borel-Leroy transform of index $p$ is defined as
\begin{equation}
f(a)=a^{-p}\int_0^\infty dt e^{-t/a}\,t^{p-1}\,
\cB_p(t), \qquad
\cB_p(t)=\sum_{n\ge0}\frac{c_n}{\Gamma(n+p)}\,t^n,
\end{equation}
the same ansatz $c_n \sim C\,A^{\,n}\,n!\,n^{-r}$ gives, using $\Gamma(n+p)\sim n^{\,p-1}n!$, 
\begin{equation}
\cB_p(t)\sim C\sum_{n\ge0}(A t)^n n^{-(p+r-1)}
= C\,\mathrm{Li}_{r+p-1}(A t).
\end{equation}
again with $t_0=1/A$. Thus the Borel-Leroy
transform has the same singularity location $t_0=1/A$ as the Borel transform but of strength $r+p-1$ instead of $r$. For $p=1$ the two agree. 